\documentclass[jmp,amsmath,amssymb]{revtex4-2}
\usepackage[T1]{fontenc}
\usepackage[latin9]{inputenc}
\usepackage{color}
\usepackage{mathrsfs}
\usepackage{amsmath}
\usepackage{amsthm}
\usepackage{amssymb}
\usepackage{graphicx}
\usepackage[unicode=true,
 bookmarks=true,bookmarksnumbered=true,bookmarksopen=true,bookmarksopenlevel=3,
 breaklinks=false,pdfborder={0 0 1},backref=section,colorlinks=true]
 {hyperref}
\hypersetup{pdftitle={Your Title},
 pdfauthor={Your Name},
 pdfpagelayout=OneColumn, pdfnewwindow=true, pdfstartview=XYZ, plainpages=false,hypertex,bookmarksnumbered=true,backref=true}

\makeatletter

\providecommand{\tabularnewline}{\\}

\numberwithin{figure}{section}
\numberwithin{equation}{section}
\numberwithin{table}{section}
\theoremstyle{plain}
\newtheorem{thm}{\protect\theoremname}
\theoremstyle{remark}
\newtheorem{rem}[thm]{\protect\remarkname}

\makeatother

\providecommand{\remarkname}{Remark}
\providecommand{\theoremname}{Theorem}

\begin{document}
\title{Factorized form of the dispersion relations of a traveling wave tube}
\author{Alexander Figotin}
\email{afigotin@uci.edu}

\affiliation{Department of Mathematics, University of California, Irvine, CA 92697,
USA}
\begin{abstract}
The traveling tube (TWT) design in a nutshell comprises of a pencil-like
electron beam (e-beam) in vacuum interacting with guiding it slow-wave
structure (SWS). In our prior studies the e-beam was represented by
one-dimensional electron flow and SWS was represented by a transmission
line (TL). We extend in this paper our previously constructed field
theory for TWTs as well the celebrated Pierce theory by replacing
there the standard transmission line (TL) with its generalization
allowing for the low frequency cutoff. Both the standard TL and generalized
transmission line (GTL) feature uniformly distributed shunt capacitance
and serial inductance, but the GTL in addition to that has uniformly
distributed serial capacitance. We remind the reader that the standard
TL represents a waveguide operating at the so-called TEM mode with
no low frequency cutoff. In contrast, the GTL represents a waveguide
operating at the so-called TM mode featuring the low frequency cutoff.
We develop all the details of the extended TWT field theory and using
a particular choice of the TWT parameters we derive a physically appealing
factorized form of the TWT dispersion relations. This form has two
factors that represent exactly the dispersion functions of non-interacting
GTL and the e-beam. We also find that the factorized dispersion relations
comes with a number of interesting features including: (i) focus points
that belong to each dispersion curve as TWT principle parameter varies;
(ii) formation of ``hybrid'' branches of the TWT dispersion curves
parts of which can be traced to non-interacting GTL and the e-beam.
\end{abstract}
\maketitle

\section{Introduction}

\begin{figure}[h]
\centering{}\includegraphics[scale=0.5]{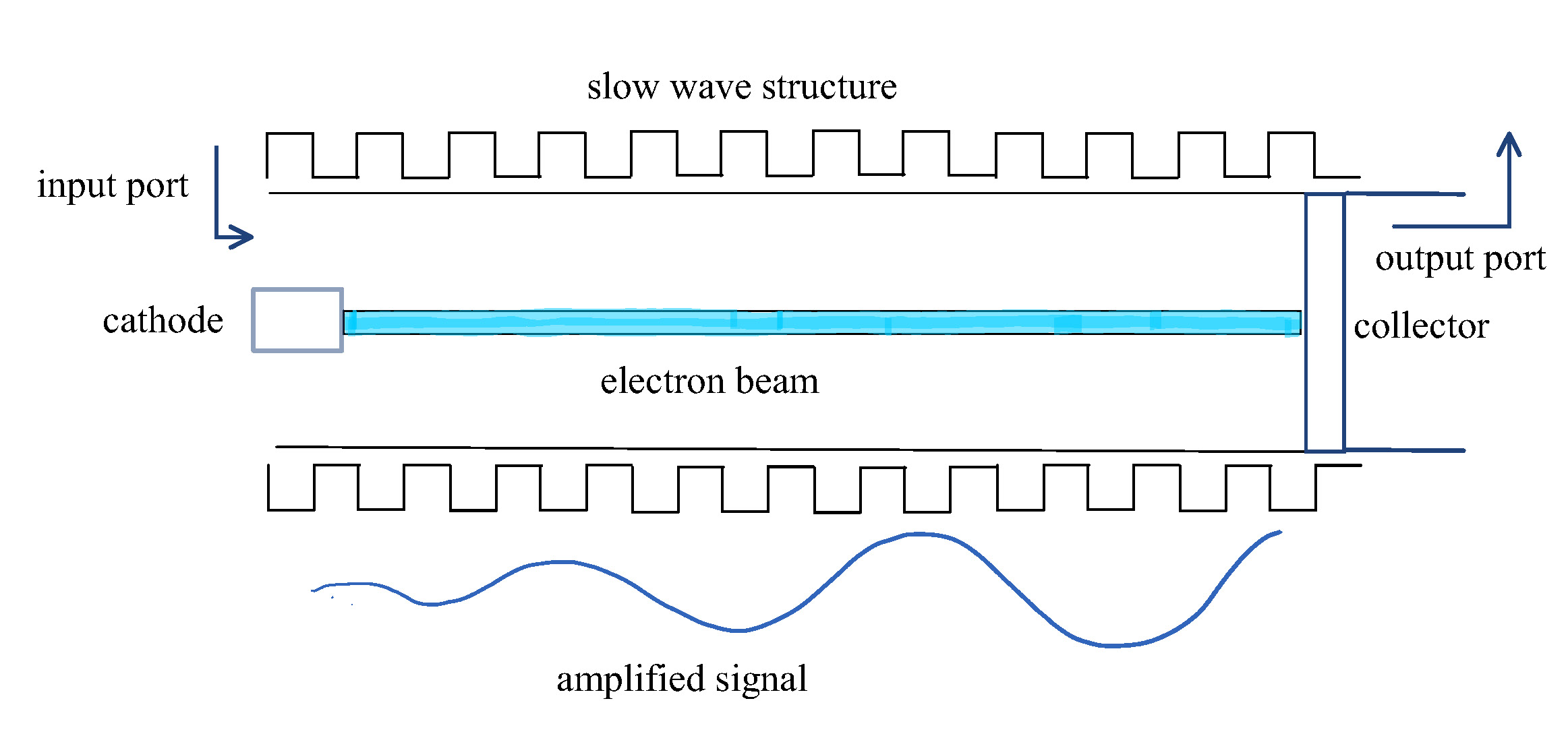}\caption{\label{fig:TWT1} The upper picture is a schematic presentation of
a traveling wave tube. The lower picture shows an RF perturbation
in the form of a space-charge wave that gets amplified exponentially
as it propagates through the traveling wave tube.}
\end{figure}

We first review concisely the basics of traveling wave tubes. Traveling
wave tube (TWT) utilizes the energy of the pencil-like electron beam
(e-beam) as a flow of free electrons in a vacuum and converts it into
an RF signal, see Fig. \ref{fig:TWT1}. To facilitate the energy conversion
and the signal amplification, the electron beam is enclosed in the
so-called \emph{slow wave structure} (SWS), which supports waves that
are slow enough to effectively interact with the e-beam. As a result
of this interaction, the kinetic energy of electrons is converted
into the electromagnetic energy stored in the field, \citep{Gilm1},
\citet{Tsimring}, \citet[Sec. 2.2]{BBLN}, \citet[Sec. 4]{SchaB}.
Consequently, the \emph{key operational principle of a TWT is a positive
feedback interaction between the slow-wave structure and the flow
of electrons}. The physical mechanism of radiation generation and
its amplification is the electron bunching caused by the acceleration
and deceleration of electrons along the e-beam. 

A schematic sketch of typical TWT is shown in Fig. \ref{fig:TWT1}.
Such a typical TWT consists of a vacuum tube containing an e-beam
that passes down the middle of an SWS such as the so-called RF circuit.
It operates as follows. The left end of the RF circuit is fed with
a low-powered RF signal to be amplified. The SWS electromagnetic field
acts upon the e-beam causing electron bunching and the formation of
the so-called \emph{space-charge wave}. In turn, the electromagnetic
field generated by the space charge wave induces more current back
into the RF circuit with a consequent enhancement of electron bunching.
As a result, the EM field is amplified as the RF signal passes down
the structure until a saturation regime is reached and a large RF
signal is collected at the output. The role of the SWS is to provide
slow-wave modes to match up with the velocity of the electrons in
the e-beam. This velocity is usually a fraction of the speed of light.
Importantly, synchronism is required for an effective in-phase interaction
between the SWS and the e-beam with optimal extraction of the kinetic
energy of the electrons. A typical simple SWS is the helix, which
reduces the speed of propagation according to its pitch. The TWT is
designed so that the RF signal travels along the tube at nearly the
same speed as electrons in the e-beam to facilitate effective coupling.
A big picture on the flow of electrons in TWT is that they move with
nearly constant velocity and loose some of their kinetic energy to
the EM wave. This kind of energy transfer is observed in the Cherenkov
radiation phenomenon that can be viewed then as a physical foundation
for the convection instability and consequent RF signal amplification,
\citep[Sec. 1.1-1.2, 4.4, 4.8-4.9, 7.3, 7.6; Chap. 8]{BenSweSch}.

Technical details on the designs and operation of TWTs can be found
in \citep{Gilm1}, \citet[Sec. 4]{BBLN} \citep{PierTWT}, \citet{Tsimring}.
As for a rich and interesting history of traveling wave tubes, we
refer the reader to \citep{MAEAD} and references therein.

An effective mathematical model for a TWT interacting with the e-beam
was introduced by Pierce \citet[Sec. I]{Pier51}, \citep{PierTWT}.
The celebrated Pierce model is one-dimensional; it accounts for the
wave amplification, energy extraction from the e-beam and its conversion
into microwave radiation in the TWT \citep{Gilm1}, \citep{Gilm},
\citet[Sec. 4]{BBLN}, \citet[Sec. 4]{SchaB}, \citet{Tsimring}.
This model captures remarkably well significant features of the wave
amplification and the beam-wave energy transfer, and is still used
for basic design estimates. The Pierce theory assumes: (i) an idealized
one-dimensional flow of electrons responding linearly to perturbations;
(ii) a lossless transmission line (TL) representing the relevant eigenmode
of the SWS that interacts with the e-beam; (iii) the TL is assumed
to be spatially homogeneous with uniformly distributed shunt capacitance
and serial inductance. In our paper \citet{FR}, we have constructed
a Lagrangian field theory by generalizing and extending the Pierce
theory to the case of a possibly inhomogeneous MTL coupled to the
e-beam. This work was extended to an analytic theory of multi-stream
electron beams in traveling wave tubes in \citep{FigTWTbk}.

In this paper we extend first our previously constructed field theory
for TWTs as well the celebrated Pierce theory by replacing there the
standard transmission line (TL) with its generalization allowing for
the low frequency cutoff. Both the standard TL and generalized transmission
line (GTL) feature uniformly distributed shunt capacitance and serial
inductance but the GTL in addition to that has uniformly distributed
serial capacitance. We remind the reader that the standard TL represents
a waveguide operating at the so-called TEM mode with no low frequency
cutoff. In contrast, the GTL represents a waveguide operating at the
so-called TM mode featuring the low frequency cutoff.

Second, using a particular choice of the TWT parameters we derive
a physically appealing factorized form of the TWT dispersion relations.
This form has two factors that represent exactly the dispersion functions
of non-interacting GTL and the e-beam. We also find that the factorized
dispersion relations imply a number of interesting features (i) focus
points that belong to different dispersion curves as TWT principle
parameter varies; (ii) formation of ``hybrid'' branches of the TWT
dispersion curves parts of which can be traced to non-interacting
GTL and the e-beam.

The paper is organized as follows. In Section \ref{sec:mainfac} we
review the main results of this paper. In Section \ref{sec:twt-mod}
we introduce the GTL and construct the Lagrangian framework the extended
analytic model for traveling wage tubes including the Euler-Lagrange
field equations. In Section \ref{sec:disp-fac} we introduce the factorized
form of the dispersion relations and discuss its structure. In Section
\ref{sec:dim-setup} we introduce ``natural'' units for our TWT
theory and convert all important equations in their dimensionless
form. Section \ref{sec:disp-dom} is devoted to: (i) finding of special
points as well so-called ``cross-points'' (at which both dispersion
relations for the GTL and the e-beam are satisfied) for the TWT dispersion
relations and the graphs the TWT dispersion relations; (ii) identification
of a specific domain containing its graph based on the factorized
of the TWT dispersion relations. In Section \ref{sec:disp-curv} we
show an exhibit of plots of the TWT dispersion curves for different
values of the TWT principle parameter $\gamma$ demonstrating different
topological patterns that depend on TWT parameters $\rho$ and $\chi$.
In Section \ref{sec:disp-instab} we (i) consider a ``big'' picture
of the TWT instabilities; (ii) introduce the concept of dispersion-instability
graph and show a number of such dispersion-instability graphs demonstrating
their qualitative dependence on the TWT main parameters $\gamma$,
$\rho$ and $\chi$. In Section \ref{sec:facImk} we consider the
plots of the imaginary part $\Im\left\{ k\left(\omega\right)\right\} $
of the frequency dependent dependent wavenumber $k\left(\omega\right)$
that satisfies the TWT dispersion relations. In Section \ref{sec:crospomod}
we introduce and study the ``cross-point model'' for the factorized
dispersion relation defining a cross-point as a point $\left(k,\omega\right)$
at which the dispersion relations for the GTL and the e-beam hold
simultaneously. This model is designed to demonstrate qualitative
features of the TWT dispersion curve near the cross-points in the
case of small coupling, that is when $\gamma\ll1$. Finally, in Appendix
we collect some additional information needed for out theoretical
arguments.

Whenever possible we use figures to visualize features of our analytical
developments.

\section{Review of main results\label{sec:mainfac}}

Leaving the detailed formulation the TWT theory including its variational
Lagrangian framework to the following sections we review here concisely
our main results.

The list of primary TWT parameters needed includes: (i) the GTL phase
velocity $w$ in the high frequency limit; (ii) the GTL low cutoff
frequency $\omega_{\mathrm{c}}$; (iii) the e-beam stationary velocity
$\mathring{v}$; (iv) the so-called \emph{reduced plasma frequency}
$\omega_{\mathrm{rp}}$; (v) the \emph{TWT principal parameter} $\gamma$
defined by, \citep[4, 24]{FigTWTbk} 
\begin{equation}
\gamma=\frac{b^{2}}{C}\frac{\sigma_{\mathrm{B}}}{4\pi}\omega_{\mathrm{rp}}^{2},\label{eq:mainfac1a}
\end{equation}
where $b$ is a dimensionless phenomenological coupling constant,
and $C$ is the GTL shunt capacitance per of length and $\sigma_{\mathrm{B}}$
is the area of the cross-section of the e-beam. As formula (\ref{eq:mainfac1a})
indicates the TWT principal parameter $\gamma$ is an integral TWT
parameter. Further analysis shows that it appears naturally in factorized
form of the TWT dispersion relations as a coupling parameter, and
that can be explained by formula (\ref{eq:mainfac1a}) showing that
$\gamma$ is proportional to the square of the original coupling parameter
$b$.
\begin{figure}[h]
\begin{centering}
\hspace{-0.1cm}\includegraphics[scale=0.3]{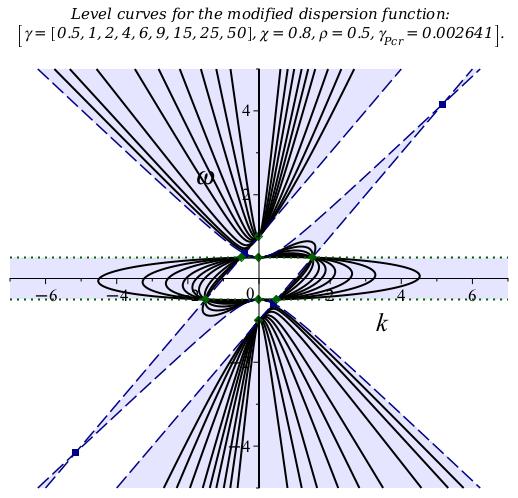}\hspace{0.1cm}\includegraphics[scale=0.3]{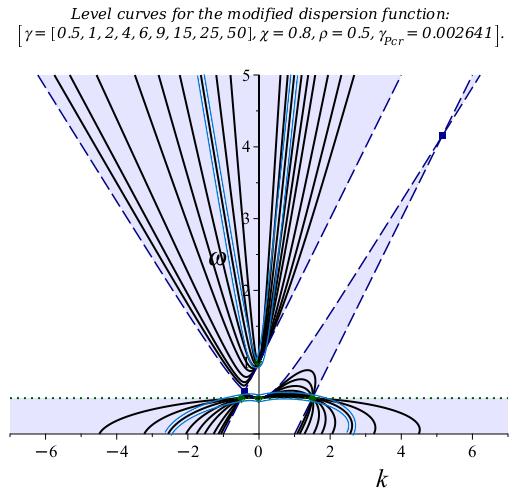}\hspace{0.1cm}\includegraphics[scale=0.3]{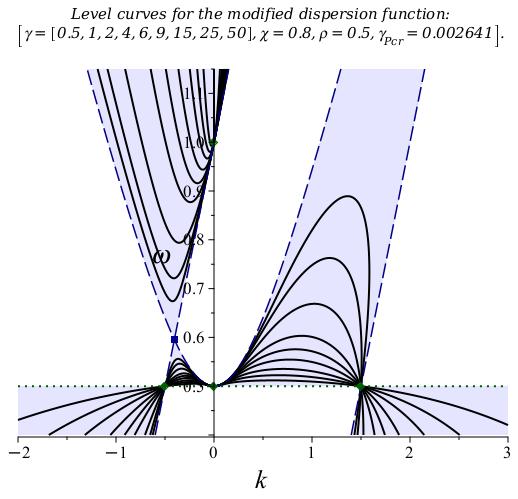}
\par\end{centering}
\centering{}(a)\hspace{5.5cm}(b)\hspace{5.5cm}(c)\caption{\label{fig:dis-Lev1} Dispersion curves of the TWT dispersion relations
(\ref{eq:mainfac2a}), (\ref{eq:mainfac2d}) for $\chi=0.8$, $\rho=0.5<1$
and $\gamma=0.5,1,2,4,6,9,15,25,50$, $\gamma_{\mathrm{Pcr}}\protect\cong0.002641$:
(a) complete plot in the designated window; (b) a zoomed fragment
of (a); (c) a zoomed fragment of (b). Solid (black) curves represent
the TWT dispersion curves for indicated values of $\gamma$; dashed
(blue) curves represent the dispersion curves $\mathrm{Gr}_{\mathrm{T}}$
and $\mathrm{Gr}_{\mathrm{B}}$ of non-interacting GTL and the e-beam.
Shaded area identifies the dispersion domain $\mathbb{D}_{\mathrm{TB}}$,
that is where $R_{\mathrm{TB}}\left(k,\omega\right)>0$. Doted (blue)
horizontal straight lines represents points $\left(k,\omega_{\mathrm{c}}\right)$.
Note the dispersion curves $\mathrm{Gr}_{\mathrm{TB}}\left(\gamma\right)$
pass through focal points defined by equations (\ref{eq:mainfac2e})
marked as circle (green) dots. Square (blue) dots mark the TWT cross-points
(see equations (\ref{eq:disGTBr1b}), (\ref{eq:disGTBr1c})). Note
also that the smaller $\gamma$ gets the closer graph $\mathrm{Gr}_{\mathrm{TB}}\left(\gamma\right)$
gets to $\mathrm{Gr}_{\mathrm{TB}}\left(0\right)=\mathrm{Gr}_{\mathrm{T}}\bigcup\mathrm{Gr}_{\mathrm{B}}$,
whereas the larger $\gamma$ gets the closer graph $\mathrm{Gr}_{\mathrm{TB}}\left(\gamma\right)$
gets to $\mathrm{Gr}_{\mathrm{TB}}\left(\infty\right)$ defined by
equation (\ref{eq:disdom1f}).}
\end{figure}
\begin{figure}[h]
\begin{centering}
\hspace{-0.1cm}\includegraphics[scale=0.3]{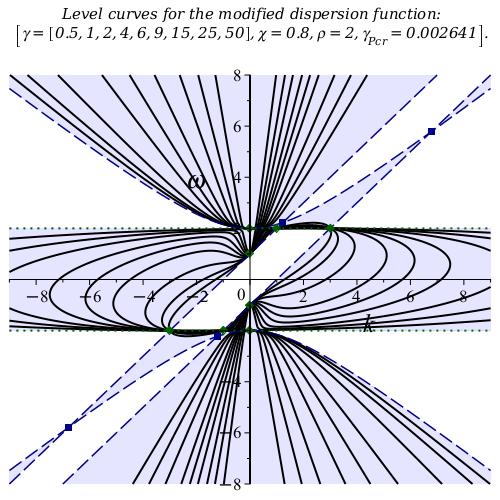}\hspace{0.1cm}\includegraphics[scale=0.3]{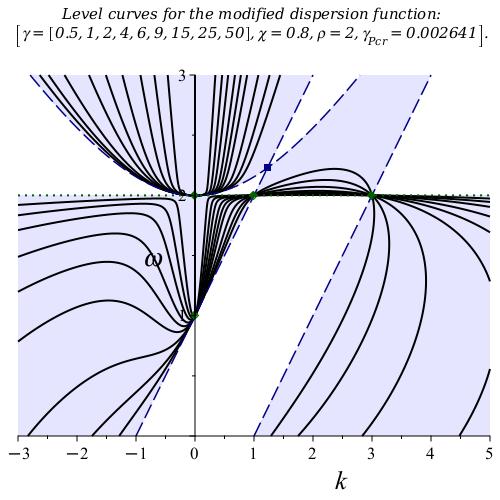}\hspace{0.1cm}\includegraphics[scale=0.3]{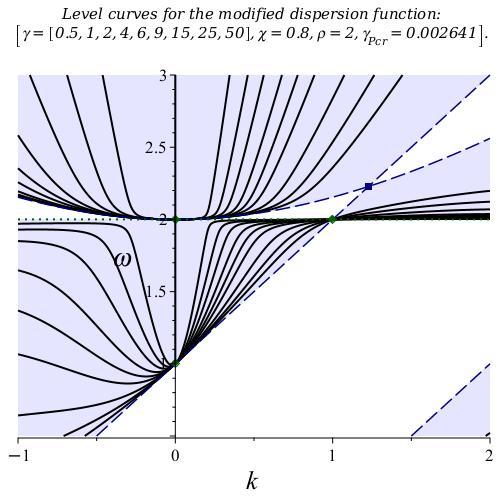}
\par\end{centering}
\centering{}(a)\hspace{5cm}(b)\hspace{5cm}(c)\caption{\label{fig:dis-Lev1L} Dispersion curves of the TWT dispersion relations
(\ref{eq:mainfac2a}), (\ref{eq:mainfac2d}) for $\chi=0.8$, $\rho=2>1$
and $\gamma=0.5,1,2,4,6,9,15,25,50$, $\gamma_{\mathrm{Pcr}}\protect\cong0.002641$:
(a) complete plot in the designated window; (b) a zoomed fragment
of (a); (c) a zoomed fragment of (b). Solid (black) curves represent
the TWT dispersion curves for indicated values of $\gamma$; dashed
(blue) curves represent the dispersion curves of non-interacting TL
and the e-beam. Shaded area identifies the dispersion domain $\mathbb{D}_{\mathrm{TB}}$,
that is where $R_{\mathrm{TB}}\left(k,\omega\right)>0$. Doted (blue)
horizontal straight lines represents points $\left(k,\omega_{\mathrm{c}}\right)$.
Note the dispersion curves $\mathrm{Gr}_{\mathrm{TB}}\left(\gamma\right)$
pass through focal points defined by equations (\ref{eq:mainfac2e})
marked as circle (green) dots. Square (blue) dots mark the TWT cross-points
(see equations (\ref{eq:disGTBr1b}), (\ref{eq:disGTBr1c})). Note
also that the smaller $\gamma$ gets the closer graph $\mathrm{Gr}_{\mathrm{TB}}\left(\gamma\right)$
gets to $\mathrm{Gr}_{\mathrm{TB}}\left(0\right)=\mathrm{Gr}_{\mathrm{T}}\bigcup\mathrm{Gr}_{\mathrm{B}}$,
whereas the larger $\gamma$ gets the closer graph $\mathrm{Gr}_{\mathrm{TB}}\left(\gamma\right)$
gets to $\mathrm{Gr}_{\mathrm{TB}}\left(\infty\right)$ defined by
equation (\ref{eq:disdom1f}).}
\end{figure}

In order to facilitate a dimensionless version of our theory we use
the following three dimensionless parameters
\begin{equation}
\chi=\frac{w}{\mathring{v}},\quad\rho=\frac{\omega_{\mathrm{c}}}{\omega_{\mathrm{rp}}},\quad\gamma^{\prime}=\frac{\gamma}{\mathring{v}^{2}}.\label{eq:mainfac1b}
\end{equation}
\emph{We use consistently throughout this section the dimensionless
settings omitting ``prime'' for $\gamma^{\prime}$ to have less
cluttered equations.}

The TWT system configuration is described by quantities $q\left(z,t\right)$
and $Q\left(z,t\right)$\emph{ }which are position and time dependent
charges associated with the e-beam and the GTL. They are defined as
time integrals of the corresponding e-beam and GTL currents. We setup
in Section \ref{sec:twt-mod} the TWT Lagrangian and obtain the Euler-Lagrange
equations for $q\left(z,t\right)$ and $Q\left(z,t\right)$. We consider
then the TWT system eigenmodes represented as follows:
\begin{equation}
Q\left(z,t\right)=\hat{Q}\left(k,\omega\right)\mathrm{e}^{-\mathrm{i}\left(\omega t-kz\right)},\quad q\left(z,t\right)=\hat{q}\left(k,\omega\right)\mathrm{e}^{-\mathrm{i}\left(\omega t-kz\right)}.\label{eq:mainfac1c}
\end{equation}
where $\omega$ and $k=k\left(\omega\right)$ are the frequency and
the wavenumber, respectively. The dependence of $k\left(\omega\right)$
on the frequency $\omega$ can be found from the TWT dispersion relations
analyzed in Section \ref{sec:disp-fac}.

The Fourier transformation (see Appendix \ref{sec:four}) in time
$t$ and space variable $z$ of the TWT Euler-Lagrange equations can
be written in the following matrix form:

\begin{equation}
M_{k\omega}x=0,\quad M_{k\omega}=\left[\begin{array}{rr}
k^{2}+\frac{\rho^{2}-\omega^{2}}{\chi^{2}} & bk^{2}\\
bk^{2} & b^{2}\left[k^{2}-\frac{\left(\omega-k\right)^{2}-1}{\gamma}\right]
\end{array}\right],\quad x=\left[\begin{array}{r}
\hat{Q}\\
\hat{q}
\end{array}\right].\label{eq:mainfac1d}
\end{equation}
Note that equations (\ref{eq:mainfac1d}) can be viewed as an eigenvalue
type problem for $k$ and $x$ assuming that $\omega$ and other parameters
are fixed. The condition for equations (\ref{eq:mainfac1d}) to have
nontrivial nonzero solutions $x$ is $\det\left\{ M_{k\omega}\right\} =0$.
The later equation after algebraic transformation turns into

\begin{equation}
R_{\mathrm{TB}}\left(k,\omega\right)=\left[\frac{1}{k^{2}}-\frac{\chi^{2}}{\omega^{2}-\rho^{2}}\right]\left[\left(\omega-k\right)^{2}-1\right]=\gamma,\label{eq:mainfac2a}
\end{equation}
and we refer to it as the \emph{TWT dispersion relations}. The TWT
dispersion function $R_{\mathrm{TB}}\left(k,\omega\right)$ is evidently
a rational function which is the product of two factors, namely 
\begin{equation}
R_{\mathrm{T}}\left(k,\omega\right)\stackrel{\mathrm{def}}{=}\frac{1}{k^{2}}-\frac{\chi^{2}}{\omega^{2}-\rho^{2}},\quad R_{\mathrm{B}}\left(k,\omega\right)\stackrel{\mathrm{def}}{=}\left(\omega-k\right)^{2}-1,\label{eq:mainfa2b}
\end{equation}
where $R_{\mathrm{T}}\left(k,\omega\right)$ and $R_{\mathrm{B}}\left(k,\omega\right)$
are determined entirely respectively by the GTL and by the e-beam.
In addition to that, if we set in equation (\ref{eq:mainfac2a}) $\gamma=0$
we it turns into two equations:
\begin{equation}
R_{\mathrm{T}}\left(k,\omega\right)=\frac{1}{k^{2}}-\frac{\chi^{2}}{\omega^{2}-\rho^{2}}=0,\quad R_{\mathrm{B}}\left(k,\omega\right)=\left(\omega-k\right)^{2}-1,\label{eq:mainfac2c}
\end{equation}
which are evidently the dispersion relations of the GTL and the e-beam
respectively when they do not interact. Points $\left(k,\omega\right)$
that solve simultaneously equations (\ref{eq:mainfac2c}) are referred
to as\emph{ cross-points}, and the exact formulas for those points
are provided in Section \ref{subsec:crospo}. \emph{The cross-points
are a commonly used in the TWT design since one might expect the interaction
between the GTL and the e-beam to be the strongest in a vicinity of
these points}.

\emph{In summary, the factorized TWT dispersion relations (\ref{eq:mainfac2a})
integrate naturally into it the dispersion relations of the non-interacting
GTL and the e-beam. The TWT dispersion curves can be viewed as the
``levels'' of the rational function $R_{\mathrm{TB}}\left(k,\omega\right)$
determined by equation $R_{\mathrm{TB}}\left(k,\omega\right)=\gamma$.}

The factorized TWT dispersion relations (\ref{eq:mainfac1d}) can
be readily recast into its polynomial form, that is

\begin{equation}
\left[\omega^{2}-\left(\chi^{2}k^{2}+\rho^{2}\right)\right]\left[\left(k-\omega\right)^{2}-1\right]=\gamma k^{2}\left(\omega^{2}-\rho^{2}\right),\label{eq:mainfac2d}
\end{equation}
The dispersion relations in the form (\ref{eq:mainfac2d}) are also
naturally factorized. 

The TWT dispersion relations (\ref{eq:mainfac2a}), (\ref{eq:mainfac2d})
are studied first for the case when the both $\omega$ and $k$ are
real, and that leads to the conventional dispersion curves. An important
consequence of the factorized form of the TWT dispersion relations
(\ref{eq:mainfac2a}), (\ref{eq:mainfac2d}) is that in generic case
$\rho\neq1$ there are exactly eight (or 10 if the multiplicity is
counted) points $\left(k,\omega\right)$ that always satisfy them:
\begin{gather}
\left(0,\rho\right),\quad\left(0,\rho\right),\quad\left(0,1\right),\quad\left(\rho-1,\rho\right),\quad\left(\rho+1,\rho\right),\label{eq:mainfac2e}\\
\left(0,-\rho\right),\quad\left(0,-\rho\right),\quad\left(0,-1\right),\quad\left(-\left(\rho-1\right),-\rho\right),\quad\left(-\left(\rho+1\right),-\rho\right).\nonumber 
\end{gather}
We refer to points (\ref{eq:mainfac2e}) as \emph{focal points} since
for any $\gamma>0$ the corresponding TWT dispersion curve passes
through these points, see Figures \ref{fig:dis-Lev1} and \ref{fig:dis-Lev1L}
and more in Section \ref{sec:disp-curv}.
\begin{figure}[h]
\begin{centering}
\hspace{-0.5cm}\includegraphics[scale=0.33]{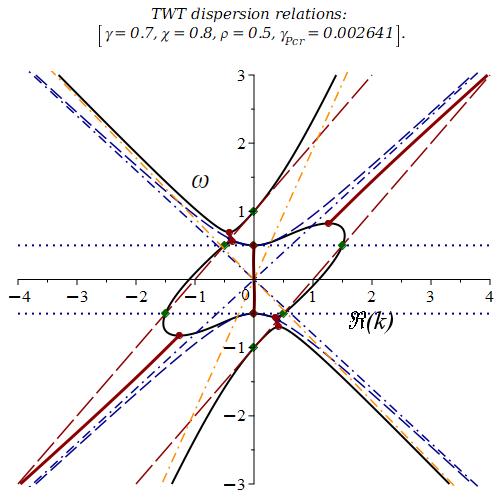}\hspace{1cm}\includegraphics[scale=0.33]{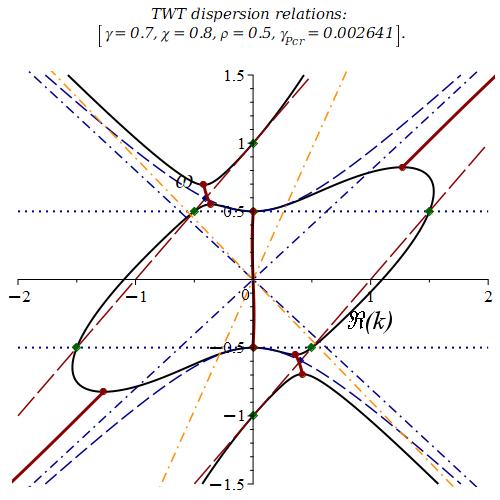}
\par\end{centering}
\centering{}\hspace{-1cm}(a)\hspace{7cm}(b)\caption{\label{fig:dis-instab1} Dispersion-instability graph for $\chi=0.8$,
$\rho=0.5<1$ and $\gamma=0.7>\gamma_{\mathrm{Pcr}}\protect\cong0.002641$:
(a) larger version; (b) zoomed fragment of (a). Solid (black) curves
represent the dispersion curves, dashed (blue) curves represent the
dispersion curves for $\gamma=0$ as a reference. Dash-doted straight
lines represented high frequency asymptotics for $\gamma=0.7$ (orange)
and for $\gamma=0$ (blue). Doted (blue) horizontal straight lines
represents points $\left(k,\pm\rho\right)$. Diamond (green) dots
represent focal points defined by equations (\ref{eq:mainfac2e}),
diamond (blue) dots represent the cross-points $\mathrm{Gr}_{\mathrm{T}}\bigcap\mathrm{Gr}_{\mathrm{B}}$.
The circular dots (red) identify transition to instability points
(see Section \ref{subsec:trans-instab}). The bold, solid (red) curves
represent branches of points $\left(\Re\left\{ k\right\} ,\omega\right)$
with real $\omega$ and $\Im\left\{ k\right\} \protect\neq0$ which
are points of the convection instability (see Section \ref{subsec:trans-instab}).}
\end{figure}
\begin{figure}[h]
\begin{centering}
\hspace{-2cm}\includegraphics[scale=0.33]{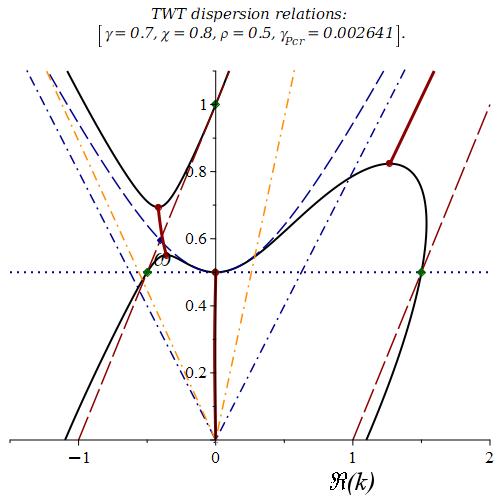}\hspace{1cm}\includegraphics[scale=0.33]{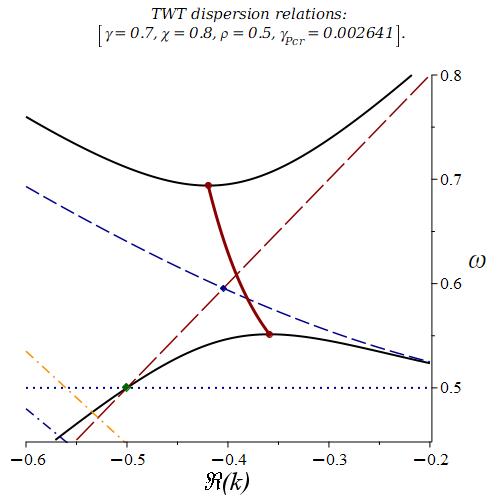}
\par\end{centering}
\centering{}\hspace{-2cm}(a)\hspace{7cm}(b)\caption{\label{fig:dis-instab1f} Zoomed fragments of the dispersion-instability
graph in Fig. \ref{fig:dis-instab1} for $\chi=0.8$, $\rho=0.5<1$
and $\gamma=0.7>\gamma_{\mathrm{Pcr}}\protect\cong0.002641$: (a)
zoomed fragment of Fig. \ref{fig:dis-instab1}(b); (b) zoomed fragment
of (a). Solid (black) curves represent the dispersion curves, dashed
(blue) curves represent the dispersion curves for $\gamma=0$ as a
reference. Dash-doted straight lines represented high frequency asymptotics
for $\gamma=0.7$ (orange) and for $\gamma=0$ (blue). Doted (blue)
horizontal straight lines represents points $\left(k,\pm\rho\right)$.
Diamond (green) dots represent focal points defined by equations (\ref{eq:mainfac2e}),
diamond (blue) dots represent the cross-points $\mathrm{Gr}_{\mathrm{T}}\bigcap\mathrm{Gr}_{\mathrm{B}}$.
The circular dots (red) identify transition to instability points
(see Section \ref{subsec:trans-instab}). The bold, solid (red) curves
represent branches of points $\left(\Re\left\{ k\right\} ,\omega\right)$
with real $\omega$ and $\Im\left\{ k\right\} \protect\neq0$ which
are points of the convection instability.}
\end{figure}

\begin{figure}[h]
\begin{centering}
\hspace{-0.5cm}\includegraphics[scale=0.33]{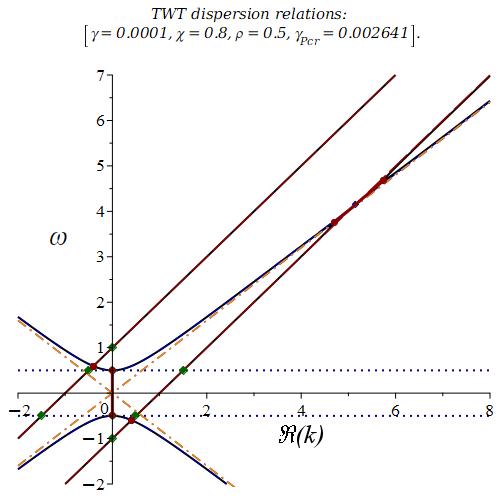}\hspace{1cm}\includegraphics[scale=0.33]{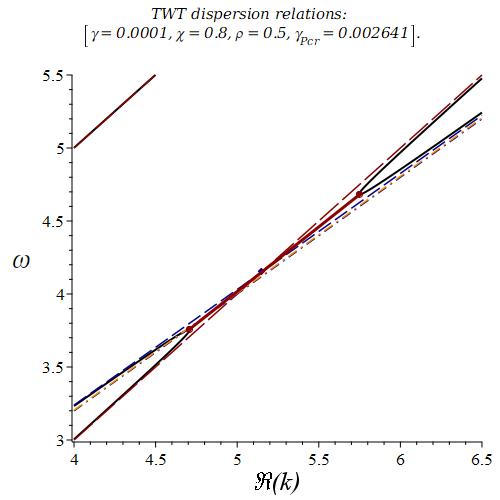}
\par\end{centering}
\centering{}\hspace{0.5cm}(a)\hspace{7cm}(b)\caption{\label{fig:dis-instab3} Dispersion-instability graph and its zoomed
fragment for $\chi=0.8$, $\rho=0.5<1$ and $\gamma=0.001<\gamma_{\mathrm{Pcr}}\protect\cong0.002641$:
(a) larger scale version; (b) zoomed fragment of (a). Solid (black)
curves represent the dispersion curves, dashed (blue) curves represent
the dispersion curves for $\gamma=0$ as a reference. Dash-doted straight
lines represented high frequency asymptotics for $\gamma=0.7$ (orange)
and for $\gamma=0$ (blue). Doted (blue) horizontal straight lines
represents points $\left(k,\pm\rho\right)$. Diamond (green) dots
represent focal points defined by equations (\ref{eq:disomfac2a}),
diamond (blue) dots represent the cross-points $\mathrm{Gr}_{\mathrm{T}}\bigcap\mathrm{Gr}_{\mathrm{B}}$.
The circular dots (red) identify transition to instability points
(see Section \ref{subsec:trans-instab}). The bold, solid (red) curves
represent branches of points $\left(\Re\left\{ k\right\} ,\omega\right)$
with real $\omega$ and $\Im\left\{ k\right\} \protect\neq0$ which
are points of the convection instability.}
\end{figure}

\begin{figure}[h]
\begin{centering}
\hspace{-0.1cm}\includegraphics[scale=0.32]{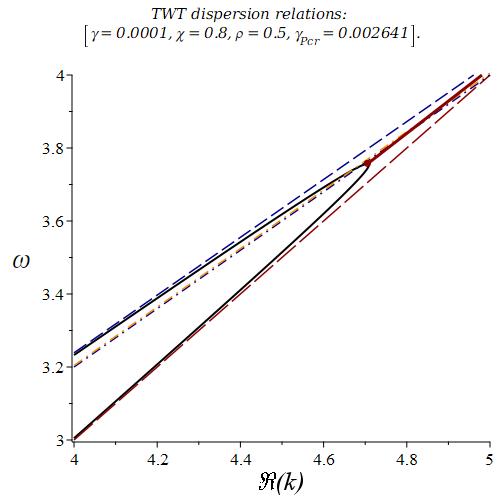}\hspace{0.1cm}\includegraphics[scale=0.32]{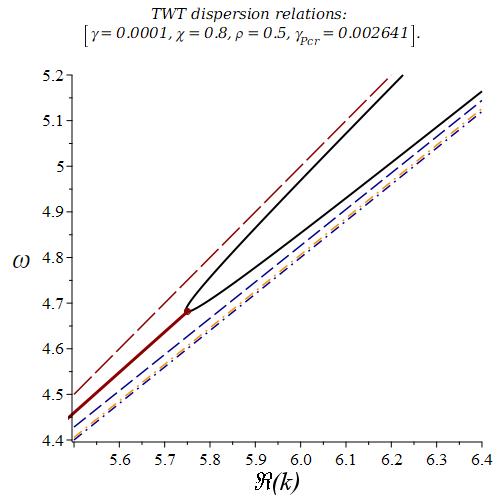}\hspace{0.1cm}\includegraphics[scale=0.32]{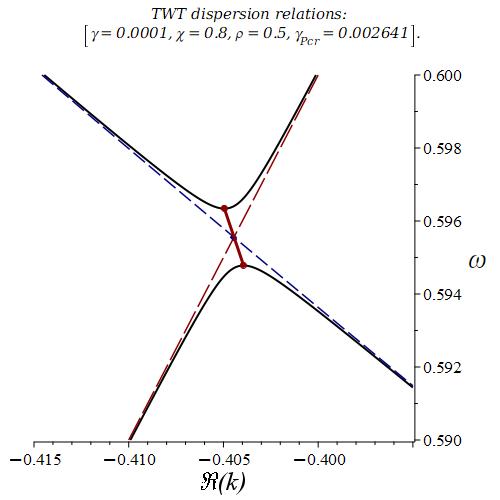}
\par\end{centering}
\centering{}(a)\hspace{5cm}(b)\hspace{5cm}(c)\caption{\label{fig:dis-instab3f} Zoomed fragments of the dispersion-instability
graph in Fig. \ref{fig:dis-instab3} for $\chi=0.8$, $\rho=0.5<1$
and $\gamma=0.001<\gamma_{\mathrm{Pcr}}\protect\cong0.002641$: (a)
zoomed fragment of Fig. \ref{fig:dis-instab3}(b); (b) another zoomed
fragment of Fig. \ref{fig:dis-instab3}(b); (c) zoomed fragment of
Fig. \ref{fig:dis-instab3}(a) for $\Re\left\{ k\right\} <0$. Solid
(black) curves represent the dispersion curves, dashed (blue) curves
represent the dispersion curves for $\gamma=0$ as a reference. Dash-doted
straight lines represented high frequency asymptotics for $\gamma=0.7$
(orange) and for $\gamma=0$ (blue). Doted (blue) horizontal straight
lines represents points $\left(k,\pm\rho\right)$. Diamond (green)
dots represent focal points defined by equations (\ref{eq:disomfac2a}),
diamond (blue) dots represent the cross-points $\mathrm{Gr}_{\mathrm{T}}\bigcap\mathrm{Gr}_{\mathrm{B}}$.
The circular dots (red) identify transition to instability points
(see Section \ref{subsec:trans-instab}). The bold, solid (red) curves
represent branches of points $\left(\Re\left\{ k\right\} ,\omega\right)$
with real $\omega$ and $\Im\left\{ k\right\} \protect\neq0$ which
are points of the convection instability.}
\end{figure}

Fig. \ref{fig:dis-Lev1}, \ref{fig:dis-instab3} and \ref{fig:dis-instab3f}
show the TWT dispersion curves denoted as graphs $\mathrm{Gr}_{\mathrm{TB}}\left(\gamma\right)$.
We also consider graphs $\mathrm{Gr}_{\mathrm{T}}$ and $\mathrm{Gr}_{\mathrm{B}}$
associated with dispersion equations (\ref{eq:mainfac2c}) as natural
reference frames. Graphs $\mathrm{Gr}_{\mathrm{TB}}\left(\gamma\right)$
are generated for a number of different values of $\gamma$ as the
corresponding level curves for the TWT dispersion function $R_{\mathrm{TB}}\left(k,\omega\right)$.
Note that all dispersion curves $\mathrm{Gr}_{\mathrm{TB}}\left(\gamma\right)$
pass through the focal points described by equations (\ref{eq:mainfac2e}). 

Figures \ref{fig:dis-instab1}- \ref{fig:dis-instab3f} display the
TWT dispersion-instability graphs with many details described in the
captions. In particular, in Figures \ref{fig:dis-instab3} and \ref{fig:dis-instab3f}
one can see the low and the high frequency cutoff for the convection
instability that are similar to what was found in \citet{SchaFig}
in the special case $\rho=0$.

We address at it is commonly done the important for TWT theory issues
of instability and consequent amplification by considering complex-valued
$\omega$ and $k$ solutions to the TWT dispersion relations (\ref{eq:mainfac2a}),
(\ref{eq:mainfac2d}). More exactly, to identify two commonly studied
instabilities - convective and absolute - one considers the TWT eigenmodes
$f_{\omega,k}\left(z,t\right)$ of the exponential form $f_{\omega,k}\left(z,t\right)\sim\exp\left\{ -\mathrm{i}\left(\omega t-kz\right)\right\} $
where possibly complex-valued $\omega$ and $k$ must satisfy the
TWT dispersion relations (\ref{eq:mainfac2a}), (\ref{eq:mainfac2d}).
In the case when $\omega$ is real and $\Im\left(k\right)\neq0$ function
$\left|f_{\omega k}\left(z,t\right)\right|$ grows or decays exponentially
if $z\rightarrow\pm\infty$ and we refer to this situation as convection
instability associated with amplification regimes. In the case $k$
is real but $\Im\left(\omega\right)\neq0$ function $\left|f_{\omega k}\left(z,t\right)\right|$
grows or decays exponentially if $t\rightarrow\pm\infty$ and we refer
to this situation as absolute instability associated with (exponentially
growing) oscillations regimes. We focus mostly on the case of the
convection instability. See Section \ref{sec:disp-instab} for more
details.

The conventional plot of the TWT dispersion relation as shown in Fig.
\ref{fig:dis-Lev1} represents real-valued $\omega$ and $k$ that
satisfies the TWT dispersion relations (\ref{eq:mainfac2a}), (\ref{eq:mainfac2d}).
To help to visualize at least partly also the complex-valued valued
solutions associated with instabilities we use the concept of \emph{dispersion-instability
graph (plot)} that we have developed in \citet[Chap. 7]{FigTWTbk}.
In a nutshell in the case of the convection instability, when $\omega$
is real and $k$ is complex-valued, we consider a solution $\left(k,\omega\right)$
to the TWT dispersion relations (\ref{eq:mainfac2a}), (\ref{eq:mainfac2d})
assuming that $k=k\left(\omega\right)$ and depict it as point $\left(\Re\left\{ k\left(\omega\right)\right\} ,\omega\right)$
in $k\omega$-plane.

To distinguish graphically points $\left(k\left(\omega\right),\omega\right)$
associated with oscillatory modes when $k\left(\omega\right)$ is
real-valued from points $\left(\Re\left\{ k\left(\omega\right)\right\} ,\omega\right)$
associated with unstable modes when $k\left(\omega\right)$ is complex-valued
with $\Im\left\{ k\left(\omega\right)\right\} \neq0$ we show points
$\Im\left\{ k\left(\omega\right)\right\} =0$ in black color whereas
points with $\Im\left\{ k\left(\omega\right)\right\} \neq0$ are shown
in red color. The corresponding curves are shown respectively as the
solid (black) and solid (red) curves. We remind that every point $\left(\Re\left\{ k\left(\omega\right)\right\} ,\omega\right)$
with $\Im\left\{ k\left(\omega\right)\right\} \neq0$ represents exactly
two complex conjugate convectively unstable modes associated with
$\pm\Im\left\{ k\left(\omega\right)\right\} $.

Similarly, in the case of the absolute instability, when $k$ is real
and $\omega$ is complex-valued, the corresponding solution $\left(k,\omega\right)$
to the TWT dispersion relations (\ref{eq:mainfac1d}), (\ref{eq:mainfac2d})
is depicted as point $\left(k,\Re\left\{ \omega\left(k\right)\right\} \right)$
in $k\omega$-plane. To distinguish graphically points $\left(k,\omega\left(k\right)\right)$
associated with oscillatory modes when $\omega\left(k\right)$ is
real-valued from points $\left(k,\Re\left\{ \omega\left(k\right)\right\} \right)$
associated with absolutely unstable modes when $\omega\left(k\right)$
is complex-valued with $\Im\left\{ \omega\left(k\right)\right\} \neq0$
we show points with $\Im\left\{ \omega\left(k\right)\right\} =0$
in in black color whereas points with $\Im\left\{ \omega\left(k\right)\right\} \neq0$
are shown in green color. 

\begin{figure}[h]
\begin{centering}
\includegraphics[scale=0.5]{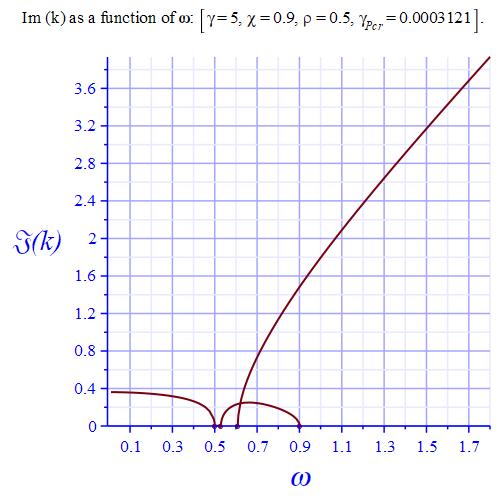}
\par\end{centering}
\centering{}\caption{\label{fig:fac-Imk1} The plot shows the imaginary part $\Im\left\{ k\right\} $
of the wavenumber $k=k\left(\omega\right)$ as a function of frequency
$\omega$ for $\gamma=5\gg\gamma_{\mathrm{Pcr}}\protect\cong0.0003121$,
$\chi=0.9<1$ and $\rho=0.5$. Diamond square dots (red) on the $\omega$-axis
mark the location of the low and the high frequency cutoffs. Note
that in the case there is no high frequency cutoff.}
\end{figure}

Fig. \ref{fig:fac-Imk1} shows the plot shows the imaginary part $\Im\left\{ k\right\} $
of the wavenumber $k=k\left(\omega\right)$ as a function of frequency
$\omega$ for the selected values of the TWT parameters.

\section{An extended analytic model of the traveling wave tube\label{sec:twt-mod}}

We extend here an analytic model of traveling wave tube introduced
and studied in our monograph \citep[4, 24]{FigTWTbk}. This extended
analytic model of TWT features a generalized transmission line (GTL)
that can have a non-zero low frequency cutoff. More precisely, this
model represents an ideal TWT as a system of a single-stream electron
beam coupled to the GTL. Just as our simpler model this model is a
generalization of the celebrated Pierce model \citep[I]{Pier51},
\citep{PierTWT}. Let us start first with non-interacting e-beam and
the GTL.

\subsection{The non-interacting electron beam and the generalized transmission
line\label{subsec:coldTB}}

The main parameter describing the e-beam is the \emph{e-beam intensity}
\begin{equation}
\beta=\frac{\sigma_{\mathrm{B}}}{4\pi}R_{\mathrm{sc}}^{2}\omega_{\mathrm{p}}^{2}=\frac{e^{2}}{m}R_{\mathrm{sc}}^{2}\sigma_{\mathrm{B}}\mathring{n},\quad\omega_{\mathrm{p}}^{2}=\frac{4\pi\mathring{n}e^{2}}{m},\label{eq:exTBbet1a}
\end{equation}
where $-e$ is electron charge with $e>0$, $m$ is the electron mass,
$\omega_{\mathrm{p}}$ is the e-beam \emph{plasma frequency}, $\sigma_{\mathrm{B}}$
is the area of the cross-section of the e-beam, the constant $R_{\mathrm{sc}}$
is the\emph{ plasma frequency reduction factor} that accounts phenomenologically
for finite dimensions of the e-beam cylinder as well as geometric
features of the slow-wave structure, \citep{BraMih}, \citep[Sec. 9.2]{Gilm1},
\citet[Sec. 3.3.3]{BBLN}. \citep[41, 63]{FigTWTbk}, and $\mathring{n}$
is the density of the number of electrons. It is assumed the electron
flow of the e-beam has steady velocity $\mathring{v}>0$. Another
important parameter related to e-beam when it interacts with containing
it waveguide slow-wave structure is the so-called \emph{reduced plasma
frequency}, \citep[9.2]{Gilm1}, \citep[Sec. 11.3.1]{Cart}
\begin{equation}
\omega_{\mathrm{rp}}=R_{\mathrm{sc}}\omega_{\mathrm{p}},\quad\omega_{\mathrm{p}}^{2}=\frac{4\pi\mathring{n}e^{2}}{m}.\label{eq:exTBbet1b}
\end{equation}
Note that the the interaction between the GTL and e-beam modifies
the features of the e-beam by reducing the conventional plasma frequency
$\omega_{\mathrm{p}}$ to the reduced plasma frequency $\omega_{\mathrm{rp}}$,
and this reduction can be significant, \citep[Sec. 11.3.1]{Cart}.
Equations (\ref{eq:LaBTq1a}) and (\ref{eq:LaBTq1b}) readily imply
\begin{equation}
\beta=\frac{\sigma_{\mathrm{B}}}{4\pi}\omega_{\mathrm{rp}}^{2},\quad\omega_{\mathrm{rp}}^{2}=R_{\mathrm{sc}}^{2}\frac{4\pi\mathring{n}e^{2}}{m}.\label{eq:exTBbet1aa}
\end{equation}

We would like to point out that there are two spatial scales related
to the e-beam. The first one is
\begin{equation}
\lambda_{\mathrm{rp}}=\frac{2\pi\mathring{v}}{\omega_{\mathrm{rp}}},\quad\omega_{\mathrm{rp}}=R_{\mathrm{sc}}\omega_{\mathrm{p}}=R_{\mathrm{sc}}\sqrt{\frac{4\pi\mathring{n}e^{2}}{m}},\label{eq:Bvom1a}
\end{equation}
which is the distance passed by an electron for the time period $\frac{2\pi}{\omega_{\mathrm{rp}}}$
associated with the plasma oscillations at the reduced plasma frequency
$\omega_{\mathrm{rp}}$. This scale is well known in the theory of
klystrons and is referred to as \emph{the electron plasma wavelength},
\citep[9.2]{Gilm1}, \citet{FigMCK}. Another spatial scale related
to the e-beam that arises in our analysis is, \citet{FigCCTWT}
\begin{equation}
g_{\mathrm{B}}=\frac{\sigma_{\mathrm{B}}}{4\lambda_{\mathrm{rp}}},\label{eq:Bvom1ba}
\end{equation}
and we will refer to it as\emph{ e-beam spatial scale}. 

Let us turn now to the GTL. We remind the reader that the standard
transmission line (TL) has two important parameters: (i) (distributed)
shunt capacitance $C>0$ per unit of length and (ii) (distributed)
inductance $L>0$ per unit of length. It is well known that there
are two important physical quantities phase associated with such a
TL: (i) the phase velocity $w$ and (ii) the characteristic impedance
$Z_{0}$ defined by the following equality, \citet[Sec. 7.2]{Franc},
\citet[Chap. 4.2]{MiaMaf}, \citet[Chap. 4, 5]{Chip}, \citealt[Sec. 7]{Pain}:

\begin{equation}
w\stackrel{\mathrm{def}}{=}\frac{1}{\sqrt{CL}},\quad Z_{0}\stackrel{\mathrm{def}}{=}\sqrt{\frac{L}{C}}.\label{eq:exTBbet1c}
\end{equation}
It is also known that the regular TL represents the so-called TEM
mode.

To obtain a model of a wave-guided structure featuring a non-zero
low frequency cutoff $\omega_{\mathrm{c}}$ one can extend the standard
TL by adding to it the distributed serial capacitance $C_{\mathrm{c}}$
per unit of length, \citet[Sec. 5.1, Fig. 5.2]{MilSchw}. The subindex
``$\mathrm{c}$'' in $C_{\mathrm{c}}$ is to remind that this serial
capacitance is an origin of the low frequency cutoff. We refer to
this extension of the standard TL as \emph{generalized transmission
line (GTL)}. We introduce also \emph{GTL cutoff wave number} $k_{\mathrm{c}}$
and the corresponding \emph{GTL cutoff frequency} $\omega_{\mathrm{c}}$by
the following formulas
\begin{equation}
k_{\mathrm{c}}\stackrel{\mathrm{def}}{=}\sqrt{\frac{C}{C_{\mathrm{c}}}},\quad\omega_{\mathrm{c}}\stackrel{\mathrm{def}}{=}wk_{\mathrm{c}}=w\sqrt{\frac{C}{C_{\mathrm{c}}}}=\frac{1}{\sqrt{C_{\mathrm{c}}L}}.\label{eq:cofdisp1a}
\end{equation}

We use the Gaussian system of units of the physical dimensions. For
the reader's convenience we collect (i) all significant parameters
related to the e-beam and the GTL and (ii) the relevant ``natural''
units in Tables \ref{tab:ebeam-dim}, \ref{tab:ebeam-unit} and \ref{tab:GTL-dim}.
\begin{table}[tbh]
\centering{}%
\begin{tabular}{|r||r||r|}
\hline 
\noalign{\vskip\doublerulesep}
$i$ & current & $\frac{\left[\text{charge}\right]}{\left[\text{time}\right]}$\tabularnewline[0.2cm]
\hline 
\noalign{\vskip\doublerulesep}
$q$ & charge & $\left[\text{charge}\right]$\tabularnewline[0.2cm]
\hline 
\noalign{\vskip\doublerulesep}
$\mathring{n}$ & number of electrons p/u of volume & $\frac{\left[\text{1}\right]}{\left[\text{length}\right]^{3}}$\tabularnewline[0.2cm]
\hline 
\noalign{\vskip\doublerulesep}
$\lambda_{\mathrm{rp}}=\frac{2\pi\mathring{v}}{\omega_{\mathrm{rp}}},\:\omega_{\mathrm{rp}}=R_{\mathrm{sc}}\omega_{\mathrm{p}}$ & the electron plasma wavelength & $\left[\text{length}\right]$\tabularnewline[0.2cm]
\hline 
\noalign{\vskip\doublerulesep}
$g_{\mathrm{B}}=\frac{\sigma_{\mathrm{B}}}{4\lambda_{\mathrm{rp}}}$ & the e-beam spatial scale & $\left[\text{length}\right]$\tabularnewline[0.2cm]
\hline 
\noalign{\vskip\doublerulesep}
$\beta=\frac{\sigma_{\mathrm{B}}}{4\pi}R_{\mathrm{sc}}^{2}\omega_{\mathrm{p}}^{2}=\frac{e^{2}}{m}R_{\mathrm{sc}}^{2}\sigma_{\mathrm{B}}\mathring{n}$ & e-beam intensity & $\frac{\left[\text{length}\right]^{2}}{\left[\text{time}\right]^{2}}$\tabularnewline[0.2cm]
\hline 
\end{tabular}\vspace{0.3cm}
\caption{\label{tab:ebeam-dim}Physical dimensions of the e-beam parameters.
Abbreviations: dimensionless \textendash{} dim-less, p/u \textendash{}
per unit.}
\end{table}
\begin{table}[h]
\centering{}%
\begin{tabular}{|l||r||r|r|}
\hline 
\noalign{\vskip\doublerulesep}
Frequency & Reduced plasma frequency & $\omega_{\mathrm{rp}}=R_{\mathrm{sc}}\omega_{\mathrm{p}}=R_{\mathrm{sc}}\sqrt{\frac{4\pi\mathring{n}e^{2}}{m}}$ & $\frac{1}{\left[\text{time}\right]}$\tabularnewline[0.2cm]
\hline 
\noalign{\vskip\doublerulesep}
Velocity & e-beam velocity & $\mathring{v}$ & $\frac{\left[\text{length}\right]}{\left[\text{time}\right]}$\tabularnewline[0.2cm]
\hline 
\noalign{\vskip\doublerulesep}
Wavenumber & Plasma oscillations wavenumber & $k_{\mathrm{rp}}=\frac{\omega_{\mathrm{rp}}}{\mathring{v}}=\frac{R_{\mathrm{sc}}\omega_{\mathrm{p}}}{\mathring{v}}$ & $\frac{1}{\left[\text{length}\right]}$\tabularnewline[0.2cm]
\hline 
\noalign{\vskip\doublerulesep}
Length & Plasma oscillations wavelength & $\lambda_{\mathrm{rp}}=\frac{2\pi}{k_{\mathrm{rp}}}=\frac{2\pi\mathring{v}}{\omega_{\mathrm{rp}}}$ & $\left[\text{length}\right]$\tabularnewline[0.2cm]
\hline 
\noalign{\vskip\doublerulesep}
Time & Plasma oscillations time period & $\tau_{\mathrm{rp}}=\frac{2\pi}{\omega_{\mathrm{rp}}}$ & $\left[\text{time}\right]$\tabularnewline[0.2cm]
\hline 
\end{tabular}\vspace{0.3cm}
\caption{\label{tab:ebeam-unit}Natural units relevant to the e-beam parameters.}
\end{table}
\begin{table}[tbh]
\centering{}%
\begin{tabular}{|r||r||r|}
\hline 
\noalign{\vskip\doublerulesep}
$I$ & Current & $\frac{\left[\text{charge}\right]}{\left[\text{time}\right]}$\tabularnewline
\hline 
\noalign{\vskip\doublerulesep}
$Q$ & Charge & $\left[\text{charge}\right]$\tabularnewline
\hline 
\noalign{\vskip\doublerulesep}
$C$ & Shunt capacitance p/u of length & $\left[\text{dim-less}\right]$\tabularnewline
\hline 
\noalign{\vskip\doublerulesep}
$C_{\mathrm{c}}$ & Serial capacitance p/u of length & $\left[\text{length}\right]^{2}$\tabularnewline
\hline 
\noalign{\vskip\doublerulesep}
$L$ & Series inductance p/u of length & $\frac{\left[\text{time}\right]^{2}}{\left[\text{length}\right]^{2}}$\tabularnewline
\hline 
\noalign{\vskip\doublerulesep}
$w=\frac{1}{\sqrt{CL}}$ & TL phase velocity & $\frac{\left[\text{length}\right]}{\left[\text{time}\right]}$\tabularnewline
\hline 
\noalign{\vskip\doublerulesep}
$k_{\mathrm{c}}=\sqrt{\frac{C}{C_{\mathrm{c}}}}$ & GTL cutoff wavenumber & $\frac{1}{\left[\text{length}\right]}$\tabularnewline
\hline 
\noalign{\vskip\doublerulesep}
$\omega_{\mathrm{c}}=wk_{\mathrm{c}}=\frac{1}{\sqrt{C_{\mathrm{c}}L}}$ & GTL cutoff frequency & $\frac{1}{\left[\text{time}\right]}$\tabularnewline
\hline 
\end{tabular}\vspace{0.3cm}
\caption{\label{tab:GTL-dim}Physical dimensions of the GTL related quantities.
Abbreviations: dimensionless \textendash{} dim-less, p/u \textendash{}
per unit.}
\end{table}

The e-beam featuring space-charge effects has the following Lagrangian,
\citep[Chap. 24]{FigTWTbk}
\begin{equation}
\mathcal{L}_{\mathrm{B}}=\frac{1}{2\beta}\left(\partial_{t}q+\mathring{v}\partial_{z}q\right)^{2}-\frac{2\pi}{\sigma_{\mathrm{B}}}q^{2}\label{eq:LaBTq1a}
\end{equation}
whereas the GTL Lagrangian is
\begin{equation}
\mathcal{L}_{\mathrm{T}}=\frac{L}{2}\left(\partial_{t}Q\right)^{2}-\frac{1}{2C}\left(\partial_{z}Q\right)^{2}-\frac{1}{2C_{\mathrm{c}}}Q^{2},\label{eq:LaBTq1b}
\end{equation}
where $q\left(z,t\right)$ and $Q\left(z,t\right)$\emph{ }are position
and time dependent charges associated with the e-beam and the GTL.
They are defined as time integrals of the corresponding e-beam current
$i(z,t)$ and the GTL current $I(z,t)$, that is
\begin{equation}
q(z,t)=\int^{t}i(z,t^{\prime})\,\mathrm{d}t^{\prime},\quad.Q(z,t)=\int^{t}I(z,t^{\prime})\,\mathrm{d}t^{\prime}.\label{eq:LaTBq1c}
\end{equation}
Note the GTL Lagrangian $\mathcal{L}_{\mathrm{T}}$ defined by equation
(\ref{eq:LaBTq1b}) has term $\frac{1}{2C_{\mathrm{c}}}Q^{2}$ that
accounts for the energy stored in the GTL serial capacitance. This
term has been added up to the TL Lagrangian introduced in \citep[Chap. 24]{FigTWTbk}.
Note also that parameters $C$ and $C_{\mathrm{c}}$ representing
respectively the shunt and the serial capacitances for the TL have
different dimensions, namely the following identity holds for their
dimensions: $\left[\frac{C}{C_{\mathrm{c}}}\right]=\left[\mathrm{L}^{2}\right]$
where $\mathrm{L}$ represents ``length''.

The Euler-Lagrange (EL) equations corresponding to Lagrangians (\ref{eq:LaBTq1a})
and (\ref{eq:LaBTq1b}) are the following second-order differential
equations:
\begin{gather}
L\partial_{t}^{2}Q-\frac{1}{C}\partial_{z}^{2}Q+\frac{1}{C_{\mathrm{c}}}Q=0,\label{eq::LaTBnon1a}\\
\frac{1}{\beta}\left(\partial_{t}+\mathring{v}\partial_{z}\right)^{2}q+\frac{4\pi}{\sigma_{\mathrm{B}}}q=0,\quad\beta=\frac{\sigma_{\mathrm{B}}}{4\pi}\omega_{\mathrm{rp}}^{2}.\label{eq:LaTBnon1b}
\end{gather}
The Fourier transformation (see Section \ref{sec:four}) in time $t$
and space variable $z$ of equations (\ref{eq::LaTBnon1a}) and (\ref{eq:LaTBnon1b})
yields
\begin{gather}
\left(\frac{k^{2}}{C}-\omega^{2}L+\frac{1}{C_{\mathrm{c}}}\right)\hat{Q}=0,\quad\frac{4\pi}{\sigma_{\mathrm{B}}}\left[1-\frac{\left(\omega-\mathring{v}k\right)^{2}}{\omega_{\mathrm{rp}}^{2}}\right]\hat{q}=0,\label{eq:LaTBnon1c}
\end{gather}
where $\omega$ and $k=k\left(\omega\right)$ are the frequency and
the wavenumber, respectively, and functions $\hat{Q}=\hat{Q}\left(\omega,k\right)$
and $\hat{q}=\hat{q}\left(\omega,k\right)$ are the Fourier transforms
of the system vector variables $Q\left(t,z\right)$ and $q\left(t,z\right)$,
see Appendix \ref{sec:four}. 

Assuming $\hat{Q}$ and $\hat{q}$ in respective equations (\ref{eq:LaTBnon1c})
to be non-zero we immediately obtain the dispersion relations for
the GTL and the e-beam, namely
\begin{equation}
\omega=\omega_{\mathrm{T}}\left(k\right)=\pm\sqrt{w^{2}k^{2}+\omega_{\mathrm{c}}^{2}},\quad\omega=\omega_{\mathrm{B}}\left(k\right)=\mathring{v}k\pm\omega_{\mathrm{rp}}.\label{eq:cofdisp1b}
\end{equation}
Then the corresponding \emph{GTL phase velocity $u_{\mathrm{T}}$}
and the \emph{e-beam phase velocity $u_{\mathrm{B}}$} are
\begin{equation}
u=u_{\mathrm{T}}\left(\omega\right)=\pm\frac{\omega w}{\sqrt{\omega^{2}-\omega_{\mathrm{c}}^{2}}},\quad u=u_{\mathrm{B}}\left(\omega\right)=\frac{\mathring{v}\omega}{\omega\pm\omega_{\mathrm{rp}}},\label{eq:cofdisp1u}
\end{equation}
The equations (\ref{eq:cofdisp1b}), (\ref{eq:cofdisp1u}) imply the
following asymptotic formulas for $\omega_{\mathrm{T}}\left(k\right)$
and $u_{\mathrm{T}}\left(\omega\right)$:
\begin{equation}
\omega=\omega_{\mathrm{T}}\left(k\right)=\pm\left[wk+\frac{\omega_{\mathrm{c}}^{2}}{2wk}-\frac{\omega_{\mathrm{c}}^{4}}{8\left(wk\right)^{3}}+\frac{\omega_{\mathrm{c}}^{6}}{16\left(wk\right)^{5}}+O\left(\frac{1}{k^{7}}\right)\right],\quad k\rightarrow\infty,\label{eq:cofdisp3a}
\end{equation}
\begin{equation}
u=u_{\mathrm{T}}\left(\omega\right)=w\left[1+\frac{\omega_{\mathrm{c}}^{2}}{2\omega^{2}}+\frac{3\omega_{\mathrm{c}}^{4}}{8\omega^{4}}+\frac{5\omega_{\mathrm{c}}^{6}}{16\omega^{6}}+O\left(\frac{1}{\omega^{8}}\right)\right],\quad\omega\rightarrow\infty.\label{eq:cofdisp3b}
\end{equation}
In particular, it follows from equation (\ref{eq:cofdisp3b}) that
\begin{equation}
\lim_{\omega\rightarrow\infty}u_{\mathrm{T}}\left(\omega\right)=\pm w=\pm\frac{1}{\sqrt{CL}}.\label{eq:cofdisp3c}
\end{equation}
Equations (\ref{eq:cofdisp3a})-(\ref{eq:cofdisp3c}) indicate that
at high frequency and wavenumbers the difference between the TL and
the GTL becomes negligibly small.

\subsection{The TWT Lagrangian and the evolution equations\label{subsec:two-lag-ev}}

The TWT-system Lagrangian $\mathcal{L}{}_{\mathrm{TB}}$ is defined
similarly to its expression in \citep[4, 24]{FigTWTbk} with the only
difference that there is an additional term related to serial capacitance
$C_{\mathrm{c}}$ as in equation (\ref{eq:LaBTq1b}), namely
\begin{gather}
\mathcal{L}{}_{\mathrm{TB}}=\mathcal{L}_{\mathrm{B}}+\mathcal{L}_{\mathrm{Tb}},\;\mathcal{L}_{\mathrm{B}}=\frac{1}{2\beta}\left(\partial_{t}q+\mathring{v}\partial_{z}q\right)^{2}-\frac{2\pi}{\sigma_{\mathrm{B}}}q^{2},\label{eq:LaTBq2a}\\
\mathcal{L}_{\mathrm{Tb}}=\frac{L}{2}\left(\partial_{t}Q\right)^{2}-\frac{1}{2C}\left(\partial_{z}Q+b\partial_{z}q\right)^{2}-\frac{1}{2C_{\mathrm{c}}}Q^{2},\nonumber 
\end{gather}
where $b$ is the so-called coupling constant which is a dimensionless
phenomenological parameter and other parameters are discussed in Section
\ref{subsec:coldTB}. Constant $b$ is assumed often to satisfy $0<b\leq1$
effectively reducing the inductive input of the e-beam current into
the shunt current, see \citet[Chap. 3]{FigTWTbk} for more details.
Note that coupling between the GTL and e-beam is introduced through
term $-\frac{1}{2C}\left(\partial_{z}Q+b\partial_{z}q\right)^{2}$indicating
that the GTL distributed shunt capacitance $C$ is shared with e-beam.
Following to developments in \citep[4, 24]{FigTWTbk} as well equations
(\ref{eq:exTBbet1aa}) we introduce the \emph{TWT principal parameter}
$\gamma$ defined by
\begin{equation}
\gamma=\frac{b^{2}}{C}\beta=\frac{b^{2}}{C}\frac{\sigma_{\mathrm{B}}}{4\pi}\omega_{\mathrm{rp}}^{2},\quad\omega_{\mathrm{rp}}^{2}=R_{\mathrm{sc}}^{2}\frac{4\pi\mathring{n}e^{2}}{m}.\label{eq:LaBTq2b}
\end{equation}

The Euler-Lagrange (EL) equations corresponding to the Lagrangian
$\mathcal{L}{}_{\mathrm{TB}}$ defined by equations (\ref{eq:LaTBq2a})
are the following system of the second-order differential equations
\begin{gather}
L\partial_{t}^{2}Q-\frac{1}{C}\partial_{z}^{2}\left(Q+bq\right)+\frac{1}{C_{\mathrm{c}}}Q^{2}=0,\label{eq::LaTBq2c}\\
\frac{1}{\beta}\left(\partial_{t}+\mathring{v}\partial_{z}\right)^{2}q+\frac{4\pi}{\sigma_{\mathrm{B}}}q-\frac{b}{C}\partial_{z}^{2}\left(Q+bq\right)=0,\quad\beta=\frac{\sigma_{\mathrm{B}}}{4\pi}\omega_{\mathrm{rp}}^{2}.\label{eq:LaTBq2d}
\end{gather}
The Fourier transformation (see Appendix \ref{sec:four}) in time
$t$ and space variable $z$ of equations (\ref{eq::LaTBq2c}) and
(\ref{eq:LaTBq2d}) yields
\begin{gather}
\left(\frac{k^{2}}{C}-\omega^{2}L+\frac{1}{C_{\mathrm{c}}}\right)\hat{Q}+k^{2}\frac{b}{C}\hat{q}=0,\label{eq:LaTBq3a}\\
\frac{bk^{2}}{C}\hat{Q}+\left\{ \frac{b^{2}k^{2}}{C}+\frac{4\pi}{\sigma_{\mathrm{B}}}\left[1-\frac{\left(\omega-\mathring{v}k\right)^{2}}{\omega_{\mathrm{rp}}^{2}}\right]\right\} \hat{q}=0,\label{eq:LaTBq3b}
\end{gather}
where functions $\hat{Q}=\hat{Q}\left(k,\omega\right)$ and $\hat{q}=\hat{q}\left(k,\omega\right)$
are the Fourier transforms of the system variables $Q\left(t,z\right)$
and $q\left(t,z\right)$. We will refer to equations (\ref{eq:LaTBq3a}),
(\ref{eq:LaTBq3b}) as\emph{ transformed EL equations}. The TWT-system
eigenmodes are naturally assumed to be of the form
\begin{equation}
Q\left(z,t\right)=\hat{Q}\left(k,\omega\right)\mathrm{e}^{-\mathrm{i}\left(\omega t-kz\right)},\quad q\left(z,t\right)=\hat{q}\left(k,\omega\right)\mathrm{e}^{-\mathrm{i}\left(\omega t-kz\right)},\label{eq:LaTBq3c}
\end{equation}
where $\omega$ and $k=k\left(\omega\right)$ are the frequency and
the wavenumber, respectively. The dependence of $k\left(\omega\right)$
on the frequency $\omega$ can be found from the TWT dispersion relations
analyzed in Section \ref{sec:disp-fac}.

Multiplying the EL equations (\ref{eq:LaTBq3a}), (\ref{eq:LaTBq3b})
by $C$ we can recast them into the following matrix form:
\begin{equation}
M_{k\omega}x=0,\quad M_{k\omega}=\left[\begin{array}{rr}
k^{2}-\frac{\omega^{2}}{w^{2}}+\frac{C}{C_{\mathrm{c}}} & bk^{2}\\
bk^{2} & b^{2}k^{2}+\frac{4\pi C}{\sigma_{\mathrm{B}}}\left[1-\frac{\left(\omega-\mathring{v}k\right)^{2}}{\omega_{\mathrm{rp}}^{2}}\right]
\end{array}\right],\quad x=\left[\begin{array}{r}
\hat{Q}\\
\hat{q}
\end{array}\right].\label{eq:LaTBq4a}
\end{equation}
Note that equations (\ref{eq:LaTBq4a}) can be viewed as an eigenvalue
type problem for $k$ and $x$ assuming that $\omega$ and other parameters
are fixed.

Taking into account expressions (\ref{eq:exTBbet1c}), (\ref{eq:exTBbet1c})
for $w$ and $\omega_{\mathrm{c}}$ as well as expression (\ref{eq:LaBTq2b})
for the TWT principle parameter $\gamma$ we can rewrite equations
(\ref{eq:LaTBq4a}) as
\begin{equation}
M_{k\omega}x=0,\quad M_{k\omega}=M_{k\omega}\left(b\right)=\left[\begin{array}{rr}
k^{2}+\frac{\omega_{\mathrm{c}}^{2}-\omega^{2}}{w^{2}} & bk^{2}\\
bk^{2} & b^{2}\left[k^{2}+\frac{\omega_{\mathrm{rp}}^{2}-\left(\omega-\mathring{v}k\right)^{2}}{\gamma}\right]
\end{array}\right],\quad x=\left[\begin{array}{r}
\hat{Q}\\
\hat{q}
\end{array}\right].\label{eq:LaTBq4b}
\end{equation}
Yet another equivalent form of equations (\ref{eq:LaTBq4b}) can be
obtained by using phase velocity $u=\frac{\omega}{k}$ instead of
wavenumber $k$ in equation (\ref{eq:LaTBq4b}), namely
\begin{gather}
M_{u\omega}x=0,\quad M_{u\omega}=M_{u\omega}\left(b\right)=\left[\begin{array}{rr}
\frac{\omega^{2}}{u^{2}}+\frac{\omega_{\mathrm{c}}^{2}-\omega^{2}}{w^{2}} & \frac{b\omega^{2}}{u^{2}}\\
\frac{b\omega^{2}}{u^{2}} & \left[\frac{\omega^{2}}{u^{2}}+\frac{1}{\gamma}\left(\omega_{\mathrm{rp}}^{2}-\frac{\omega^{2}\left(u-\mathring{v}\right)^{2}}{u^{2}}\right)\right]b^{2}
\end{array}\right],\quad x=\left[\begin{array}{r}
\hat{Q}\\
\hat{q}
\end{array}\right],\label{eq:LaTBq5a}
\end{gather}
where we use once again the principal TWT parameter $\gamma=\frac{b^{2}}{C}\beta=\frac{b^{2}}{C}\frac{\sigma_{\mathrm{B}}}{4\pi}\omega_{\mathrm{rp}}^{2}$
defined by equations (\ref{eq:LaBTq2b}).

Note that matrices $M_{k\omega}\left(b\right)$ and $M_{u\omega}\left(b\right)$
satisfy the following factorized representations
\begin{equation}
M_{k\omega}\left(b\right)=D_{b}M_{k\omega}\left(1\right)D_{b},\quad M_{u\omega}\left(b\right)=D_{b}M_{u\omega}\left(1\right)D_{b}\quad D_{b}=\left[\begin{array}{rr}
1 & 0\\
0 & b
\end{array}\right],\label{eq:LaBTq5b}
\end{equation}
where matrices $M_{k\omega}\left(1\right)$ and $M_{u\omega}\left(1\right)$
evidently do not depend on $b$.

\section{The dispersion relations in factorized form\label{sec:disp-fac}}

As it was already mentioned the matrix form of the EL equations (\ref{eq:LaTBq4b})
and (\ref{eq:LaTBq5a}) can be viewed as a kind of eigenvalue problem
for $k$ and $u$ with other involved parameters including frequency
$\omega$ considered as being fixed. \emph{The condition for these
equations to have nontrivial nonzero solutions $x$ is that the determinants
of matrices $M_{k\omega}$ and $M_{u\omega}$ defined respectively
by equations (\ref{eq:LaTBq4b}) and (\ref{eq:LaTBq5a}) must be zero},
that is
\begin{equation}
\det\left\{ M_{k\omega}\right\} =0,\quad\det\left\{ M_{u\omega}\right\} =0.\label{eq:dispkuom1a}
\end{equation}
Equations (\ref{eq:dispkuom1a}) establish relations between $k$
or $u$ and $\omega$ and consequently they can be viewed respectively
as \emph{the dispersion relation or the velocity dispersion relation}.
After tedious algebraic transformations equations (\ref{eq:dispkuom1a})
can be turned respectively into the following factorized forms: 
\begin{equation}
\left[\frac{1}{k^{2}}-\frac{w^{2}}{\omega^{2}-\omega_{\mathrm{c}}^{2}}\right]\left[\left(\omega-\mathring{v}k\right)^{2}-\omega_{\mathrm{rp}}^{2}\right]=\gamma,\quad\gamma=\frac{b^{2}}{C}\frac{\sigma_{\mathrm{B}}}{4\pi}\omega_{\mathrm{rp}}^{2}\label{eq:dispkuom1b}
\end{equation}

\begin{equation}
\left[u^{2}-\frac{w^{2}\omega^{2}}{\omega^{2}-\omega_{\mathrm{c}}^{2}}\right]\left[\frac{\left(u-\mathring{v}\right)^{2}}{u^{2}}-\frac{\omega_{\mathrm{rp}}^{2}}{\omega^{2}}\right]=\gamma,\quad u=\frac{\omega}{k}.\label{eq:dispkuom1c}
\end{equation}
We refer to equations (\ref{eq:dispkuom1b}) and (\ref{eq:dispkuom1c})
respectively as \emph{the factorized form of the TWT dispersion relations
and the TWT velocity dispersion relations.} These equations suggest
a \emph{natural interpretation of the TWT principle parameter $\gamma$
as a coupling parameter between GTL and the e-beam}. Note that equations
(\ref{eq:dispkuom1b}) and (\ref{eq:dispkuom1c}) do not involve parameter
$b$ explicitly but rather through the TWT principle parameter $\gamma$,
and that is explained by identities (\ref{eq:LaBTq5b}).

Sometimes it is advantageous to use the following polynomial equation
which is evidently equivalent to the TWT dispersion relations (\ref{eq:dispkuom1b}):
\begin{equation}
\left[\omega^{2}-\left(w^{2}k^{2}+\omega_{\mathrm{c}}^{2}\right)\right]\left[\left(\omega-\mathring{v}k\right)^{2}-\omega_{\mathrm{rp}}^{2}\right]=\gamma k^{2}\left(\omega^{2}-\omega_{\mathrm{c}}^{2}\right).\label{eq:disomfac1a}
\end{equation}
For every fixed $\omega$ equation (\ref{eq:disomfac1a}) is exactly
forth degree polynomial equation for $k=k\left(\omega\right)$, and
for every fixed $k$ equation (\ref{eq:disomfac1a}) is exactly forth
degree polynomial equation for $\omega=\omega\left(k\right)$. \emph{The
factorized form (\ref{eq:disomfac1a}) of the TWT dispersion relations
naturally integrates into it the dispersion relations (\ref{eq:cofdisp1b})
of the GTL and the e-beam.} Equation (\ref{eq:disomfac1a}) is instrumental
for our analysis of analytical properties the TWT dispersion relations.

To emphasize the structure of the factorized form (\ref{eq:disomfac1a})
of the TWT dispersion relation we introduce
\begin{equation}
G_{\mathrm{T}}\left(k,\omega\right)\stackrel{\mathrm{def}}{=}\omega^{2}-\left(w^{2}k^{2}+\omega_{\mathrm{c}}^{2}\right);\quad G_{\mathrm{B}}\left(k,\omega\right)\stackrel{\mathrm{def}}{=}\left(\omega-\mathring{v}k\right)^{2}-\omega_{\mathrm{rp}}^{2};\quad G_{\mathrm{cT}}\left(k,\omega\right)\stackrel{\mathrm{def}}{=}k^{2}\left(\omega^{2}-\omega_{\mathrm{c}}^{2}\right),\label{eq:disomfac1b}
\end{equation}
where $G_{\mathrm{T}}\left(k,\omega\right)$ and $G_{\mathrm{B}}\left(k,\omega\right)$
are the dispersion functions related to respectively the GTL and the
e-beam. Then the TWT dispersion relation (\ref{eq:disomfac1a}) can
be readily recast as
\begin{equation}
G_{\mathrm{T}}\left(k,\omega\right)G_{\mathrm{B}}\left(k,\omega\right)=\gamma G_{\mathrm{cT}}\left(k,\omega\right),\label{eq:disomfac1c}
\end{equation}
where $\gamma$ \emph{can be naturally interpreted as the coupling
parameter} and $G_{\mathrm{cT}}\left(k,\omega\right)$ as a \emph{supplementary
coupling function} that dependents only on the GTL through its cutoff
frequency $\omega_{\mathrm{c}}$. 

Using equations (\ref{eq:disomfac1b}) we can recast the dispersion
relation in the form (\ref{eq:dispkuom1b}) as follows
\begin{gather}
R_{\mathrm{TB}}\left(k,\omega\right)\stackrel{\mathrm{def}}{=}\frac{G_{\mathrm{T}}\left(k,\omega\right)G_{\mathrm{B}}\left(k,\omega\right)}{G_{\mathrm{cT}}\left(k,\omega\right)}=\gamma,\quad k\neq0,\quad\omega^{2}\neq\omega_{\mathrm{c}}^{2},\label{eq:disomfac3a}\\
R_{\mathrm{TB}}\left(k,\omega\right)=\left[\frac{1}{k^{2}}-\frac{w^{2}}{\omega^{2}-\omega_{\mathrm{c}}^{2}}\right]\left[\left(\omega-\mathring{v}k\right)^{2}-\omega_{\mathrm{rp}}^{2}\right].\label{eq:disomfac3b}
\end{gather}
The significance of the TWT dispersion relation in the form (\ref{eq:disomfac3a}),
(\ref{eq:disomfac3b}) is that its right-hand side is just a coupling
constant $\gamma$. Consequently, these equations represent the TWT
dispersion curves as the \emph{level curves} of the factorized function
$R_{\mathrm{TB}}\left(k,\omega\right)$ with its first and the second
factors associated respectively with the GTL and the e-beam. In particular,
the coupling parameter $\gamma$ appears in equation (\ref{eq:disomfac3a})
as the ``level'' of function $R_{\mathrm{TB}}\left(k,\omega\right)$.
The factorized form (\ref{eq:disomfac3b}) for function $R_{\mathrm{TB}}\left(k,\omega\right)$
and equation (\ref{eq:disomfac3a}) demonstrate once again that parameter
$\gamma$ plays a role of a coupling parameter between the GTL and
the e-beam. 

If we set in equation (\ref{eq:disomfac1a}) $\gamma=0$ the equation
turns readily into the dispersion relations (\ref{eq:disomfac1d})
of non-interacting GTL and the e-beam, namely
\begin{equation}
G_{\mathrm{T}}\left(k,\omega\right)=\omega^{2}-\left(w^{2}k^{2}+\omega_{\mathrm{c}}^{2}\right)=0,\quad G_{\mathrm{B}}\left(k,\omega\right)=\left(\omega-\mathring{v}k\right)^{2}-\omega_{\mathrm{rp}}^{2}=0,\quad\gamma=0.\label{eq:disomfac1d}
\end{equation}
It is worth noting that the dispersion relations (\ref{eq:disomfac1d})
for non-interacting GTL and e-beam clearly identify \emph{two natural
frequency scales: (i) the cutoff frequency $\omega_{\mathrm{c}}$
associated with the GTL; (ii) the plasma frequency $\omega_{\mathrm{p}}$
associated with the e-beam.}

Note also that if $\gamma\rightarrow\infty$ then dividing both sides
of equation (\ref{eq:disomfac1c}) by $\gamma$ we obtain in the limit
\begin{equation}
G_{\mathrm{cT}}\left(k,\omega\right)=k^{2}\left(\omega^{2}-\omega_{\mathrm{c}}^{2}\right)=0,\quad\gamma\rightarrow\infty.\label{eq:disomfac1e}
\end{equation}

As to the TWT velocity the dispersion relations (\ref{eq:dispkuom1c})
we can proceed similarly to the case of the TWT dispersion relations
(\ref{eq:dispkuom1b}). We first introduce \emph{dispersion functions}
$F_{\mathrm{T}}\left(u\right)$ and $F_{\mathrm{B}}\left(u\right)$
corresponding respectively to the GTL and the e-beam as follows:
\begin{equation}
F_{\mathrm{T}}\left(u\right)\stackrel{\mathrm{def}}{=}u^{2}-\frac{w^{2}\omega^{2}}{\omega^{2}-\omega_{\mathrm{c}}^{2}},\quad F_{\mathrm{B}}\left(u\right)\stackrel{\mathrm{def}}{=}\frac{\left(u-\mathring{v}\right)^{2}}{u^{2}}-\frac{\omega_{\mathrm{rp}}^{2}}{\omega^{2}},\quad u=\frac{\omega}{k},\label{eq:cofdisp1c}
\end{equation}
where $u$ is the phase velocity. Then the TWT velocity dispersion
relations (\ref{eq:dispkuom1c}) takes the form
\begin{equation}
F_{\mathrm{T}}\left(u\right)F_{\mathrm{B}}\left(u\right)=\left[u^{2}-\frac{w^{2}\omega^{2}}{\omega^{2}-\omega_{\mathrm{c}}^{2}}\right]\left[\frac{\left(u-\mathring{v}\right)^{2}}{u^{2}}-\frac{\omega_{\mathrm{rp}}^{2}}{\omega^{2}}\right]=\gamma,\quad u=\frac{\omega}{k},\label{eq:cofdisp1d}
\end{equation}
Note for $\gamma=0$ we readily recover from the TWT velocity dispersion
equation (\ref{eq:cofdisp1d}) equations
\begin{equation}
F_{\mathrm{T}}\left(u\right)=u^{2}-\frac{w^{2}\omega^{2}}{\omega^{2}-\omega_{\mathrm{c}}^{2}}=0,\quad F_{\mathrm{B}}\left(u\right)=\frac{\left(u-\mathring{v}\right)^{2}}{u^{2}}-\frac{\omega_{\mathrm{rp}}^{2}}{\omega^{2}}=0,\label{eq:cofdisp1e}
\end{equation}
that match the velocity dispersion relations (\ref{eq:cofdisp1u})
for non-interacting GTL and the e-beam.

Recall that the Pierce theory emerges from our field theory as the
high-frequency limit $\omega\rightarrow\infty$, \citet[Chap. 4.2, 29, 62]{FigTWTbk}.
Consequently the high-frequency limit of dispersion relations (\ref{eq:cofdisp1d})
is
\begin{equation}
\left(u^{2}-w^{2}\right)\frac{\left(u-\mathring{v}\right)^{2}}{u^{2}}=\gamma,\quad u=\frac{\omega}{k},\quad\omega\rightarrow\infty,\label{eq:cofdisp2c}
\end{equation}
\begin{equation}
F\left(u\right)=\frac{\left(u^{2}-{\it \chi}^{2}\right)\left(u-1\right)^{2}}{u^{2}}=\gamma>0.\label{eq:charP1a-1}
\end{equation}
which is exactly the velocity dispersion relation for the Pierce theory,
\citet[Chap. 4.2, 29, 62]{FigTWTbk}.

\section{Dimensionless set up\label{sec:dim-setup}}

We pursue here a dimensionless setup for the theory developed in previous
sections. We often use symbol ``prime'' to indicate that we are
dealing with the dimensionless version of a variable or a parameter.
It is often convenient to use the following dimensionless variables
and parameters, \citet[Chap. 3, 4, 30]{FigTWTbk}.
\[
u^{\prime}=\frac{u}{\mathring{v}},\quad\gamma^{\prime}=\frac{\gamma}{\mathring{v}{}^{2}},\quad\omega^{\prime}=\frac{\omega}{\omega_{\mathrm{rp}}},\quad\rho\stackrel{\mathrm{def}}{=}\frac{\omega}{\omega_{\mathrm{rp}}}\quad\chi\stackrel{\mathrm{def}}{=}\frac{w}{\mathring{v}},
\]
where when defining dimensionless parameter $\rho$ we tacitly assume
that $\omega_{\mathrm{rp}}>0$. The definitions of the GTL and e-beam
parameters are provided in Section \ref{subsec:coldTB}.

For the reader convenience expressions for dimensionless variables
and dimensionless TWT parameters are collected respectively in Tables
\ref{tab:ebeam-nodim} and \ref{tab:TWT-nodim}
\begin{table}[h]
\centering{}%
\begin{tabular}{|l||r|}
\hline 
\noalign{\vskip\doublerulesep}
Frequency & $\omega^{\prime}=\frac{\omega}{\omega_{\mathrm{rp}}},\quad\omega_{\mathrm{rp}}=R_{\mathrm{sc}}\omega_{\mathrm{p}},\quad\omega_{\mathrm{p}}=\sqrt{\frac{4\pi\mathring{n}e^{2}}{m}}$\tabularnewline[0.2cm]
\hline 
\noalign{\vskip\doublerulesep}
Velocity & $v^{\prime}=\frac{v}{\mathring{v}}$\tabularnewline[0.2cm]
\hline 
\noalign{\vskip\doublerulesep}
Wavenumber & $k^{\prime}=\frac{k}{k_{\mathrm{rp}}},\quad k_{\mathrm{rp}}=\frac{\omega_{\mathrm{rp}}}{\mathring{v}}=\frac{R_{\mathrm{sc}}\omega_{\mathrm{p}}}{\mathring{v}}$\tabularnewline[0.2cm]
\hline 
\noalign{\vskip\doublerulesep}
Length & $\lambda^{\prime}=\frac{\lambda}{\lambda_{\mathrm{rp}}},\quad\lambda_{\mathrm{rp}}=\frac{2\pi}{k_{\mathrm{rp}}}=\frac{2\pi\mathring{v}}{\omega_{\mathrm{rp}}}$\tabularnewline[0.2cm]
\hline 
\noalign{\vskip\doublerulesep}
Time & $t^{\prime}=\frac{t}{\tau_{\mathrm{rp}}},\quad\tau_{\mathrm{rp}}=\frac{2\pi}{\omega_{\mathrm{rp}}}$\tabularnewline[0.2cm]
\hline 
\end{tabular}\vspace{0.3cm}
\caption{\label{tab:ebeam-nodim}Dimensionless primary variables based on the
e-beam natural units, see Table \ref{tab:ebeam-unit} .}
\end{table}
The main dimensionless TWT parameters are collected in Table \ref{tab:TWT-nodim}.
\begin{table}[tbh]
\centering{}%
\begin{tabular}{|r||r|}
\hline 
\noalign{\vskip\doublerulesep}
$\chi=\frac{w}{\mathring{v}}=\frac{1}{\mathring{v}\sqrt{CL}}$ & TL dim-less phase velocity\tabularnewline
\hline 
\noalign{\vskip\doublerulesep}
$\rho=\frac{\omega_{\mathrm{c}}}{\omega_{\mathrm{rp}}}=\frac{1}{\omega_{\mathrm{rp}}\sqrt{C_{\mathrm{c}}L}}$ & GTL dimensionless cutoff frequency\tabularnewline
\hline 
\noalign{\vskip\doublerulesep}
$\beta^{\prime}=\frac{\beta}{\mathring{v}^{2}}=\frac{\pi\sigma_{\mathrm{B}}}{\lambda_{\mathrm{rp}}^{2}}=\frac{4\pi g_{\mathrm{B}}}{\lambda_{\mathrm{rp}}}$ & dimensionless e-beam intensity\tabularnewline
\hline 
\noalign{\vskip\doublerulesep}
$\gamma^{\prime}=\frac{b^{2}}{C}\beta^{\prime}=\frac{b^{2}\beta}{C\mathring{v}^{2}}=\frac{b^{2}\pi\sigma_{\mathrm{B}}}{C\lambda_{\mathrm{rp}}^{2}}$ & dimensionless TWT parameter\tabularnewline
\hline 
\end{tabular}\vspace{0.3cm}
\caption{\label{tab:TWT-nodim}Dimensionless TWT variables, see Table \ref{tab:ebeam-dim}.}
\end{table}

\emph{To avoid cluttered equations we often omit ``prime'' symbol
when writing expressions with dimensionless variables and parameters
presuming that it is clear from the context if the dimensionless variables
are used. For instance, it is the case when dimensionless parameters
$\chi$ and/or $\rho$ enter the relevant expressions.}

The dimensionless form of equations (\ref{eq:LaTBq4b}) and (\ref{eq:LaTBq5a})
is

\begin{equation}
M_{k\omega}x=0,\quad M_{k\omega}=\left[\begin{array}{rr}
k^{2}+\frac{\rho^{2}-\omega^{2}}{\chi^{2}} & bk^{2}\\
bk^{2} & b^{2}\left[k^{2}-\frac{\left(\omega-k\right)^{2}-1}{\gamma}\right]
\end{array}\right],\quad x=\left[\begin{array}{r}
\hat{Q}\\
\hat{q}
\end{array}\right]\label{eq:MELdim1a}
\end{equation}
\begin{equation}
M_{u\omega}x=0,\quad M_{u\omega}=\left[\begin{array}{rr}
\frac{\omega^{2}}{u^{2}}+\frac{\rho^{2}-\omega^{2}}{\chi^{2}} & \frac{b\omega^{2}}{u^{2}}\\
\frac{b\omega^{2}}{u^{2}} & \left[\frac{\omega^{2}}{u^{2}}+\frac{1}{\gamma}\left(1-\frac{\omega^{2}\left(u-1\right)^{2}}{u^{2}}\right)\right]b^{2}
\end{array}\right],\quad x=\left[\begin{array}{r}
\hat{Q}\\
\hat{q}
\end{array}\right]\label{eq:MELdim1b}
\end{equation}
The\emph{ dimensionless form of the TWT dispersion relations} (\ref{eq:dispkuom1b})
and (\ref{eq:disomfac1a}) is
\begin{equation}
\left[\frac{1}{k^{2}}-\frac{\chi^{2}}{\omega^{2}-\rho^{2}}\right]\left[\left(\omega-k\right)^{2}-1\right]=\gamma,\label{eq:dispdim1a}
\end{equation}

\begin{equation}
\left[\omega^{2}-\left(\chi^{2}k^{2}+\rho^{2}\right)\right]\left[\left(k-\omega\right)^{2}-1\right]=\gamma k^{2}\left(\omega^{2}-\rho^{2}\right),\label{eq:dispdim1b}
\end{equation}
The \emph{dimensionless form of the TWT velocity dispersion relation}s
(\ref{eq:dispkuom1c}) is

\begin{equation}
\left[u^{2}-\frac{\chi^{2}\omega^{2}}{\omega^{2}-\rho^{2}}\right]\left[\frac{\left(u-1\right)^{2}}{u^{2}}-\frac{1}{\omega^{2}}\right]=\gamma,\quad u=\frac{\omega}{k},\label{eq:dispdim1c}
\end{equation}
The dimensionless form of the TWT velocity dispersion relations (\ref{eq:cofdisp2c})
for the Pierce theory, which is the high-frequency limit $\omega\rightarrow\infty$
of our theory, \citet[Chap. 4.2, 29, 62]{FigTWTbk}, is
\begin{equation}
\frac{\left(u^{2}-{\it \chi}^{2}\right)\left(u-1\right)^{2}}{u^{2}}=\gamma,\quad u=\frac{\omega}{k}.\label{eq:dispdim1d}
\end{equation}
The dimensionless form of the dispersion functions (\ref{eq:disomfac1b})
is
\begin{equation}
G_{\mathrm{T}}\left(k,\omega\right)\stackrel{\mathrm{def}}{=}\omega^{2}-\left(\chi^{2}k^{2}+\rho^{2}\right);\quad G_{\mathrm{B}}\left(k,\omega\right)\stackrel{\mathrm{def}}{=}\left(k-\omega\right)^{2}-1;\quad G_{\mathrm{cT}}\left(k,\omega\right)\stackrel{\mathrm{def}}{=}k^{2}\left(\omega^{2}-\rho^{2}\right),\label{eq:dispdim2a}
\end{equation}
The dimensionless form of the dispersion relations (\ref{eq:disomfac1d})
for non-interacting GTL and the e-beam is
\begin{equation}
G_{\mathrm{T}}\left(k,\omega\right)=\omega^{2}-\left(\chi^{2}k^{2}+\rho^{2}\right)=0,\quad G_{\mathrm{B}}\left(k,\omega\right)=\left(k-\omega\right)^{2}-1=0,\quad\gamma=0.\label{eq:dispdim2b}
\end{equation}
Finally, the dimensionless form of the TWT dispersion relations (\ref{eq:disomfac3a})
and (\ref{eq:disomfac3b}) is
\begin{gather}
R_{\mathrm{TB}}\left(k,\omega\right)\stackrel{\mathrm{def}}{=}\frac{G_{\mathrm{T}}\left(k,\omega\right)G_{\mathrm{B}}\left(k,\omega\right)}{G_{\mathrm{cT}}\left(k,\omega\right)}=\gamma,\quad k\neq0,\quad\omega^{2}\neq\rho^{2},\label{eq:dispdim2c}\\
R_{\mathrm{TB}}\left(k,\omega\right)=\left[\frac{1}{k^{2}}-\frac{\chi^{2}}{\omega^{2}-\rho^{2}}\right]\left[\left(k-\omega\right)^{2}-1\right].\label{eq:dispdim2d}
\end{gather}

\section{Dispersion graphs and domains\label{sec:disp-dom}}

In this section we introduce and study geometric and topological concepts
related to the TWT dispersion relations (\ref{eq:dispdim1a}), (\ref{eq:dispdim1b}),
(\ref{eq:dispdim2c}) and (\ref{eq:dispdim2d}) as well as the GTL
and e-beam dispersion relations (\ref{eq:dispdim2b}). The topological
concepts we have in mind are the structural features of the connected
components of the relevant dispersion curves. Importantly their qualitative
properties the those features dependent on values of parameters $\gamma$,
$\chi$ and $\rho$ . In particular, we find that topological properties
of the dispersion curves depend significantly on whether $\chi<1$
or $\chi>1$, and on whether $\rho<1$ or $\rho>1$. Our studies of
geometric and topological properties of the dispersion curves allow
to see how the features of the non-interacting GTL and the e-beam
are related to and integrated into the TWT dispersion relations. As
it was already pointed out the TWT principle parameter $\gamma$ plays
also a role of the coupling constant. As $\gamma$ gets smaller some
parts of the TWT dispersion curves get closer to the GTL and the e-beam
dispersion curves as one might expect.

\subsection{Special points of the dispersion curves\label{subsec:specpo}}

Note that if a point $\left(k,\omega\right)$ being on the graph of
the TWT dispersion relations (\ref{eq:dispdim1b}) belongs also at
least one of the graphs of the dispersion relations of the TL or the
e-beam then the right-hand side $\gamma k^{2}\left(\omega^{2}-\rho^{2}\right)$
of equation (\ref{eq:dispdim1b}) must vanish. Then at least one of
equations $k=0$, $\omega=\pm\rho$ must hold. On the other hand if
either $k=0$ or $\omega=\pm\rho$ then at least of equations (\ref{eq:dispdim2b})
must hold. These observations identify in generic case $\rho\neq1$
the following eight (or 10 if the multiplicity is counted) points
$\left(k,\omega\right)$ that always satisfy the TWT dispersion relation
(\ref{eq:dispdim1b}): 
\begin{gather}
\left(0,\rho\right),\quad\left(0,\rho\right),\quad\left(0,1\right),\quad\left(\rho-1,\rho\right),\quad\left(\rho+1,\rho\right),\quad\rho=\frac{\omega_{\mathrm{c}}}{\omega_{\mathrm{rp}}}\label{eq:disomfac2a}\\
\left(0,-\rho\right),\quad\left(0,-\rho\right),\quad\left(0,-1\right),\quad\left(-\left(\rho-1\right),-\rho\right),\quad\left(-\left(\rho+1\right),-\rho\right).\nonumber 
\end{gather}
We refer to points (\ref{eq:disomfac2a}) as \emph{focal points} for
the TWT dispersion curves for all $\gamma>0$ pass through this points,
see Section \ref{sec:disp-curv} and Figures \ref{fig:dis-Lev1},
\ref{fig:dis-Lev1L}, \ref{fig:dis-Lev1LC}-\ref{fig:dis-Lev7}

Note also that for $\omega=0$ equation (\ref{eq:dispdim1b}) yields
the following equation for $k$
\begin{equation}
\left(\chi^{2}k^{2}+\rho^{2}\right)\left(k^{2}-1\right)=-\gamma k^{2}\rho^{2},\quad\omega=0.\label{eq:disomfac2b}
\end{equation}
Solving the above equations for $k^{2}$ we obtain the following expressions
for the solutions
\begin{equation}
k^{2}=\frac{\rho^{2}\left(\gamma-1\right)+\chi^{2}+\sqrt{\left(\rho^{2}\left(\gamma-1\right)+\chi^{2}\right)^{2}+4\chi^{2}\rho^{2}}}{2\chi^{2}},\label{eq:disomfac2c}
\end{equation}
\begin{equation}
k^{2}=-\frac{2\rho^{2}}{\rho^{2}\left(\gamma-1\right)+\chi^{2}+\sqrt{\left(\rho^{2}\left(\gamma-1\right)+\chi^{2}\right)^{2}+4\chi^{2}\rho^{2}}}.\label{eq:disomfac2d}
\end{equation}
In particular in case of $\rho=0$ the equations (\ref{eq:disomfac2c})
and (\ref{eq:disomfac2d}) are reduced to
\begin{equation}
k^{2}=1,\quad k^{2}=0.\label{eq:disomfac2e}
\end{equation}

\subsection{Cross-points\label{subsec:crospo}}

Another set of important points that are commonly considered in the
theory of TWTs are the \emph{cross-points} (intersection points) of
the graphs of the GTL and e-beam dispersion relations (\ref{eq:dispdim2b})
defined as solutions to the following system of equations:
\begin{equation}
G_{\mathrm{T}}\left(k,\omega\right)=\omega^{2}-\left(\chi^{2}k^{2}+\rho^{2}\right)=0,\quad G_{\mathrm{B}}\left(k,\omega\right)=\left(k-\omega\right)^{2}-1=0.\label{eq:disGTBr1a}
\end{equation}
\begin{figure}[h]
\begin{centering}
\includegraphics[scale=0.5]{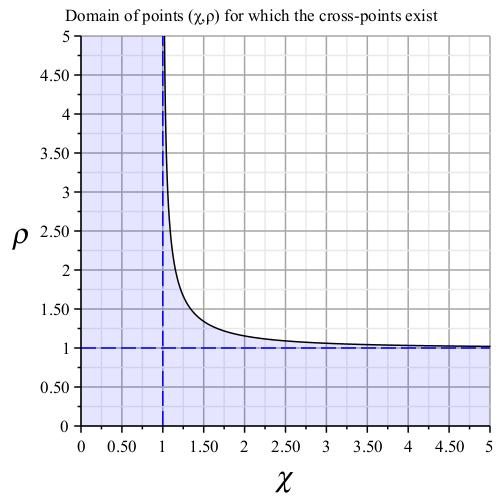}
\par\end{centering}
\centering{}\caption{\label{fig:rhor-cr} The plot of the critical value function $\rho_{\mathrm{cr}}\left(\chi\right)=\frac{\chi}{\sqrt{\chi^{2}-1}}$
for $\chi>1$. The shaded area represent all points $\left(\chi,\rho\right)$
for which there are exactly 4 cross-points. The blue dashed lines
represent the asymptotes of $\rho_{\mathrm{cr}}\left(\chi\right)$
as $\chi\rightarrow1$ and $\chi\rightarrow\infty$.}
\end{figure}
In other words, a cross-point is a point $\left(k,\omega\right)$
at which the dispersion relations for the GTL and the e-beam hold
simultaneously. The solutions to the system of equations (\ref{eq:disGTBr1a})
can be readily found as follows. We solve first equation $G_{\mathrm{B}}\left(k,\omega\right)=0$
yielding two solutions $\omega=k\pm1$. Then we substitute for $\omega$
in the equation $G_{\mathrm{T}}\left(k,\omega\right)=0$ these two
solutions obtaining two quadratic equations for $k$. Solving those
quadratic equations for $k$ we find exactly four (counting the multiplicity)
solutions (cross-points) to the system of equations (\ref{eq:disGTBr1a}):

\begin{gather}
\left(\frac{\pm1+d}{\chi^{2}-1},\frac{\pm1+d}{\chi^{2}-1}\pm1\right),\quad\left(\frac{\pm1-d}{\chi^{2}-1},\frac{\pm1-d}{\chi^{2}-1}\pm1\right),\label{eq:disGTBr1b}\\
d\stackrel{\mathrm{def}}{=}d\left(\chi,\rho\right)=\sqrt{\rho^{2}+\left(1-\rho^{2}\right)\chi^{2}}=\sqrt{\rho^{2}\left(1-\chi^{2}\right)+\chi^{2}}.\label{eq:disGTBr1c}
\end{gather}
We note then either all four points defined by equations (\ref{eq:disGTBr1b})
and (\ref{eq:disGTBr1c}) have real components or they have complex-valued
components depending on whether or not $d^{2}\left(\chi,\rho\right)\geq0$.
If $0<\chi\leq1$ then it readily follows from equation (\ref{eq:disGTBr1c})
that $d\geq0$ and consequently all four solutions have real-valued
components implying that we have exactly four (counting multiplicity)
cross-points.

\emph{The cross-points are a commonly used in the TWT design since
as one might expect the interaction between the GTL and the e-beam
to be the strongest in a vicinity of these points}.

If $\chi>1$ then the condition $d^{2}\left(\chi,\rho\right)\geq0$
can be recast as
\begin{equation}
\rho\leq\rho_{\mathrm{cr}}\left(\chi\right)\stackrel{\mathrm{def}}{=}\frac{\chi}{\sqrt{\chi^{2}-1}},\quad\chi>1.\label{eq:disGTBr1d}
\end{equation}
Evidently function $\rho_{\mathrm{cr}}\left(\chi\right)$ defined
in relations (\ref{eq:disGTBr1d}) determines the critical value $\rho_{\mathrm{cr}}\left(\chi\right)$
such that for $\chi>1$ the validity of inequality $\rho\leq\rho_{\mathrm{cr}}\left(\chi\right)$
determines if there are exactly four (counting multiplicity) cross-points
or there are none. Since $\rho=\frac{\omega_{\mathrm{c}}}{\omega_{\mathrm{rp}}}$
we may view $\rho_{\mathrm{cr}}\left(\chi\right)$ can be viewed as
\emph{critical vale of the relative low frequency cutoff} $\rho=\frac{\omega_{\mathrm{c}}}{\omega_{\mathrm{rp}}}$.
In other words $\rho=\frac{\omega_{\mathrm{c}}}{\omega_{\mathrm{rp}}}$
has to be smaller than $\rho_{\mathrm{cr}}\left(\chi\right)$ for
the cross-points to exist.

Fig. \ref{fig:rhor-cr} shows the plot of the critical low frequency
cutoff $\rho_{\mathrm{cr}}\left(\chi\right)$. The corresponding critical
value $\omega_{\mathrm{ccr}}$ of the the low cutoff frequency $\omega_{\mathrm{c}}$
is
\begin{equation}
\omega_{\mathrm{ccr}}=\frac{\chi}{\sqrt{\chi^{2}-1}}\omega_{\mathrm{rp}},\label{eq:GTB2ra}
\end{equation}
and $\omega_{\mathrm{c}}$ must be smaller than $\omega_{\mathrm{ccr}}$
for the cross-points to exist. Note also the straight line defined
by equation
\begin{equation}
\omega=k+\frac{\rho\sqrt{\chi^{2}-1}}{\chi},\quad\chi>1,\label{eq:GTB2rb}
\end{equation}
 is tangent to the GTL dispersion curve at the following point $\left(k,\omega\right):$
\begin{equation}
\left(k,\omega\right)=\left(\frac{\rho}{\chi\sqrt{\chi^{2}-1}},\frac{\rho\chi}{\sqrt{\chi^{2}-1}}\right),\quad\chi>1.\label{eq:GTB2rc}
\end{equation}

\subsection{Dispersion graphs\label{subsec:disp-graph}}

We start with representing geometrically the dispersion relations
(\ref{eq:dispdim2b}) and (\ref{eq:dispdim2c}) for the TWT, the GTL
and the e-beam as follows:
\begin{equation}
\mathrm{Gr}_{\mathrm{TB}}\left(\gamma\right)\stackrel{\mathrm{def}}{=}\left\{ \left(k,\omega\right):\left[\omega^{2}-\left(\chi^{2}k^{2}+\rho^{2}\right)\right]\left[\left(k-\omega\right)^{2}-1\right]=\gamma k^{2}\left(\omega^{2}-\rho^{2}\right)\right\} ,\quad\gamma>0,\label{eq:disdom1a}
\end{equation}
\begin{equation}
\mathrm{Gr}_{\mathrm{T}}\stackrel{\mathrm{def}}{=}\left\{ \left(k,\omega\right):G_{\mathrm{T}}\left(k,\omega\right)=\omega^{2}-\left(\chi^{2}k^{2}+\rho^{2}\right)=0\right\} ,\label{eq:disdom1b}
\end{equation}
\begin{equation}
\mathrm{Gr}_{\mathrm{B}}\stackrel{\mathrm{def}}{=}\left\{ \left(k,\omega\right):G_{\mathrm{T}}\left(k,\omega\right)=\left(k-\omega\right)^{2}-1=0\right\} ,\label{eq:disdom1c}
\end{equation}
\begin{equation}
\mathrm{Gr}_{\mathrm{cT}}\stackrel{\mathrm{def}}{=}\left\{ \left(k,\omega\right):G_{\mathrm{cT}}\left(k,\omega\right)=k^{2}\left(\omega^{2}-\rho^{2}\right)=0\right\} .\label{eq:disdom1d}
\end{equation}
\begin{equation}
\mathrm{Gr}_{\mathrm{TB}}\left(0\right)\stackrel{\mathrm{def}}{=}\mathrm{Gr}_{\mathrm{T}}\bigcup\mathrm{Gr}_{\mathrm{B}},\label{eq:disdom1e}
\end{equation}
\begin{equation}
\mathrm{Gr}_{\mathrm{TB}}\left(\infty\right)\stackrel{\mathrm{def}}{=}\left\{ \left(0,\omega\right):\left|\omega\right|\geq1\right\} \bigcup\left\{ \left(k,\rho\right),\left(k,-\rho\right):k\in\mathbb{R}\right\} .\label{eq:disdom1f}
\end{equation}
\begin{figure}[h]
\begin{centering}
\hspace{-0.5cm}\includegraphics[scale=0.38]{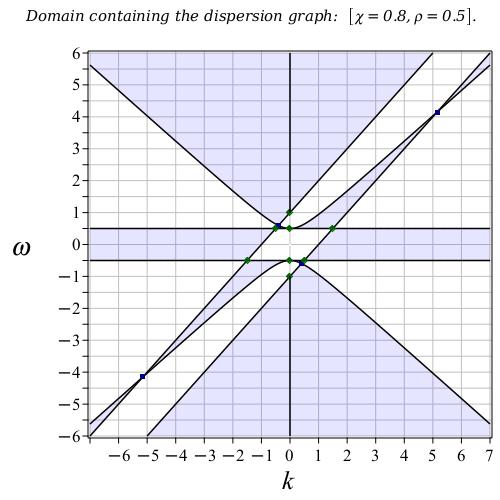}\hspace{2.5cm}\includegraphics[scale=0.38]{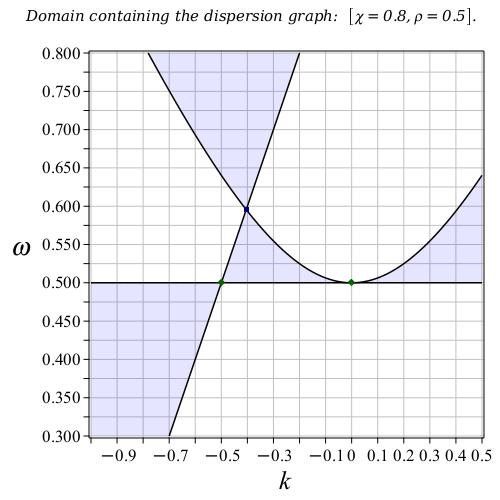}
\par\end{centering}
\centering{}\hspace{1cm}(a)\hspace{9cm}(b)\caption{\label{fig:disp-dom1} Dispersion domain $\mathbb{D}_{\mathrm{TB}}$
(where $R_{\mathrm{TB}}\left(k,\omega\right)>0$) is shown as shadowed
(blue) area for $\chi=0.8$ and $\rho=0.5<1$: (a) complete plot in
the designated window; (b) a zoomed fragment of the plot (a). Solid
(black) curves represent the boundary $\partial\mathbb{D}_{\mathrm{TB}}$
of dispersion domain $\mathbb{D}_{\mathrm{TB}}$, that is the union
$\mathrm{Gr}_{\mathrm{T}}\bigcup\mathrm{Gr}_{\mathrm{B}}\bigcup\mathrm{Gr}_{\mathrm{cT}}$
of graphs defined by equations (\ref{eq:disdom1b}), (\ref{eq:disdom1c})
and (\ref{eq:disdom1d}). Diamond (green) dots represent focal points
defined by equations (\ref{eq:disomfac2a}), square (blue) dots represent
the cross-points $\mathrm{Gr}_{\mathrm{T}}\bigcap\mathrm{Gr}_{\mathrm{B}}$.
For the chosen values of parameters the dispersion domain $\mathbb{D}_{\mathrm{TB}}$
has four unbounded connected components and four bounded ones for
$\left|\omega\right|>\left|\rho\right|$.}
\end{figure}
\begin{figure}[h]
\begin{centering}
\hspace{-0.5cm}\includegraphics[scale=0.42]{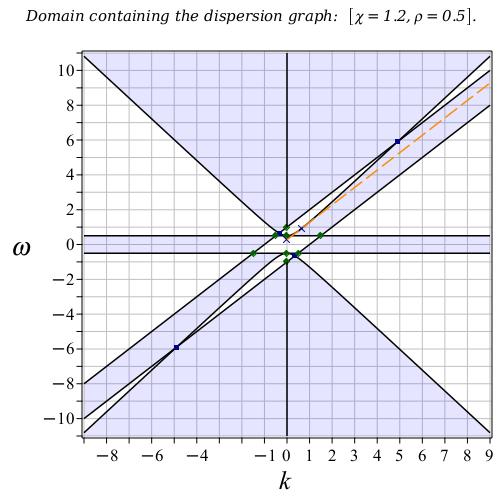}\hspace{1cm}\includegraphics[scale=0.42]{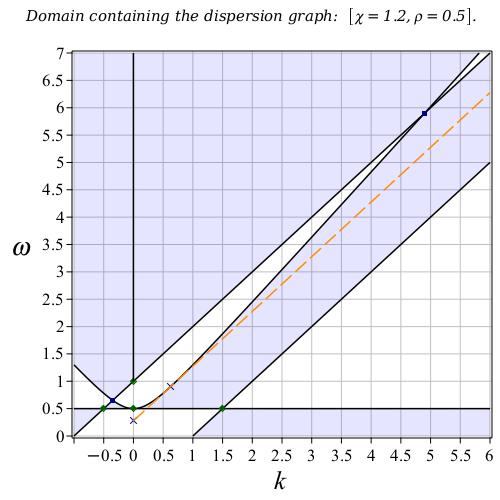}
\par\end{centering}
\centering{}(a)\hspace{8cm}(b)\caption{\label{fig:disp-dom2} Dispersion domain $\mathbb{D}_{\mathrm{TB}}$
(where $R_{\mathrm{TB}}\left(k,\omega\right)>0$) is shown as shadowed
(blue) area for $\chi=1.2$ and $\rho=0.5<1$: (a) complete plot in
the designated window; (b) a zoomed fragment of the plot (a). Solid
(black) curves represent the boundary $\partial\mathbb{D}_{\mathrm{TB}}$
of dispersion domain $\mathbb{D}_{\mathrm{TB}}$, that is the union
$\mathrm{Gr}_{\mathrm{T}}\bigcup\mathrm{Gr}_{\mathrm{B}}\bigcup\mathrm{Gr}_{\mathrm{cT}}$
of graphs defined by equations (\ref{eq:disdom1b}), (\ref{eq:disdom1c})
and (\ref{eq:disdom1d}). Diamond (green) dots represent focal points
defined by equations (\ref{eq:disomfac2a}), square (blue) dots represent
the cross-points $\mathrm{Gr}_{\mathrm{T}}\bigcap\mathrm{Gr}_{\mathrm{B}}$.
The orange dashed straight line described by equation (\ref{eq:GTB2rb})
is parallel to the e-beam lines $\omega=k\pm1$ and it is tangent
to the graph $\mathrm{Gr}_{\mathrm{T}}$. The corresponding diagonal
cross dot represents its intersection with $\omega$-axis described
by (\ref{eq:GTB2rc}). For the chosen values of parameters the dispersion
domain $\mathbb{D}_{\mathrm{TB}}$ has four unbounded connected components
and no bounded ones for $\left|\omega\right|>\left|\rho\right|$.}
\end{figure}
\begin{figure}[h]
\begin{centering}
\hspace{-0.5cm}\includegraphics[scale=0.42]{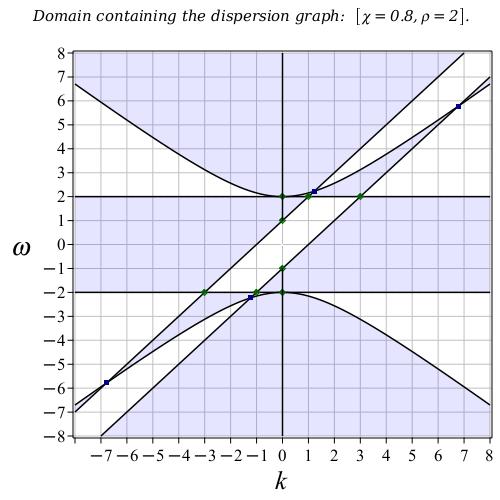}\hspace{1cm}\includegraphics[scale=0.42]{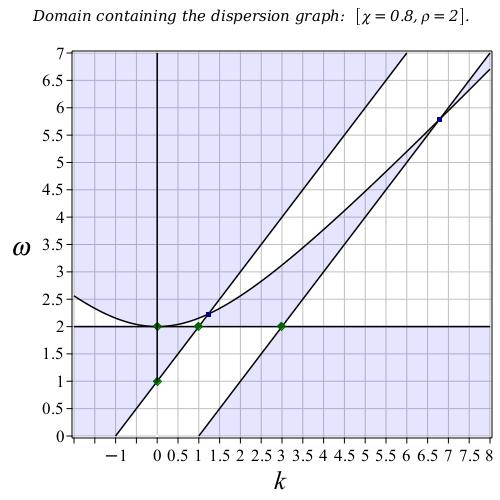}
\par\end{centering}
\centering{}(a)\hspace{8.5cm}(b)\caption{\label{fig:disp-dom3} Dispersion domain $\mathbb{D}_{\mathrm{TB}}$
(where $R_{\mathrm{TB}}\left(k,\omega\right)>0$) is shown as shadowed
(blue) area for $\chi=0.8$ and $\rho=2>1$: (a) complete plot in
the designated window; (b) a zoomed fragment of the plot (a). Solid
(black) curves represent the boundary $\partial\mathbb{D}_{\mathrm{TB}}$
of dispersion domain $\mathbb{D}_{\mathrm{TB}}$, that is the union
$\mathrm{Gr}_{\mathrm{T}}\bigcup\mathrm{Gr}_{\mathrm{B}}\bigcup\mathrm{Gr}_{\mathrm{cT}}$
of graphs defined by equations (\ref{eq:disdom1b}), (\ref{eq:disdom1c})
and (\ref{eq:disdom1d}). Diamond (green) dots represent focal points
defined by equations (\ref{eq:disomfac2a}), square (blue) dots represent
the cross-points $\mathrm{Gr}_{\mathrm{T}}\bigcap\mathrm{Gr}_{\mathrm{B}}$.
For the chosen values of parameters the dispersion domain $\mathbb{D}_{\mathrm{TB}}$
has four unbounded connected components and two bounded ones for $\left|\omega\right|>\left|\rho\right|$.}
\end{figure}
\begin{figure}[h]
\begin{centering}
\hspace{-0.5cm}\includegraphics[scale=0.42]{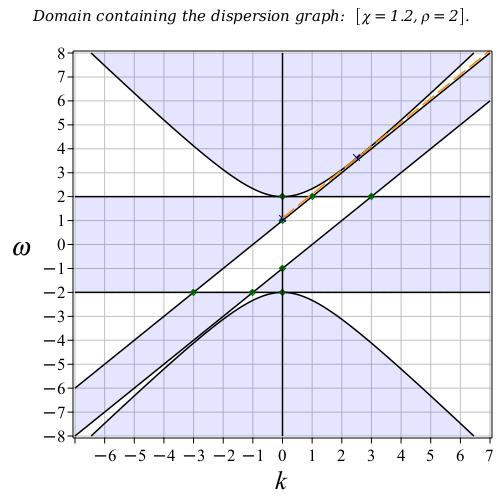}\hspace{1cm}\includegraphics[scale=0.42]{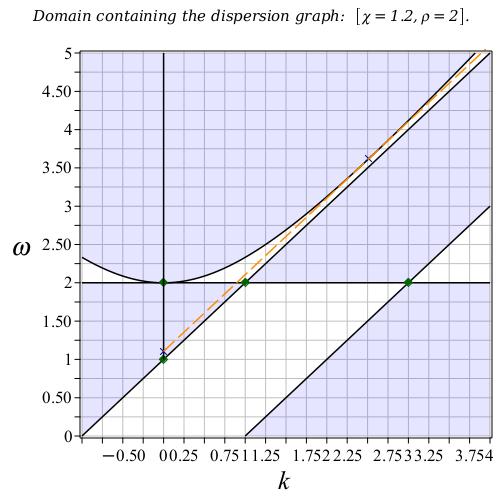}
\par\end{centering}
\centering{}\hspace{0.5cm}(a)\hspace{8cm}(b)\caption{\label{fig:disp-dom4} Dispersion domain $\mathbb{D}_{\mathrm{TB}}$
(where $R_{\mathrm{TB}}\left(k,\omega\right)>0$) is shown as shadowed
(blue) area for $\chi=1.2$ and $\rho=2>1$: (a) complete plot in
the designated window; (b) a zoomed fragment of the plot (a). Solid
(black) curves represent the boundary $\partial\mathbb{D}_{\mathrm{TB}}$
of dispersion domain $\mathbb{D}_{\mathrm{TB}}$, that is the union
$\mathrm{Gr}_{\mathrm{T}}\bigcup\mathrm{Gr}_{\mathrm{B}}\bigcup\mathrm{Gr}_{\mathrm{cT}}$
of graphs defined by equations (\ref{eq:disdom1b}), (\ref{eq:disdom1c})
and (\ref{eq:disdom1d}). Diamond (green) dots represent focal points
defined by equations (\ref{eq:disomfac2a}), square (blue) dots represent
the cross-points $\mathrm{Gr}_{\mathrm{T}}\bigcap\mathrm{Gr}_{\mathrm{B}}$.
The orange dashed straight line described by equation (\ref{eq:GTB2rb})
is parallel to the e-beam lines $\omega=k\pm1$ and it is tangent
to the graph $\mathrm{Gr}_{\mathrm{T}}$. The corresponding diagonal
cross dot represents its intersection with $\omega$-axis described
by (\ref{eq:GTB2rc}). For the chosen values of parameters the dispersion
domain $\mathbb{D}_{\mathrm{TB}}$ has four unbounded connected components
and no bounded ones for $\left|\omega\right|>\left|\rho\right|$.}
\end{figure}
Domain definitions (\ref{eq:disdom1d}) and (\ref{eq:disdom1f}) readily
imply that
\begin{equation}
\mathrm{Gr}_{\mathrm{TB}}\left(\infty\right)\subseteq\mathrm{Gr}_{\mathrm{cT}}.\label{eq:disdom1g}
\end{equation}

It is instructive to see what happens with the graphs $\mathrm{Gr}_{\mathrm{TB}}\left(\gamma\right)$
as $\gamma\rightarrow0$ and $\gamma\rightarrow+\infty$. The analysis
of the plots of the graphs suggests that for any square $\mathrm{Sq}\left(a\right)=\left[-a,a\right]\times\left[-a,a\right]$,
in $k\omega$-plane we have
\begin{equation}
\lim_{\gamma\rightarrow+0}\mathrm{Gr}_{\mathrm{TB}}\left(\gamma\right)\bigcap\mathrm{Sq}\left(a\right)=\left[\mathrm{Gr}_{\mathrm{T}}\bigcup\mathrm{Gr}_{\mathrm{B}}\right]\bigcap\mathrm{Sq}\left(a\right),\quad a>0,\label{eq:disdom2a}
\end{equation}
\begin{equation}
\lim_{\gamma\rightarrow+\infty}\mathrm{Gr}_{\mathrm{TB}}\left(\gamma\right)\bigcap\mathrm{Sq}\left(a\right)=\mathrm{Gr}_{\mathrm{TB}}\left(\infty\right)\bigcap\mathrm{Sq}\left(a\right),\quad a>0,\label{eq:disdom2b}
\end{equation}
that is points of the graph $\mathrm{Gr}_{\mathrm{TB}}\left(\gamma\right)$
residing also in square $\mathrm{Sq}\left(a\right)$ approach the
union of graphs $\mathrm{Gr}_{\mathrm{T}}\bigcup\mathrm{Gr}_{\mathrm{B}}$
as $\gamma\rightarrow+0$ and these points approach graph $\mathrm{Gr}_{\mathrm{TB}}\left(\infty\right)$
as $\gamma\rightarrow+\infty$. Indeed, since $\left|k\right|,\left|\omega\right|\leq a$
equations (\ref{eq:dispdim1b}) and (\ref{eq:dispdim2b}) imply limit
relations (\ref{eq:disdom2a}) and (\ref{eq:disdom2b}).

Let us introduce the following parallelogram domain
\begin{equation}
\Pi_{\mathrm{TB}}=\left\{ \left(k,\omega\right):k-1<\omega<k+1,\quad-\rho<\omega<\rho\right\} .\label{eq:graphkom1a}
\end{equation}
Note that based on an elementary analysis of the signs of the factors
$G_{\mathrm{T}}\left(k,\omega\right)$, $G_{\mathrm{B}}\left(k,\omega\right)$
and $G_{\mathrm{cT}}\left(k,\omega\right)$ and the definition (\ref{eq:dispdim2c}),
(\ref{eq:dispdim2d}) of function $R_{\mathrm{TB}}\left(k,\omega\right)$
we can obtain the following inequality
\begin{equation}
R_{\mathrm{TB}}\left(k,\omega\right)<0,\quad\left(k,\omega\right)\in\Pi_{\mathrm{TB}}.\label{eq:graphkom1b}
\end{equation}
Consequently, there no real-valued solutions to the dispersion equation
$R_{\mathrm{TB}}\left(k,\omega\right)=\gamma>0$ in parallelogram
$\Pi_{\mathrm{TB}}$, that is
\begin{equation}
\mathrm{Gr}_{\mathrm{TB}}\left(\gamma\right)\bigcap\Pi_{\mathrm{TB}}=\emptyset.\label{eq:graphkom1c}
\end{equation}
Note also that for $k=0$ the dispersion equation (\ref{eq:dispdim1b})
turns into
\begin{equation}
\left[\omega^{2}-\rho^{2}\right]\left[\omega^{2}-1\right]=0,\label{eq:graphkom1d}
\end{equation}
readily implying that it has no real-valued solutions for $\omega\neq\pm1$
with an exception of $\omega=\pm\rho$ in case when $\rho\neq1$.

\subsection{Dispersion domains\label{subsec:disp-dom}}

In this section we identify domains that contain the TWT dispersion
graphs $\mathrm{Gr}_{\mathrm{TB}}\left(\gamma\right)$ defined by
equations (\ref{eq:disdom1a}) based on the signs of the factors $G_{\mathrm{T}}\left(k,\omega\right)$,
$G_{\mathrm{B}}\left(k,\omega\right)$ and $G_{\mathrm{cT}}\left(k,\omega\right)$
entering the TWT dispersion relation (\ref{eq:dispdim1b}), (\ref{eq:dispdim1b})
, (\ref{eq:dispdim2c}) and (\ref{eq:dispdim2d}). We proceed with
introducing \emph{dispersion domains $\mathbb{D}_{\mathrm{TB}}$,
$\mathbb{D}_{\mathrm{T}}\left(\sigma\right)$} and $\mathbb{D}_{\mathrm{B}}\left(\sigma\right)$
for $\sigma=\pm1$ related respectively to the TWT, the GTL and the
e-beam as follows:
\begin{equation}
\mathbb{D}_{\mathrm{TB}}\stackrel{\mathrm{def}}{=}\left\{ \left(k,\omega\right)\in\mathbb{R}^{2}:R_{\mathrm{TB}}\left(k,\omega\right)>0\right\} =\bigcup_{\gamma>0}\mathrm{Gr}_{\mathrm{TB}}\left(\gamma\right),\label{eq:disdom3a}
\end{equation}
\begin{equation}
\mathbb{D}_{\mathrm{T}}\left(\sigma\right)\stackrel{\mathrm{def}}{=}\left\{ \left(k,\omega\right):\mathrm{sign}\left[G_{\mathrm{T}}\left(k,\omega\right)\right]=\mathrm{sign}\left[\omega^{2}-\left(\chi^{2}k^{2}+\rho^{2}\right)\right]=\sigma\right\} ,\quad\sigma=\pm1,\label{eq:disdom3b}
\end{equation}
\begin{equation}
\mathbb{D}_{\mathrm{B}}\left(\sigma\right)\stackrel{\mathrm{def}}{=}\left\{ \left(k,\omega\right):\mathrm{sign}\left[G_{\mathrm{B}}\left(k,\omega\right)\right]=\mathrm{sign}\left[\left(k-\omega\right)^{2}-1\right]=\sigma\right\} ,\quad\sigma=\pm1.\label{eq:disdom3c}
\end{equation}
In addition to that, we make use of domain
\begin{equation}
\mathbb{D}_{\mathrm{cT}}\left(\sigma\right)\stackrel{\mathrm{def}}{=}\left\{ \left(k,\omega\right):\mathrm{sign}\left[G_{\mathrm{cT}}\left(k,\omega\right)\right]=\mathrm{sign}\left[\omega^{2}-\rho^{2}\right]=\sigma\right\} ,\quad\sigma=\pm1.\label{eq:disdom3d}
\end{equation}
Based on definitions (\ref{eq:disdom1b}), (\ref{eq:disdom1c}) and
(\ref{eq:disdom1d}) we obtain
\begin{equation}
\partial\mathbb{D}_{\mathrm{TB}}=\mathrm{Gr}_{\mathrm{T}}\bigcup\mathrm{Gr}_{\mathrm{B}}\bigcup\mathrm{Gr}_{\mathrm{cT}}.\label{eq:disdom3e}
\end{equation}

\section{Dispersion curves for varying coupling\label{sec:disp-curv}}

\begin{figure}[h]
\begin{centering}
\hspace{-0.1cm}\includegraphics[scale=0.3]{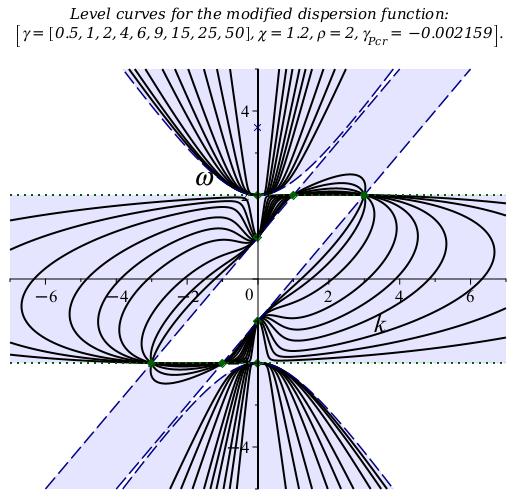}\hspace{0.1cm}\includegraphics[scale=0.3]{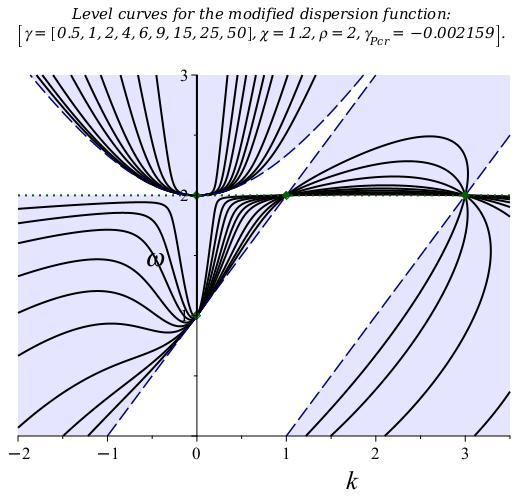}\hspace{0.1cm}\includegraphics[scale=0.3]{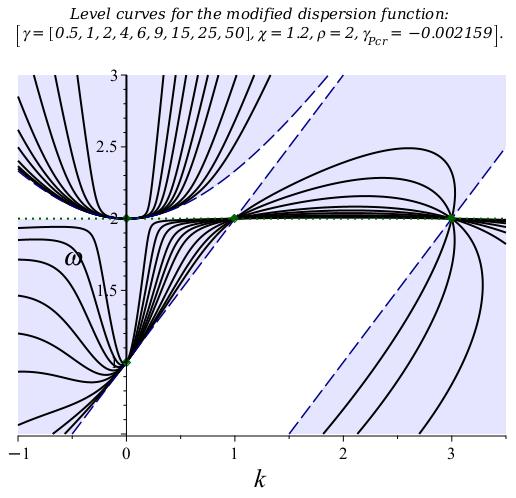}
\par\end{centering}
\centering{}(a)\hspace{5.5cm}(b)\hspace{5.5cm}(c)\caption{\label{fig:dis-Lev1LC} Dispersion curves of the TWT dispersion relations
(\ref{eq:dispdim1a}) and (\ref{eq:dispdim1b}) for $\chi=1.2$, $\rho=0.5<1$
and $\gamma=0.5,1,2,4,6,9,15,25,50$, $\gamma_{\mathrm{Pcr}}\protect\cong0.002641$:
(a) complete plot in the designated window; (b) a zoomed fragment
of (a); (c) a zoomed fragment of (b). Solid (black) curves represent
the TWT dispersion curves for indicated values of $\gamma$; dashed
(blue) curves represent the dispersion curves of non-interacting TL
and the e-beam for $\gamma=0$ as a reference. Shaded area identifies
the dispersion domain $\mathbb{D}_{\mathrm{TB}}$, that is where $R_{\mathrm{TB}}\left(k,\omega\right)>0$.
Doted (green) horizontal straight lines represents points $\left(k,\pm\rho\right)$.
Note the dispersion curves $\mathrm{Gr}_{\mathrm{TB}}\left(\gamma\right)$
pass through focal points defined by equations (\ref{eq:disomfac2a})
marked as circle (green) dots. Square (blue) dots mark the TWT cross-points
(see equations (\ref{eq:disGTBr1b}), (\ref{eq:disGTBr1c})). Note
also that the smaller $\gamma$ gets the closer graph $\mathrm{Gr}_{\mathrm{TB}}\left(\gamma\right)$
gets to $\mathrm{Gr}_{\mathrm{TB}}\left(0\right)=\mathrm{Gr}_{\mathrm{T}}\bigcup\mathrm{Gr}_{\mathrm{B}}$,
whereas the larger $\gamma$ gets the closer graph $\mathrm{Gr}_{\mathrm{TB}}\left(\gamma\right)$
gets to $\mathrm{Gr}_{\mathrm{TB}}\left(\infty\right)$ defined by
equation (\ref{eq:disdom1f}).}
\end{figure}

\begin{figure}[h]
\begin{centering}
\hspace{-0.1cm}\includegraphics[scale=0.33]{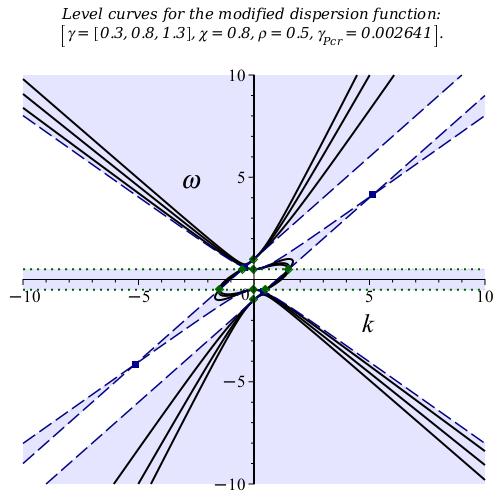}\hspace{0.1cm}\includegraphics[scale=0.33]{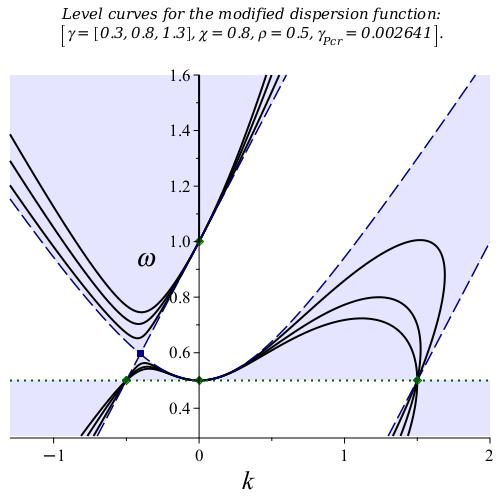}\hspace{0.1cm}\includegraphics[scale=0.33]{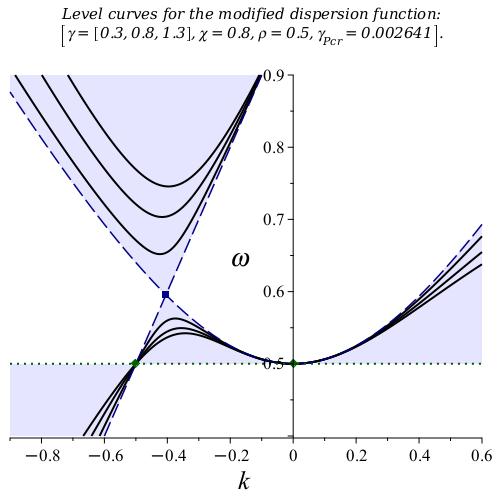}
\par\end{centering}
\centering{}(a)\hspace{5.5cm}(b)\hspace{5.5cm}(c)\caption{\label{fig:dis-Lev2} Dispersion curves of the TWT dispersion relations
(\ref{eq:dispdim1a}) and (\ref{eq:dispdim1b}) for $\chi=0.8$, $\rho=0.5<1$
and $\gamma=0.3,0.8,1.3$, $\gamma_{\mathrm{Pcr}}\protect\cong0.002641$:
(a) complete plot in the designated window; (b) a zoomed fragment
of (a); (c) a zoomed fragment of (b). Solid (black) curves represent
the TWT dispersion curves for indicated values of $\gamma$; dashed
(blue) curves represent the dispersion curves of non-interacting TL
and the e-beam for $\gamma=0$ as a reference. Shaded area identifies
the dispersion domain $\mathbb{D}_{\mathrm{TB}}$, that is where $R_{\mathrm{TB}}\left(k,\omega\right)>0$.
Doted (green) horizontal straight lines represents points $\left(k,\pm\rho\right)$.
Note the dispersion curves pass through focal points $\left(0,\pm1\right)$.
Note the dispersion curves $\mathrm{Gr}_{\mathrm{TB}}\left(\gamma\right)$
pass through focal points defined by equations (\ref{eq:disomfac2a})
marked as circle (green) dots. Square (blue) dots mark the TWT cross-points
(see equations (\ref{eq:disGTBr1b}), (\ref{eq:disGTBr1c})). Note
also that the smaller $\gamma$ gets the closer graph $\mathrm{Gr}_{\mathrm{TB}}\left(\gamma\right)$
gets to $\mathrm{Gr}_{\mathrm{TB}}\left(0\right)=\mathrm{Gr}_{\mathrm{T}}\bigcup\mathrm{Gr}_{\mathrm{B}}$,
whereas the larger $\gamma$ gets the closer graph $\mathrm{Gr}_{\mathrm{TB}}\left(\gamma\right)$
gets to $\mathrm{Gr}_{\mathrm{TB}}\left(\infty\right)$ defined by
equation (\ref{eq:disdom1f}).}
\end{figure}

\begin{figure}[h]
\begin{centering}
\hspace{-0.1cm}\includegraphics[scale=0.33]{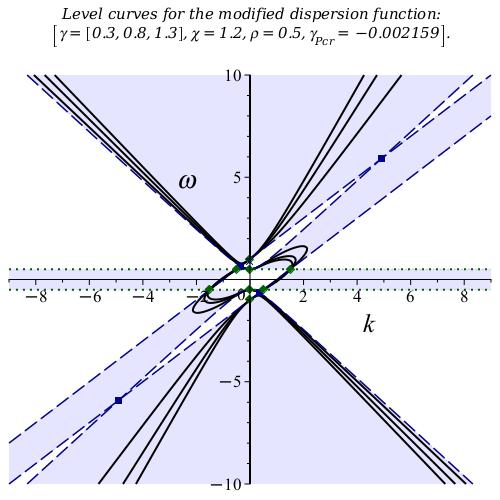}\hspace{0.1cm}\includegraphics[scale=0.33]{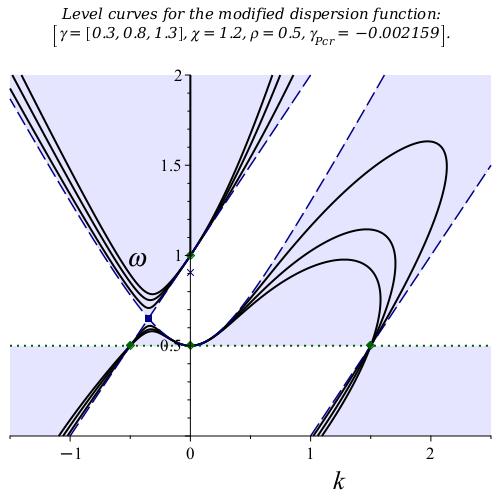}\hspace{0.1cm}\includegraphics[scale=0.33]{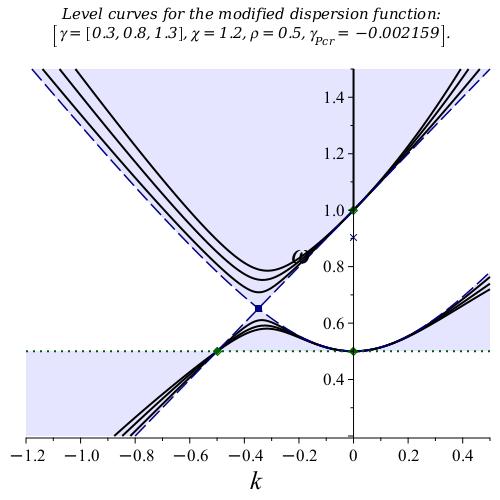}
\par\end{centering}
\centering{}(a)\hspace{5.5cm}(b)\hspace{5.5cm}(c)\caption{\label{fig:dis-Lev3} Dispersion curves of the TWT dispersion relations
(\ref{eq:dispdim1a}) and (\ref{eq:dispdim1b}) for $\chi=1.2$, $\rho=0.5<1$
and $\gamma=0.3,0.8,1.3$, $\gamma_{\mathrm{Pcr}}\protect\cong-0.002159$:
(a) complete plot in the designated window; (b) a zoomed fragment
of (a); (c) a zoomed fragment of (b). Solid (black) curves represent
the TWT dispersion curves for indicated values of $\gamma$; dashed
(blue) curves represent the dispersion curves of non-interacting TL
and the e-beam for $\gamma=0$ as a reference. Shaded area identifies
the dispersion domain $\mathbb{D}_{\mathrm{TB}}$, that is where $R_{\mathrm{TB}}\left(k,\omega\right)>0$.
Doted (green) horizontal straight lines represents points $\left(k,\pm\rho\right)$.
Note the dispersion curves pass through focal points $\left(0,\pm1\right)$.
Note the dispersion curves $\mathrm{Gr}_{\mathrm{TB}}\left(\gamma\right)$
pass through focal points defined by equations (\ref{eq:disomfac2a})
marked as circle (green) dots. Square (blue) dots mark the TWT cross-points
(see equations (\ref{eq:disGTBr1b}), (\ref{eq:disGTBr1c})). Note
also that the smaller $\gamma$ gets the closer graph $\mathrm{Gr}_{\mathrm{TB}}\left(\gamma\right)$
gets to $\mathrm{Gr}_{\mathrm{TB}}\left(0\right)=\mathrm{Gr}_{\mathrm{T}}\bigcup\mathrm{Gr}_{\mathrm{B}}$,
whereas the larger $\gamma$ gets the closer graph $\mathrm{Gr}_{\mathrm{TB}}\left(\gamma\right)$
gets to $\mathrm{Gr}_{\mathrm{TB}}\left(\infty\right)$ defined by
equation (\ref{eq:disdom1f}).}
\end{figure}

\begin{figure}[h]
\begin{centering}
\hspace{-0.5cm}\includegraphics[scale=0.42]{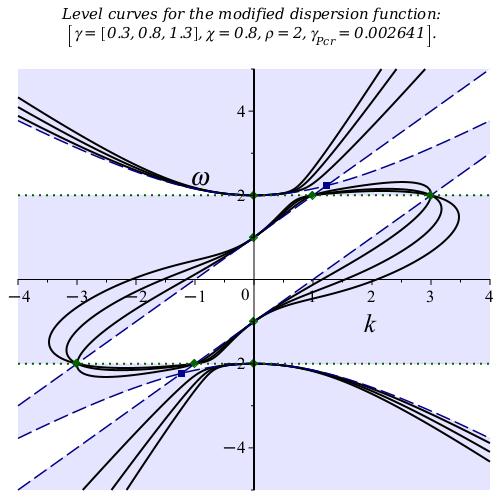}\hspace{1cm}\includegraphics[scale=0.42]{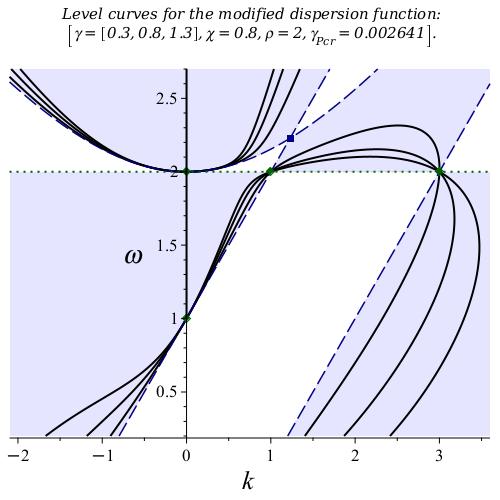}
\par\end{centering}
\centering{}(a)\hspace{7cm}(b)\caption{\label{fig:disp-Lev4} Dispersion curves of the TWT dispersion relations
(\ref{eq:dispdim1a}) and (\ref{eq:dispdim1b}) for $\chi=0.8$, $\rho=2>1$
and $\gamma=0.3,0.8,1.3$, $\gamma_{\mathrm{Pcr}}\protect\cong0.002641$:
(a) complete plot in the designated window; (b) a zoomed fragment
of (a); (c) a zoomed fragment of (b). Solid (black) curves represent
the TWT dispersion curves for indicated values of $\gamma$; dashed
(blue) curves represent the dispersion curves of non-interacting TL
and the e-beam for $\gamma=0$ as a reference. dashed (blue) curves
represent the dispersion curves of non-interacting TL and the e-beam.
Shaded area identifies the dispersion domain $\mathbb{D}_{\mathrm{TB}}$,
that is where $R_{\mathrm{TB}}\left(k,\omega\right)>0$. Doted (green)
horizontal straight lines represents points $\left(k,\pm\rho\right)$.
Note the dispersion curves passing through points focal points $\left(0,\pm1\right)$.
Note the dispersion curves $\mathrm{Gr}_{\mathrm{TB}}\left(\gamma\right)$
pass through focal points defined by equations (\ref{eq:disomfac2a})
marked as circle (green) dots. Square (blue) dots mark the TWT cross-points
(see equations (\ref{eq:disGTBr1b}), (\ref{eq:disGTBr1c})). Note
also that the smaller $\gamma$ gets the closer graph $\mathrm{Gr}_{\mathrm{TB}}\left(\gamma\right)$
gets to $\mathrm{Gr}_{\mathrm{TB}}\left(0\right)=\mathrm{Gr}_{\mathrm{T}}\bigcup\mathrm{Gr}_{\mathrm{B}}$,
whereas the larger $\gamma$ gets the closer graph $\mathrm{Gr}_{\mathrm{TB}}\left(\gamma\right)$
gets to $\mathrm{Gr}_{\mathrm{TB}}\left(\infty\right)$ defined by
equation (\ref{eq:disdom1f}).}
\end{figure}

\begin{figure}[h]
\begin{centering}
\hspace{-0.5cm}\includegraphics[scale=0.42]{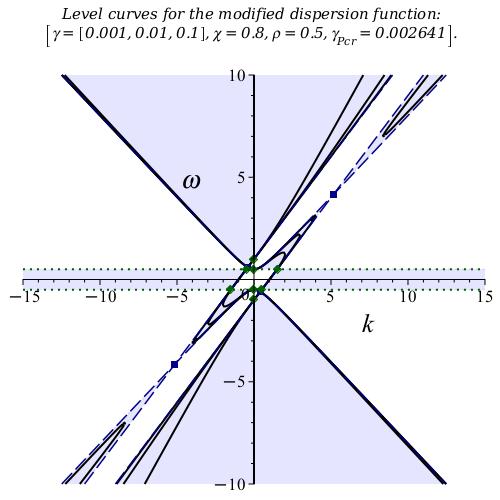}\hspace{1cm}\includegraphics[scale=0.42]{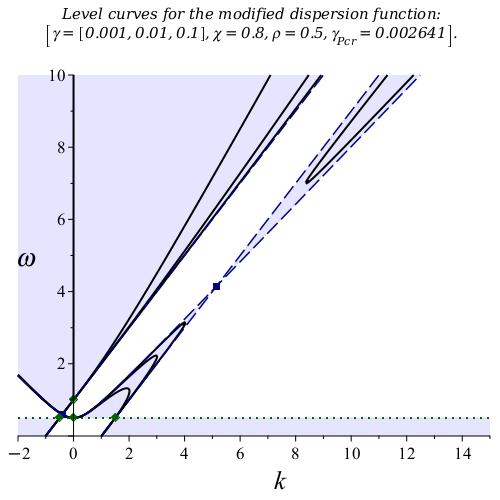}
\par\end{centering}
\centering{}(a)\hspace{8cm}(b)\caption{\label{fig:disp-Lev5} Dispersion curves of the TWT dispersion relations
(\ref{eq:dispdim1a}) and (\ref{eq:dispdim1b}) for $\chi=0.8$, $\rho=0.5<1$
and $\gamma=0.001,0.01,0.1$, $\gamma_{\mathrm{Pcr}}\protect\cong0.002641$:
(a) complete plot in the designated window; (b) a zoomed fragment
of (a); (c) a zoomed fragment of (b). Solid (black) curves represent
the TWT dispersion curves for indicated values of $\gamma$; dashed
(blue) curves represent the dispersion curves of non-interacting TL
and the e-beam for $\gamma=0$ as a reference. Shaded area identifies
the dispersion domain $\mathbb{D}_{\mathrm{TB}}$, that is where $R_{\mathrm{TB}}\left(k,\omega\right)>0$.
Doted (green) horizontal straight lines represents points $\left(k,\pm\rho\right)$.
Note the dispersion curves $\mathrm{Gr}_{\mathrm{TB}}\left(\gamma\right)$
pass through focal points defined by equations (\ref{eq:disomfac2a})
marked as circle (green) dots. Square (blue) dots mark the TWT cross-points
(see equations (\ref{eq:disGTBr1b}), (\ref{eq:disGTBr1c})). Note
also that the smaller $\gamma$ gets the closer graph $\mathrm{Gr}_{\mathrm{TB}}\left(\gamma\right)$
gets to $\mathrm{Gr}_{\mathrm{TB}}\left(0\right)=\mathrm{Gr}_{\mathrm{T}}\bigcup\mathrm{Gr}_{\mathrm{B}}$,
whereas the larger $\gamma$ gets the closer graph $\mathrm{Gr}_{\mathrm{TB}}\left(\gamma\right)$
gets to $\mathrm{Gr}_{\mathrm{TB}}\left(\infty\right)$ defined by
equation (\ref{eq:disdom1f}).}
\end{figure}

\begin{figure}[h]
\begin{centering}
\hspace{-0.1cm}\includegraphics[scale=0.33]{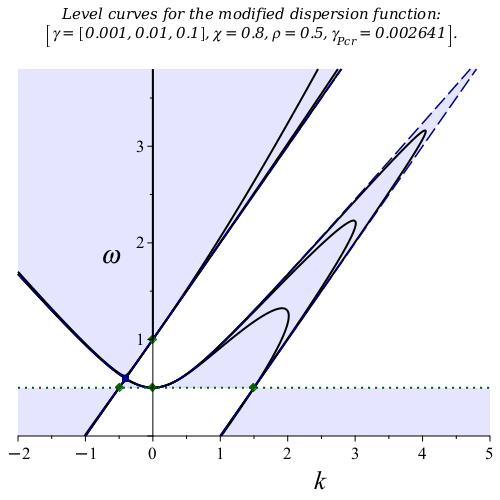}\hspace{0.1cm}\includegraphics[scale=0.33]{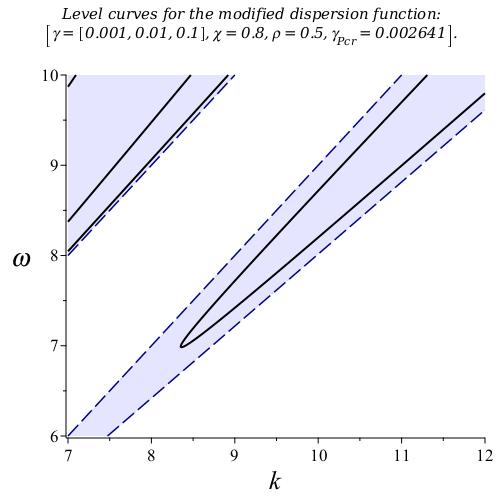}\hspace{0.1cm}\includegraphics[scale=0.33]{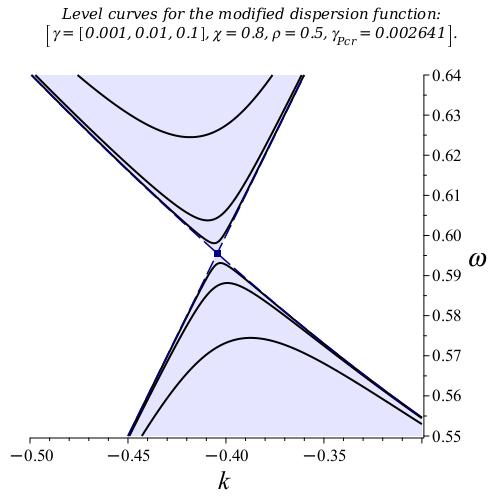}
\par\end{centering}
\centering{}(a)\hspace{5.5cm}(b)\hspace{5.5cm}(c)\caption{\label{fig:dis-Lev5f} Dispersion curves of the TWT dispersion relations
(\ref{eq:dispdim1a}) and (\ref{eq:dispdim1b}) for $\chi=0.8$, $\rho=0.5<1$
and $\gamma=0.001,0.01,0.1$, $\gamma_{\mathrm{Pcr}}\protect\cong0.002641$:
(a) zoomed fragment of Fig. \ref{fig:disp-Lev5}(b) for $k>0$; (b);
another zoomed fragment of Fig. \ref{fig:disp-Lev5}(b) for $k>0$;
(c) zoomed fragment of Fig. \ref{fig:disp-Lev5}(a) for $k<0$. Solid
(black) curves represent the TWT dispersion curves for indicated values
of $\gamma$; dashed (blue) curves represent the dispersion curves
of non-interacting TL and the e-beam for $\gamma=0$ as a reference.
Shaded area identifies the dispersion domain $\mathbb{D}_{\mathrm{TB}}$,
that is where $R_{\mathrm{TB}}\left(k,\omega\right)>0$. Doted (blue)
horizontal straight lines represents points $\left(k,\pm\rho\right)$.
Doted (green) horizontal straight lines represents points $\left(k,\pm\rho\right)$.
Note the dispersion curves $\mathrm{Gr}_{\mathrm{TB}}\left(\gamma\right)$
pass through focal points defined by equations (\ref{eq:disomfac2a})
marked as circle (green) dots. Square (blue) dots mark the TWT cross-points
(see equations (\ref{eq:disGTBr1b}), (\ref{eq:disGTBr1c})). Note
also that the smaller $\gamma$ gets the closer graph $\mathrm{Gr}_{\mathrm{TB}}\left(\gamma\right)$
gets to $\mathrm{Gr}_{\mathrm{TB}}\left(0\right)=\mathrm{Gr}_{\mathrm{T}}\bigcup\mathrm{Gr}_{\mathrm{B}}$,
whereas the larger $\gamma$ gets the closer graph $\mathrm{Gr}_{\mathrm{TB}}\left(\gamma\right)$
gets to $\mathrm{Gr}_{\mathrm{TB}}\left(\infty\right)$ defined by
equation (\ref{eq:disdom1f}).}
\end{figure}

\begin{figure}[h]
\begin{centering}
\hspace{-0.5cm}\includegraphics[scale=0.4]{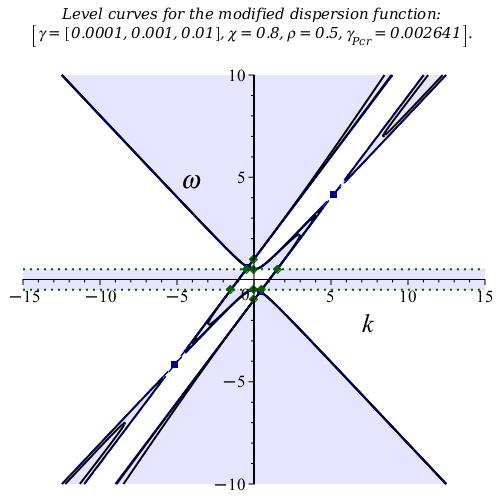}\hspace{0.5cm}\includegraphics[scale=0.4]{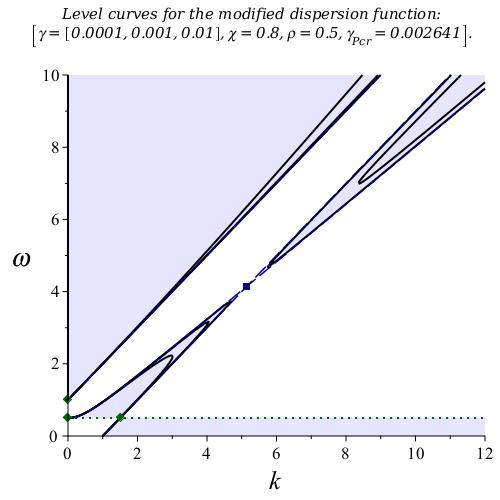}
\par\end{centering}
\centering{}(a)\hspace{7cm}(b)\caption{\label{fig:disp-Lev6} Dispersion curves of the TWT dispersion relations
(\ref{eq:dispdim1a}) and (\ref{eq:dispdim1b}) for $\chi=0.8$, $\rho=0.5<1$
and $\gamma=0.0001,0.001,0.01$, $\gamma_{\mathrm{Pcr}}\protect\cong0.002641$:
(a) complete plot in the designated window; (b) a zoomed fragment
of (a); (c) a zoomed fragment of (b). Solid (black) curves represent
the TWT dispersion curves for indicated values of $\gamma$; dashed
(blue) curves represent the dispersion curves of non-interacting TL
and the e-beam for $\gamma=0$ as a reference. Shaded area identifies
the dispersion domain $\mathbb{D}_{\mathrm{TB}}$, that is where $R_{\mathrm{TB}}\left(k,\omega\right)>0$.
Doted (green) horizontal straight lines represents points $\left(k,\pm\rho\right)$.
Note the dispersion curves $\mathrm{Gr}_{\mathrm{TB}}\left(\gamma\right)$
pass through focal points defined by equations (\ref{eq:disomfac2a})
marked as circle (green) dots. Square (blue) dots mark the TWT cross-points
(see equations (\ref{eq:disGTBr1b}), (\ref{eq:disGTBr1c})). Note
also that the smaller $\gamma$ gets the closer graph $\mathrm{Gr}_{\mathrm{TB}}\left(\gamma\right)$
gets to $\mathrm{Gr}_{\mathrm{TB}}\left(0\right)=\mathrm{Gr}_{\mathrm{T}}\bigcup\mathrm{Gr}_{\mathrm{B}}$,
whereas the larger $\gamma$ gets the closer graph $\mathrm{Gr}_{\mathrm{TB}}\left(\gamma\right)$
gets to $\mathrm{Gr}_{\mathrm{TB}}\left(\infty\right)$ defined by
equation (\ref{eq:disdom1f}).}
\end{figure}

\begin{figure}[h]
\begin{centering}
\hspace{-0.1cm}\includegraphics[scale=0.33]{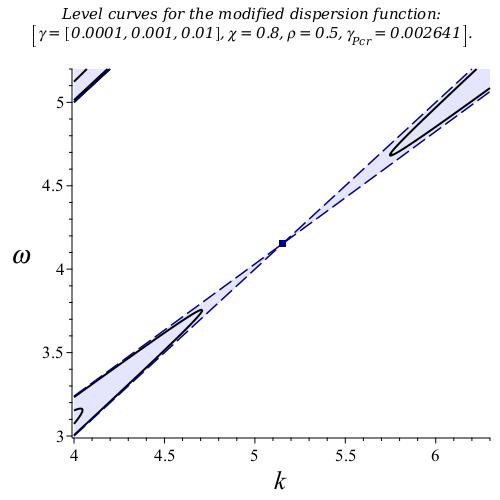}\hspace{0.1cm}\includegraphics[scale=0.33]{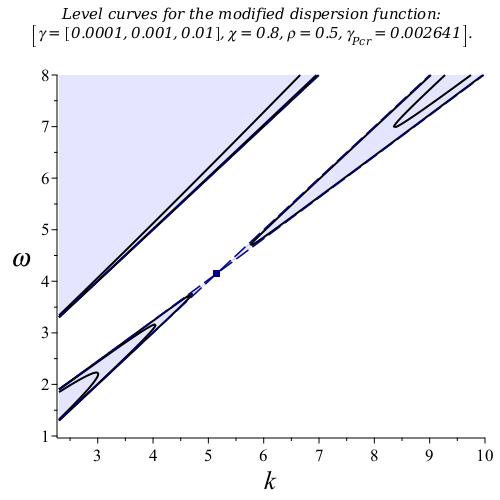}\hspace{0.1cm}\includegraphics[scale=0.33]{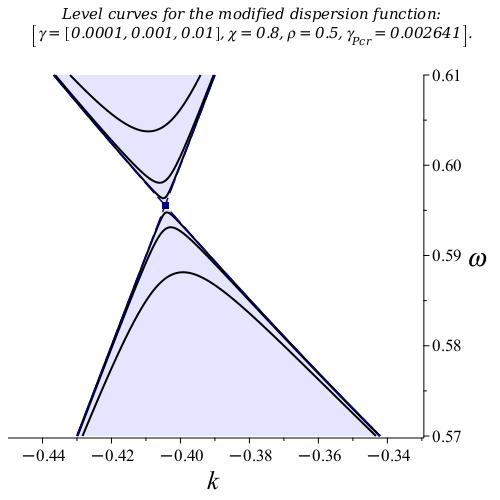}
\par\end{centering}
\centering{}(a)\hspace{5cm}(b)\hspace{5cm}(c)\caption{\label{fig:dis-Lev6f} Dispersion curves of the TWT dispersion relations
(\ref{eq:dispdim1a}) and (\ref{eq:dispdim1b}) for $\chi=0.8$, $\rho=0.5<1$
and $\gamma=0.0001,0.001,0.01$, $\gamma_{\mathrm{Pcr}}\protect\cong0.002641$:
(a) zoomed fragment of Fig. \ref{fig:disp-Lev6}(b) for $k>0$; (b)
another zoomed fragment of Fig. \ref{fig:disp-Lev6}(b) for $k>0$;
(c) zoomed fragment of Fig. \ref{fig:disp-Lev6}(a) for $k<0$. Solid
(black) curves represent the TWT dispersion curves for indicated values
of $\gamma$; dashed (blue) curves represent the dispersion curves
of non-interacting TL and the e-beam for $\gamma=0$ as a reference.
Shaded area identifies the dispersion domain $\mathbb{D}_{\mathrm{TB}}$,
that is where $R_{\mathrm{TB}}\left(k,\omega\right)>0$. Doted (green)
horizontal straight lines represents points $\left(k,\pm\rho\right)$.
Note the dispersion curves $\mathrm{Gr}_{\mathrm{TB}}\left(\gamma\right)$
pass through focal points defined by equations (\ref{eq:disomfac2a})
marked as circle (green) dots. Square (blue) dots mark the TWT cross-points
(see equations (\ref{eq:disGTBr1b}), (\ref{eq:disGTBr1c})). Note
also that the smaller $\gamma$ gets the closer graph $\mathrm{Gr}_{\mathrm{TB}}\left(\gamma\right)$
gets to $\mathrm{Gr}_{\mathrm{TB}}\left(0\right)=\mathrm{Gr}_{\mathrm{T}}\bigcup\mathrm{Gr}_{\mathrm{B}}$,
whereas the larger $\gamma$ gets the closer graph $\mathrm{Gr}_{\mathrm{TB}}\left(\gamma\right)$
gets to $\mathrm{Gr}_{\mathrm{TB}}\left(\infty\right)$ defined by
equation (\ref{eq:disdom1f}).}
\end{figure}

\begin{figure}[h]
\begin{centering}
\hspace{-0.1cm}\includegraphics[scale=0.32]{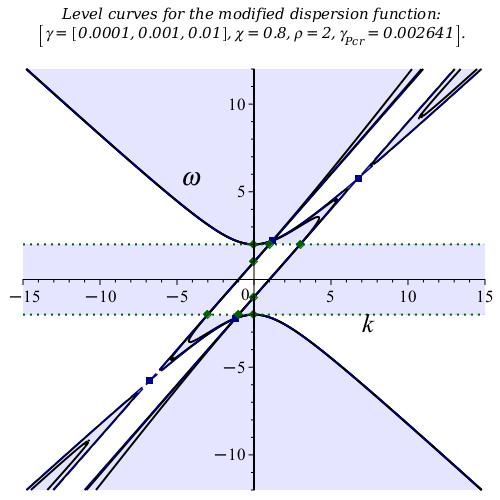}\hspace{0.1cm}\includegraphics[scale=0.32]{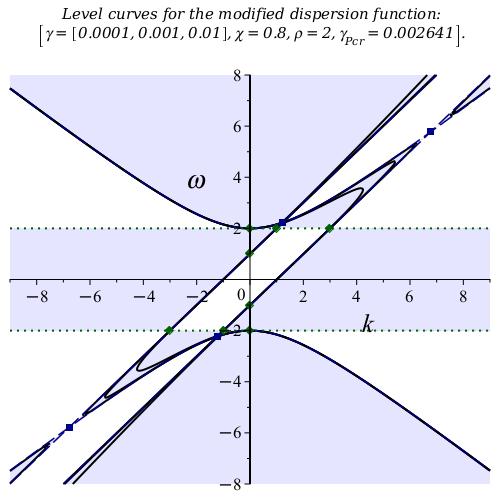}\hspace{0.1cm}\includegraphics[scale=0.32]{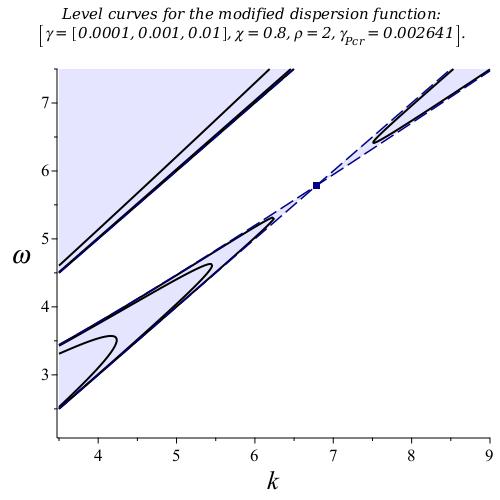}
\par\end{centering}
\centering{}(a)\hspace{5.5cm}(b)\hspace{5.5cm}(c)\caption{\label{fig:dis-Lev7} Dispersion curves of the TWT dispersion relations
(\ref{eq:dispdim1a}) and (\ref{eq:dispdim1b}) for $\chi=0.8$, $\rho=2>1$
and $\gamma=0.0001,0.001,0.01$, $\gamma_{\mathrm{Pcr}}\protect\cong0.002641$:
(a) zoomed fragment of Fig. \ref{fig:disp-Lev6}(b) for $k>0$; (b);
another zoomed fragment of Fig. \ref{fig:disp-Lev6}(b) for $k>0$;
(c) zoomed fragment of Fig. \ref{fig:disp-Lev6}(a) for $k<0$. Solid
(black) curves represent the TWT dispersion curves for indicated values
of $\gamma$; dashed (blue) curves represent the dispersion curves
of non-interacting TL and the e-beam for $\gamma=0$ as a reference.
Shaded area identifies the dispersion domain $\mathbb{D}_{\mathrm{TB}}$,
that is where $R_{\mathrm{TB}}\left(k,\omega\right)>0$. Doted (green)
horizontal straight lines represents points $\left(k,\pm\rho\right)$.
Note the dispersion curves $\mathrm{Gr}_{\mathrm{TB}}\left(\gamma\right)$
pass through focal points defined by equations (\ref{eq:disomfac2a})
marked as circle (green) dots. Square (blue) dots mark the TWT cross-points
(see equations (\ref{eq:disGTBr1b}), (\ref{eq:disGTBr1c})). Note
also that the smaller $\gamma$ gets the closer graph $\mathrm{Gr}_{\mathrm{TB}}\left(\gamma\right)$
gets to $\mathrm{Gr}_{\mathrm{TB}}\left(0\right)=\mathrm{Gr}_{\mathrm{T}}\bigcup\mathrm{Gr}_{\mathrm{B}}$,
whereas the larger $\gamma$ gets the closer graph $\mathrm{Gr}_{\mathrm{TB}}\left(\gamma\right)$
gets to $\mathrm{Gr}_{\mathrm{TB}}\left(\infty\right)$ defined by
equation (\ref{eq:disdom1f}).}
\end{figure}

We exhibit and discuss here a series of plots of the TWT dispersion
curves generated for different values of the TWT principle parameter
$\gamma$, see Figures \ref{fig:dis-Lev1}, \ref{fig:dis-Lev1L} and
Figures \ref{fig:dis-Lev1LC}-\ref{fig:dis-Lev7}. The goal we pursue
when generating these plots is to demonstrate different topological
patterns occurring for different sets of the involved parameters.
In particular, the plots show qualitatively different topological
patterns that occur when: (i) $\rho>1$ or $\rho<1$; (ii) $\chi>1$
or $\chi<1$; (iii) $\gamma$ is small or large. The point for generating
a number of dispersive curves for particular selections of sets of
different $\gamma$ is to demonstrate: (i) the closer proximity of
the TWT dispersion curves for smaller values $\gamma$ to the dispersion
curves of the non-interacting GTL and the e-beam used as the reference
frames; (ii) the focal points that belong to all the TWT dispersion
curves. Another benefit of looking into the TWT dispersive curves
for small values of $\gamma$ is that indicate their naturally hybrid
nature. Indeed, when the principle TWT parameter $\gamma$, which
also plays the role of a coupling constant in the TWT dispersion equations
(\ref{eq:dispdim1a}), gets smaller some parts each of the TWT dispersive
curves get closer to the dispersion curves of the non-interacting
GTL whereas other parts get closer to the dispersion curves of the
non-interacting e-beam. To see the hybrid natures of the TWT dispersive
curves we integrated into each plot the dispersion curves of non-interacting
GTL and the e-beam as reference frames.

We invite a curious reader to explore Figures \ref{fig:dis-Lev1},
\ref{fig:dis-Lev1L} and Figures \ref{fig:dis-Lev1LC}-\ref{fig:dis-Lev7}
displaying different patterns of the selected sets of the TWT dispersion
curves including focal points and the dispersion curves of the non-interacting
GTL and the e-beam used as the reference frames.
Figures \ref{fig:disp-Lev5}-\ref{fig:dis-Lev7} indicate in particular
the low and the high frequency cutoff for the convection instability
that are similar to what was found in \citet{SchaFig} for the special
case $\rho=0$.

\section{Instabilities and dispersion-instability graphs\label{sec:disp-instab}}

It is clear that the electron beam has to supply the exponentially
growing in space - the phenomenon known as \emph{convection instability}.
Slowing down of the EM wave facilitates more efficient interaction
with the electron flow manifested in electron bunching and synchronism
that are essential for the RF signal amplification. The flow of electrons
in TWT move with nearly constant velocity and loose some of their
kinetic energy to the EM wave. This kind of energy transfer is observed
in the Cherenkov radiation phenomenon that can be viewed then as a
physical foundation for the convection instability and consequent
RF signal amplification, \citep[Sec. 1.1-1.2, 4.4, 4.8-4.9, 7.3, 7.6; Chap. 8]{BenSweSch}.
Based on a Lagrangian field theory, we developed in our work \citet{SchaFig}
a convincing argument that the collective Cherenkov effect in TWTs
is, in fact, a convective instability, that is, amplification. We
also derive there expressions of the low- and high-frequency cutoffs
for amplification in TWTs. Relations ``wave-particle interactions''
and origins of instability is discussed in \citet[Chap. 5]{FigTWTbk}.

As to instabilities in general we remind the reader that as amplification
regimes are based on the convective instability oscillation regimes
are based on the \emph{absolute instability}, \citep{Stur58}, \citep{Brig}.
The difference between the two instabilities in a nutshell is as follows.
The convective instability is when the original pulse disturbance
convect from its origin in space as it grows in amplitude but at every
fixed point of space the disturbance is bounded by a time independent
constant. In contrast, the absolute instability is when the original
pulse disturbance grows without a bound in time at least in some points
of space. Note that the labeling of an instability is always with
respect to a particular reference frame, since a convective instability
would appear as an absolute instability to an observer moving along
with the \textquotedblright pulse\textquotedblright . 

In order to identify convection and absolutes instabilities in TWTs
one commonly considers the relevant eigenmodes of the following exponential
form

\begin{equation}
f_{\omega,k}\left(z,t\right)=a_{\omega k}\exp\left\{ -\mathrm{i}\left(\omega t-kz\right)\right\} ,\label{eq:fomkz1a}
\end{equation}
where $a_{\omega k}$ is a complex-valued constant and $\omega$ and
$k$ are respectively the frequency and wave-number that \emph{can
possibly be complex-valued}. \emph{Importantly, for $f_{\omega,k}\left(z,t\right)$
to be an eigenmode $\omega$ and $k$, that can be complex-valued
now, must satisfy the TWT dispersion relations (\ref{eq:dispkuom1b}),
(\ref{eq:disomfac1a}),} (\ref{eq:dispdim1a}) and (\ref{eq:dispdim1b})\emph{.}
In the case when $\omega$ is real and $\Im\left(k\right)\neq0$ function
$\left|f_{\omega k}\left(z,t\right)\right|$ grows or decays exponentially
if $z\rightarrow\pm\infty$ and we refer to this situation as convection
instability associated with amplification regimes. In the case $k$
is real but $\Im\left(\omega\right)\neq0$ function $\left|f_{\omega k}\left(z,t\right)\right|$
grows or decays exponentially if $t\rightarrow\pm\infty$ and we refer
to this situation as absolute instability associated with (exponentially
growing) oscillations regimes. 

Note that for complex-valued\emph{ $\omega$ and $k$} the phase velocity
$u=\frac{\omega}{\omega}$ can be also complex-valued satisfying the
TWT velocity dispersion relations (\ref{eq:dispkuom1c}), (\ref{eq:cofdisp1d}).

In Sections \ref{sec:disp-dom} and \ref{sec:disp-curv} we considered
the graphs of the conventional dispersion relations. It is useful
to integrate the information about the TWT instabilities into the
dispersion relations using the concept of \emph{dispersion-instability
graph} that we have developed in \citet[Chap. 7]{FigTWTbk}. Typical
examples of the dispersion-instability graphs are shown in Figures
\ref{fig:dis-instab1}-\ref{fig:dis-instab10}. Recall that the conventional
dispersion relations are defined as the relations between real-valued
frequency $\omega$ and real-valued wavenumber $k$ associated with
the relevant eigenmodes. In the case of the convection instability
frequency $\omega$ is real and wavenumber $k$ can be complex-valued,
whereas in the case of absolute instability $k$ is real and $\omega$
can be complex-valued. To represent the corresponding modes geometrically
as points in real $k\omega$-plane we proceed as follows, \citep[7]{FigTWTbk}.

In this paper we focus primarily on the convection instability with
the only exception made in Section \ref{sec:crospomod} where we consider
also absolutely unstable modes.

First, let us consider the case of the convection instability when
$\omega$ is real and $k$ is complex-valued. In this case we parametrize
every mode of the TWT system uniquely by the pair $\left(k\left(\omega\right),\omega\right)$
where $\omega$ is its frequency and $k\left(\omega\right)$ is its
wavenumber. If $k\left(\omega\right)$ is degenerate, it is counted
a number of times according to its multiplicity. We can naturally
partition all the system modes represented by pairs $\left(k\left(\omega\right),\omega\right)$
into two distinct classes \textendash{} oscillatory modes and (convectively)
unstable ones \textendash{} based on whether the wavenumber $k\left(\omega\right)$
is real- or complex-valued with $\Im\left\{ k\left(\omega\right)\right\} \neq0$.
We refer to a mode (eigenmode) of the system as an \emph{oscillatory
mode} if its wavenumber $k\left(\omega\right)$ is real-valued. We
associate with such an oscillatory mode point $\left(k\left(\omega\right),\omega\right)$
in the $k\omega$-plane with $k$ being the horizontal axis and $\omega$
being the vertical one. Similarly, we refer to a mode (eigenmode)
of the system as a \emph{convectively unstable mode} if its wavenumber
$k=k\left(\omega\right)$ is complex-valued with a nonzero imaginary
part, that is, $\Im\left\{ k\left(\omega\right)\right\} \neq0$. We
associate with such an unstable mode point $\left(\Re\left\{ k\left(\omega\right)\right\} ,\omega\right)$
in the $k\omega$-plane.

Based on the above considerations, we represent the set of all the
oscillatory and convectively unstable modes of the system geometrically
by the set of the corresponding modal points $\left(k\left(\omega\right),\omega\right)$
and $\left(\Re\left\{ k\left(\omega\right)\right\} ,\omega\right)$
in the $k\omega$-plane. We name this set the \emph{dispersion-instability
graph}. To distinguish graphically points $\left(k\left(\omega\right),\omega\right)$
associated with oscillatory modes when $k\left(\omega\right)$ is
real-valued from points $\left(\Re\left\{ k\left(\omega\right)\right\} ,\omega\right)$
associated with unstable modes when $k\left(\omega\right)$ is complex-valued
with $\Im\left\{ k\left(\omega\right)\right\} \neq0$ we show points
$\Im\left\{ k\left(\omega\right)\right\} =0$ in black color whereas
points with $\Im\left\{ k\left(\omega\right)\right\} \neq0$ are shown
in red color. The corresponding curves are shown respectively as the
solid (black) and solid (red) curves. We remind that every point $\left(\Re\left\{ k\left(\omega\right)\right\} ,\omega\right)$
with $\Im\left\{ k\left(\omega\right)\right\} \neq0$ represents exactly
two complex conjugate convectively unstable modes associated with
$\pm\Im\left\{ k\left(\omega\right)\right\} $.

To integrate into the dispersion-instability graph the information
about the absolute instability, we proceed similarly to the case of
the convection instability. In this case we assume $k$ to be real
whereas $\omega$ can be complex-valued. Then every relevant mode
can be represented uniquely by the pair $\left(k,\omega\left(k\right)\right)$
where $\omega\left(k\right)$ is the mode frequency and $k$ is its
wavenumber. If $\omega\left(k\right)$ is degenerate, it is counted
a number of times according to its multiplicity. Similarly to the
convection instability we partition all the modes into two distinct
classes \textendash{} oscillatory modes and unstable ones \textendash{}
based on whether the frequency $\omega\left(k\right)$ is real- or
complex-valued with $\Im\left\{ \omega\left(k\right)\right\} \neq0$.
As before we refer to a mode (eigenmode) of the system as an \emph{oscillatory
mode} if its frequency is real-valued. \emph{Note that the class of
oscillatory modes for both convection and absolute instability are
exactly the same.} We refer to a mode (eigenmode) of the system as
\emph{absolutely unstable mode} if its frequency $\omega\left(k\right)$
is complex-valued with a nonzero imaginary part, that is, $\Im\left\{ \omega\left(k\right)\right\} \neq0$.
We associate with such an absolutely unstable mode point $\left(k,\Re\left\{ \omega\left(k\right)\right\} \right)$
in $k\omega$-plane. Notice that every point $\left(k,\Re\left\{ \omega\left(k\right)\right\} \right)$
is in fact associated with two complex conjugate system modes with
$\pm\Im\left\{ \omega\left(k\right)\right\} $. Consequently, each
point on the curve associated with the absolute instability represents
exactly two complex-conjugate wave number absolutely unstable modes.

To distinguish graphically points $\left(k,\omega\left(k\right)\right)$
associated with oscillatory modes when $\omega\left(k\right)$ is
real-valued from points $\left(k,\Re\left\{ \omega\left(k\right)\right\} \right)$
associated with absolutely unstable modes when $\omega\left(k\right)$
is complex-valued with $\Im\left\{ \omega\left(k\right)\right\} \neq0$
we show points with $\Im\left\{ \omega\left(k\right)\right\} =0$
in black color whereas points with $\Im\left\{ \omega\left(k\right)\right\} \neq0$
are shown in green color.

In Section \ref{sec:crospomod} we study a simpler cross-point model
for the factorized dispersion relation. In that particular case we
studied not only the convectively unstable modes but also \emph{absolutely
unstable} ones. To integrate those absolutely unstable modes into
the dispersion-instability graph we proceed similarly to the case
of the convection instability. But in this case we assume $k$ to
be real whereas $\omega$ is allowed to be complex-valued. We refer
to a mode (eigenmode) of the system as \emph{absolutely unstable mode}
if its frequency $\omega\left(k\right)$ is complex-valued with a
nonzero imaginary part, that is, $\Im\left\{ \omega\left(k\right)\right\} \neq0$.
We associate with such an absolutely unstable mode point $\left(k,\Re\left\{ \omega\left(k\right)\right\} \right)$
in the $k\omega$-plane. Notice that every point $\left(k,\Re\left\{ \omega\left(k\right)\right\} \right)$
represents exactly two complex conjugate system modes with imaginary
part equal to $\pm\Im\left\{ \omega\left(k\right)\right\} $. Similarly
to the convection instability we add the absolutely unstable modes
to the set of oscillatory modes. \emph{Note that the set of oscillatory
modes for both convection and absolute instability are exactly the
same.} To distinguish graphically points $\left(k,\Re\left\{ \omega\left(k\right)\right\} \right)$
associated with absolutely unstable modes with $\Im\left\{ \omega\left(k\right)\right\} \neq0$
we show them in green color and the corresponding curves are shown
as the dashed (green) curves.the TWT dispersion relations (\ref{eq:dispdim1a}) 

To visualize our analytical developments for the TWT dispersion relations
and the instability branches we show here a number of dispersion-instability
graphs and their fragments. To demonstrate important features of the
dispersion relations (\ref{eq:dispdim1a}) and the instability branches
we have selected several sets of relevant TWT parameters and generated
for them the dispersion-instability graphs shown in Figures \ref{fig:dis-instab1}-
\ref{fig:dis-instab10}. Just as stated for dispersion curves in Section
\ref{sec:disp-dom} the dispersion-instability graphs depend significantly
on whether $\chi<1$ or $\chi>1$, and on whether $\rho<1$ or $\rho>1$.
They also depend significantly on if TWT principal parameter $\gamma$
is small or large. It is based on these considerations we made a selection
of the several sets of parameters used in Figures \ref{fig:dis-instab1}-
\ref{fig:dis-instab10}.

\subsection{Transition to instability points\label{subsec:trans-instab}}

Following to our prior studies on the TWT instabilities in \citet[Sec. 13, 30]{FigTWTbk}
we introduce and describe points $\left(k,\omega\right)$ that signify
the onset of the TWT convective instability. These points are identified
as points of extreme values for the TWT dispersion relation $\omega\left(k\right)$,
that is the points points $\left(k,\omega\right)$ for which $\frac{d\omega\left(k\right)}{dk}=0$. 

Since function $\omega\left(k\right)$ is defined a solution to the
TWT dispersion relations (\ref{eq:dispdim1b}) we proceed with recasting
first the polynomial TWT dispersion relations (\ref{eq:disomfac1a})
as
\begin{equation}
H_{\mathrm{TB}}\left(k,\omega\right)\stackrel{\mathrm{def}}{=}\left[\omega^{2}-\left(\chi^{2}k^{2}+\rho^{2}\right)\right]\left[\left(k-\omega\right)^{2}-1\right]-\gamma k^{2}\left(\omega^{2}-\rho^{2}\right)=0.\label{eq:transHTB1a}
\end{equation}
 Then an elementary analysis shows that the extreme points of function
$\omega\left(k\right)$ can be found as solutions to the following
system of equations
\begin{equation}
H_{\mathrm{TB}}\left(k,\omega\right)=0,\quad\frac{\partial H_{\mathrm{TB}}\left(k,\omega\right)}{\partial k}=0.\label{eq:transHTB1b}
\end{equation}
Exact expression for functions involved in equations (\ref{eq:transHTB1b})
can be readily found from the definition (\ref{eq:transHTB1a}) of
function $H_{\mathrm{TB}}$but since they are not particular enlightening
we omit them here. We refer to the extreme points $\left(k,\omega\right)$
of $\omega\left(k\right)$ that are solutions to the system of equations
(\ref{eq:transHTB1b}) as \emph{transition to (convection) instability
points} for an infinitesimally small vicinity of these points contains
points $\left(k,\omega\right)$ with $\Im\left\{ k\right\} \neq0$
satisfying the TWT dispersion relations (\ref{eq:disomfac1a}) and
consequently they are associated with the convection instability.

The case of the absolute instability can treated similarly. Namely
we consider points of extreme values for the TWT dispersion relation
written now in the form $k=k\left(\omega\right)$. The extreme points
then are points $\left(k,\omega\right)$ for which $\frac{dk\left(\omega\right)}{d\omega}=0$
and proceeding the same way as above we find that the extreme points
$\left(k,\omega\right)$ satisfy the following equations:
\begin{equation}
H_{\mathrm{TB}}\left(k,\omega\right)=0,\quad\frac{\partial H_{\mathrm{TB}}\left(k,\omega\right)}{\partial\omega}=0.\label{eq:transHTB1c}
\end{equation}
We refer to the extreme points $\left(k,\omega\right)$ satisfying
the system of equations (\ref{eq:transHTB1c}) as \emph{transition
to (absolute) instability points}. Similarly to considered case of
the convection instability an infinitesimally small vicinity of the
points satisfying equations (\ref{eq:transHTB1c}) contains points
$\left(k,\omega\right)$ with $\Im\left\{ \omega\right\} \neq0$ satisfying
the TWT dispersion relations (\ref{eq:disomfac1a}) and consequently
they are associated with the absolute instability.
\begin{rem}[transition to instability points]
 \label{rem:tran-instab} A transition to instability point satisfying
either equation (\ref{eq:transHTB1b}) or (\ref{eq:transHTB1c}) is
also an exceptional points of degeneracy (EPD). By definition an EPD
is point of a degeneracy of the relevant system matrix when not only
some eigenvalues coincide but the corresponding eigenvectors coincide
also, see \citet[Sec. II.1]{Kato}. In the case of $\omega_{\mathrm{c}}=0$
or $\rho=0$ the transition to (convective) instability points were
thoroughly studied in \citet[Sec. 4.4]{FigTWTbk} and \citet{FigtwtEPD},
where we referred to EPDs respectively as ``eigenvector degeneracy
points'' and ``nodal points''. We also developed in \citet[Sec. IV]{FigtwtEPD}
an approach of how to constructively use EPDs occurring in TWTs for
enhanced sensing. More information on the usage of EPDs for enhanced
sensing can be found in \citet{Wie,Wie1} are references therein.
\end{rem}

\subsection{Examples of the dispersion-instability graphs\label{subsec:disp-instab-graph}}

We show and discuss here a series of dispersion-instability plots
generated for different values for selected values of the TWT parameter
$\gamma$, $\rho$ and $\chi$. Just as in Section \ref{sec:disp-curv}
we selected different sets of values for the TWT parameters to demonstrate
different topological patterns of the dispersion-instability plots
that occur when: (i) $\rho>1$ or $\rho<1$; (ii) $\chi>1$ or $\chi<1$;
(iii) $\gamma$ is small or large. The TWT dispersive curves for small
values of $\gamma$ indicate their natural hybrid nature. Indeed,
one can see when the $\gamma$ gets smaller some parts the TWT dispersive
curves get closer to the dispersion curves of the non-interacting
GTL whereas other parts are closer to the dispersion curves of the
non-interacting e-beam.

\begin{figure}[h]
\begin{centering}
\hspace{-0.1cm}\includegraphics[scale=0.3]{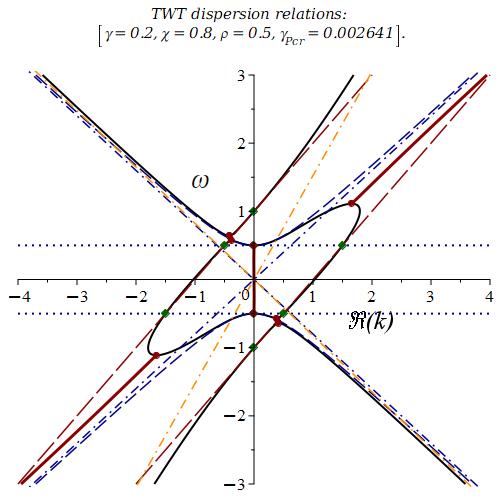}\hspace{0.1cm}\includegraphics[scale=0.3]{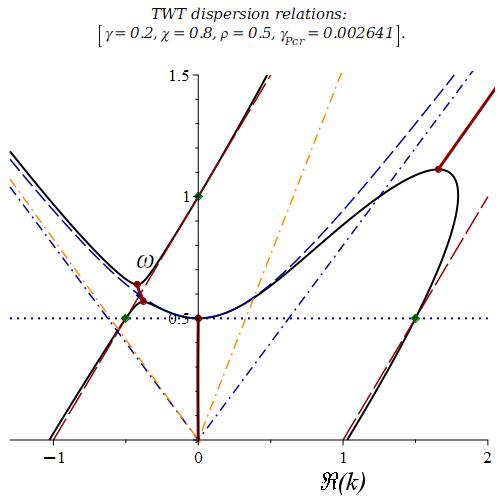}\hspace{0.1cm}\includegraphics[scale=0.3]{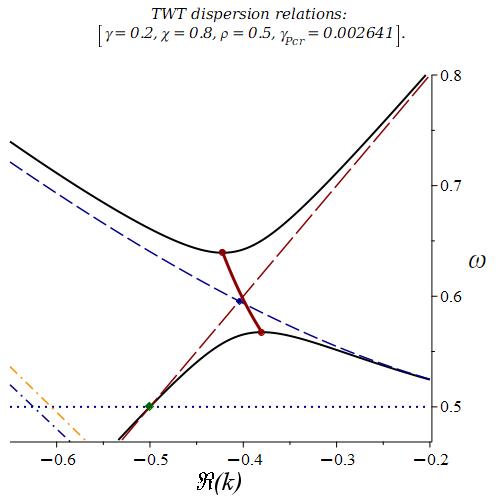}
\par\end{centering}
\centering{}(a)\hspace{5.5cm}(b)\hspace{5.5cm}(c)\caption{\label{fig:dis-instab2} Dispersion-instability graph and its zoomed
fragments for $\chi=0.8$, $\rho=0.5<1$ and $\gamma=0.2>\gamma_{\mathrm{Pcr}}\protect\cong0.002641$:
(a) larger scale version; (b) zoomed fragment of (a); (c) zoomed fragment
of (b) for $\Re\left\{ k\right\} <0$. Solid (black) curves represent
the dispersion curves, dashed (blue) curves represent the dispersion
curves for $\gamma=0$ as a reference. Dash-doted straight lines represented
high frequency asymptotics for $\gamma=0.7$ (orange) and for $\gamma=0$
(blue). Doted (blue) horizontal straight lines represents points $\left(k,\omega_{\mathrm{c}}\right)$.
Diamond (green) dots represent focal points defined by equations (\ref{eq:disomfac2a}),
diamond (blue) dots represent the cross-points $\mathrm{Gr}_{\mathrm{T}}\bigcap\mathrm{Gr}_{\mathrm{B}}$.
The circular dots (red) identify transition to instability points
(see Section \ref{subsec:trans-instab}). The bold, solid (red) curves
represent branches of points $\left(\Re\left\{ k\right\} ,\omega\right)$
with real $\omega$ and $\Im\left\{ k\right\} \protect\neq0$ which
are points of the convection instability.}
\end{figure}

\begin{figure}[h]
\begin{centering}
\hspace{-2cm}\includegraphics[scale=0.33]{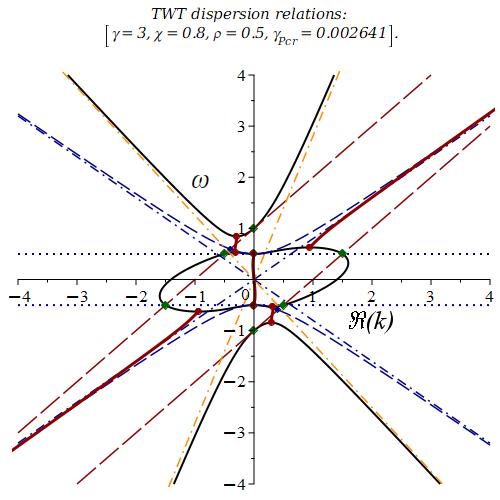}\hspace{1cm}\includegraphics[scale=0.33]{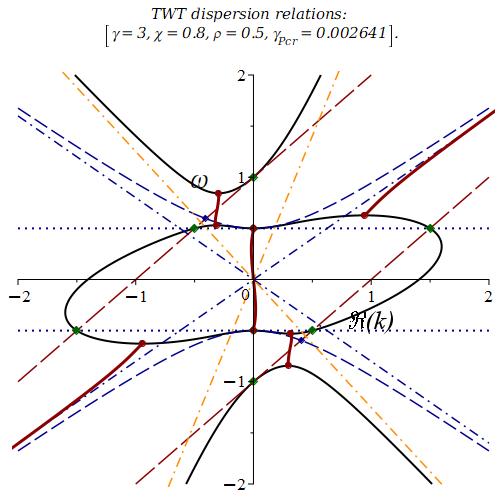}
\par\end{centering}
\centering{}\hspace{-2cm}(a)\hspace{7cm}(b)\caption{\label{fig:dis-instab4} Dispersion-instability graph for $\chi=0.8$,
$\rho=0.5<1$ and $\gamma=3\gg\gamma_{\mathrm{Pcr}}\protect\cong0.002641$:
(a) larger scale version; (b) zoomed fragment of (a). Solid (black)
curves represent the dispersion curves, dashed (blue) curves represent
the dispersion curves for $\gamma=0$ as a reference. Dash-doted straight
lines represented high frequency asymptotics for $\gamma=0.7$ (orange)
and for $\gamma=0$ (blue). Doted (blue) horizontal straight lines
represents points $\left(k,\omega_{\mathrm{c}}\right)$. Diamond (green)
dots represent focal points defined by equations (\ref{eq:disomfac2a}),
diamond (blue) dots represent the cross-points $\mathrm{Gr}_{\mathrm{T}}\bigcap\mathrm{Gr}_{\mathrm{B}}$.
The circular dots (red) identify transition to instability points
(see Section \ref{subsec:trans-instab}). The bold, solid (red) curves
represent branches of points $\left(\Re\left\{ k\right\} ,\omega\right)$
with real $\omega$ and $\Im\left\{ k\right\} \protect\neq0$ which
are points of the convection instability.}
\end{figure}

\begin{figure}[h]
\begin{centering}
\hspace{-0.5cm}\includegraphics[scale=0.33]{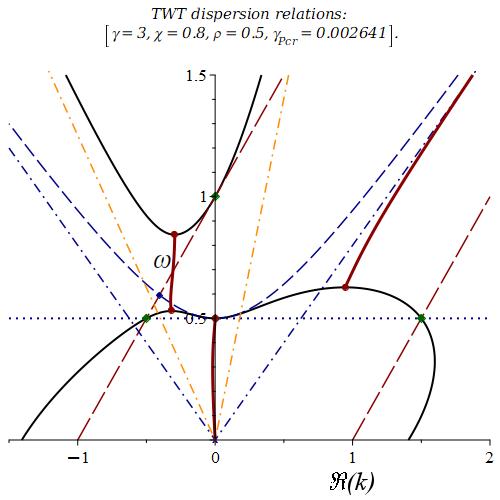}\hspace{3cm}\includegraphics[scale=0.33]{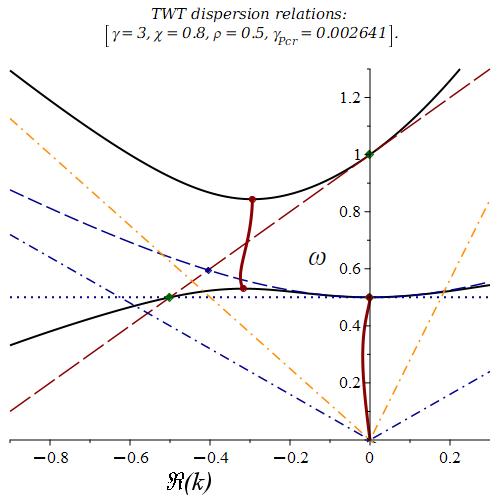}
\par\end{centering}
\centering{}\hspace{-1cm}(a)\hspace{8cm}(b)\caption{\label{fig:dis-instab4f} Zoomed fragments of the dispersion-instability
graph in Fig. \ref{fig:dis-instab4} for $\chi=0.8$, $\rho=0.5<1$
and $\gamma=3\gg\gamma_{\mathrm{Pcr}}\protect\cong0.002641$: (a)
zoomed fragment of Fig. \ref{fig:dis-instab4}(b); (b) zoomed fragment
of (a). Solid (black) curves represent the dispersion curves, dashed
(blue) curves represent the dispersion curves for $\gamma=0$ as a
reference. Dash-doted straight lines represented high frequency asymptotics
for $\gamma=0.7$ (orange) and for $\gamma=0$ (blue). Doted (blue)
horizontal straight lines represents points $\left(k,\omega_{\mathrm{c}}\right)$.
Diamond (green) dots represent focal points defined by equations (\ref{eq:disomfac2a}),
diamond (blue) dots represent the cross-points $\mathrm{Gr}_{\mathrm{T}}\bigcap\mathrm{Gr}_{\mathrm{B}}$.
The circular dots (red) identify transition to instability points
(see Section \ref{subsec:trans-instab}). The bold, solid (red) curves
represent branches of points $\left(\Re\left\{ k\right\} ,\omega\right)$
with real $\omega$ and $\Im\left\{ k\right\} \protect\neq0$ which
are points of the convection instability.}
\end{figure}

\begin{figure}[h]
\begin{centering}
\hspace{-0.1cm}\includegraphics[scale=0.3]{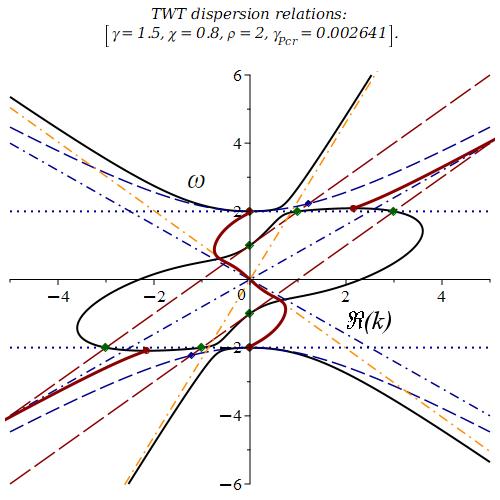}\hspace{0.1cm}\includegraphics[scale=0.3]{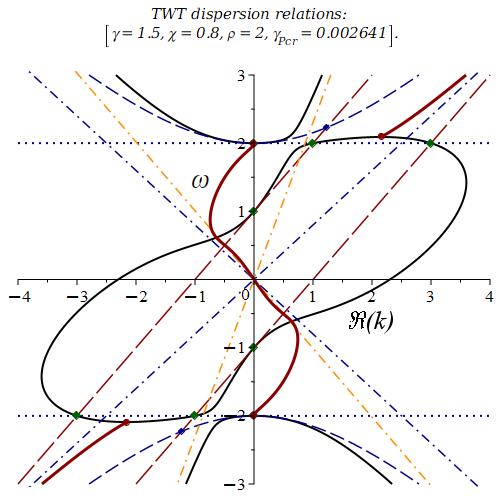}\hspace{0.1cm}\includegraphics[scale=0.3]{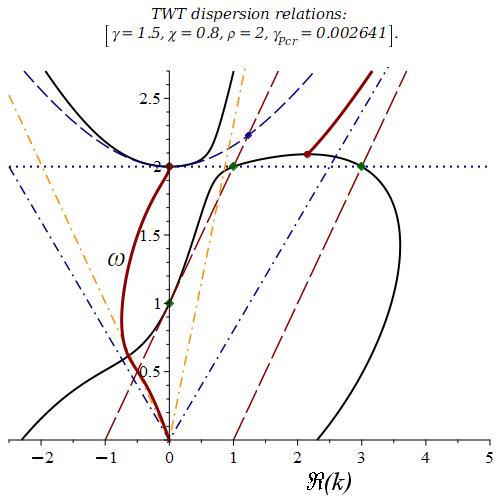}
\par\end{centering}
\centering{}(a)\hspace{5cm}(b)\hspace{5.5cm}(c)\caption{\label{fig:dis-instab9} Dispersion-instability graph and its zoomed
fragments for $\chi=0.8$, $\rho=2>1$ and $\gamma=1.3\gg\gamma_{\mathrm{Pcr}}\protect\cong0.002641$:
(a) larger scale version; (b) zoomed fragment of (a); (c) zoomed fragment
of (b). Solid (black) curves represent the dispersion curves, dashed
(blue) curves represent the dispersion curves for $\gamma=0$ as a
reference. Dash-doted straight lines represented high frequency asymptotics
for $\gamma=0.7$ (orange) and for $\gamma=0$ (blue). Doted (blue)
horizontal straight lines represents points $\left(k,\omega_{\mathrm{c}}\right)$.
Diamond (green) dots represent focal points defined by equations (\ref{eq:disomfac2a}),
diamond (blue) dots represent the cross-points $\mathrm{Gr}_{\mathrm{T}}\bigcap\mathrm{Gr}_{\mathrm{B}}$.
The circular dots (red) identify transition to instability points
(see Section \ref{subsec:trans-instab}). The bold, solid (red) curves
represent branches of points $\left(\Re\left\{ k\right\} ,\omega\right)$
with real $\omega$ and $\Im\left\{ k\right\} \protect\neq0$ which
are points of the convection instability.}
\end{figure}

\begin{figure}[h]
\begin{centering}
\hspace{-0.1cm}\includegraphics[scale=0.3]{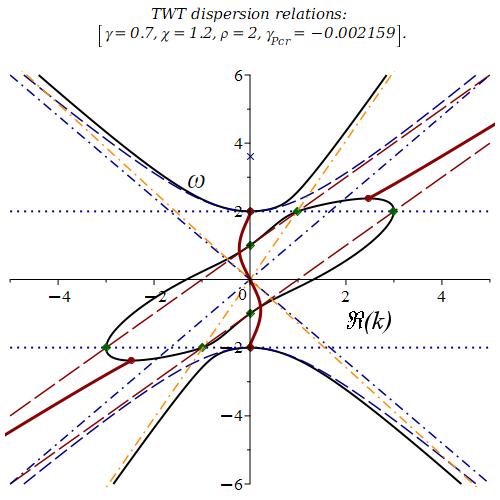}\hspace{0.1cm}\includegraphics[scale=0.3]{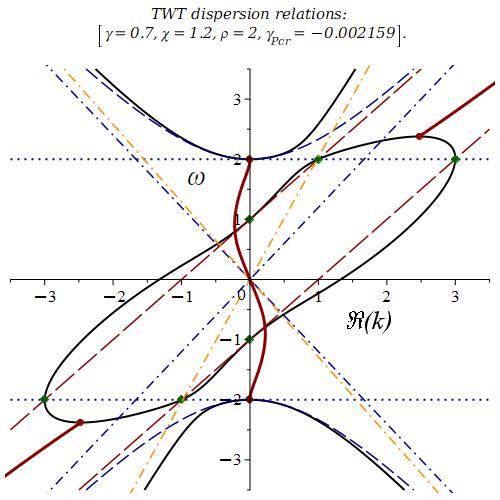}\hspace{0.1cm}\includegraphics[scale=0.3]{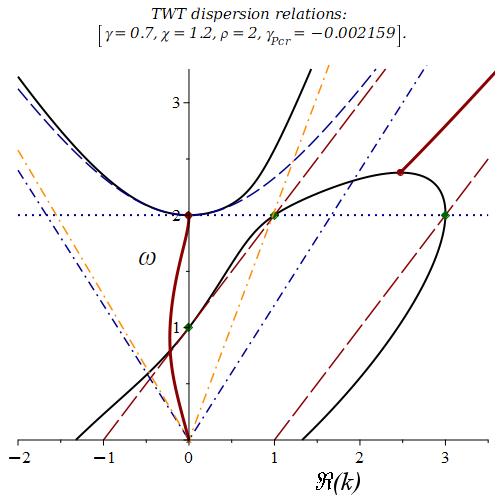}
\par\end{centering}
\centering{}(a)\hspace{5.5cm}(b)\hspace{5.5cm}(c)\caption{\label{fig:dis-instab10} Dispersion-instability graph and its zoomed
fragments for $\chi=1.2$, $\rho=2>1$ and $\gamma=0.7$: (a) larger
scale version; (b) zoomed fragment of (a); (c) zoomed fragment of
(b). Solid (black) curves represent the dispersion curves, dashed
(blue) curves represent the dispersion curves for $\gamma=0$ as a
reference. Dash-doted straight lines represented high frequency asymptotics
for $\gamma=0.7$ (orange) and for $\gamma=0$ (blue). Doted (blue)
horizontal straight lines represents points $\left(k,\omega_{\mathrm{c}}\right)$.
Diamond (green) dots represent focal points defined by equations (\ref{eq:disomfac2a}).
The circular dots (red) identify transition to instability points
(see Section \ref{subsec:trans-instab}). The bold, solid (red) curves
represent branches of points $\left(\Re\left\{ k\right\} ,\omega\right)$
with real $\omega$ and $\Im\left\{ k\right\} \protect\neq0$ which
are points of the convection instability.}
\end{figure}

We invite the reader to explore Figures \ref{fig:dis-instab2}- \ref{fig:dis-instab10}
different patterns of dispersion-instability plots where we displayed
graphically the detailed information of the TWT dispersion relations
(\ref{eq:dispdim1a}) including the dispersion curves of the non-interacting
GTL and the e-beam as reference frames, the transition to instability
points, the focal points, the branches of the TWT convective instability
and more.

\section{Imaginary part of the wave number as a measure of the amplification\label{sec:facImk}}

The dispersion-instability graphs considered in Section \ref{sec:disp-instab}
contain only partial information about the instabilities since the
information about the imaginary part $\Im\left\{ k\right\} $ is suppressed
there. In this section we complement dispersion-instability graphs
in Section \ref{sec:disp-instab} with plots of the imaginary part
$\Im\left\{ k\left(\omega\right)\right\} $ as a function of real-valued
frequency $\omega$. We remind that $\Im\left\{ k\left(\omega\right)\right\} $
is responsible for the amplification since the eigenmode dependence
of $t$ and $z$ is proportional to $\exp\left\{ -\mathrm{i}\left(\omega t-kz\right)\right\} $.

\begin{figure}[h]
\begin{centering}
\hspace{-0.1cm}\includegraphics[scale=0.32]{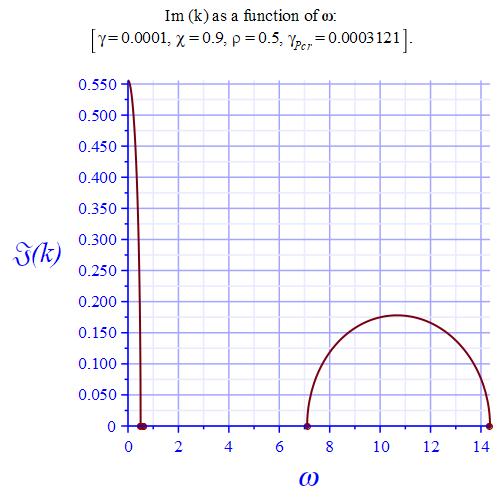}\hspace{0.3cm}\includegraphics[scale=0.32]{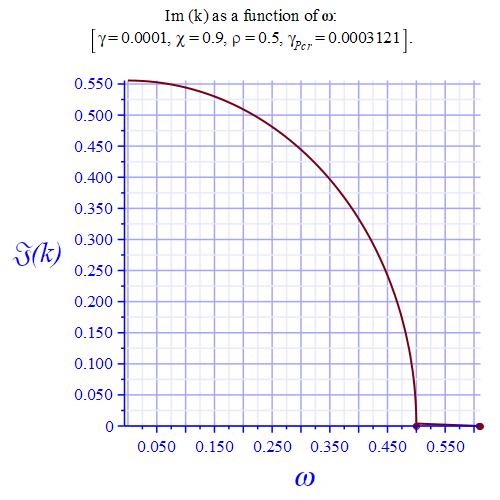}\hspace{0.3cm}\includegraphics[scale=0.32]{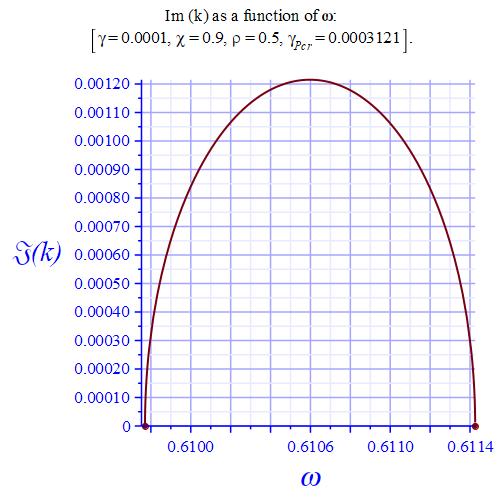}
\par\end{centering}
\centering{}\hspace{1cm}(a)\hspace{5.5cm}(b)\hspace{5.5cm}(c)\caption{\label{fig:fac-Imk2} The plot (a) shows the imaginary part $\Im\left\{ k\right\} $
of the wavenumber $k=k\left(\omega\right)$ as a function of frequency
$\omega$ for $\gamma=0.0001<\gamma_{\mathrm{Pcr}}\protect\cong0.0003121$,
$\chi=0.9<1$ and $\rho=0.5$. The plots (b) and (c) are fragments
of the plot (a). Diamond square dots (red) on the $\omega$-axis mark
the location of the low and the high frequency cutoffs. Note that
in the case there is no high frequency cutoff.}
\end{figure}

\begin{figure}[h]
\begin{centering}
\hspace{-0.1cm}\includegraphics[scale=0.32]{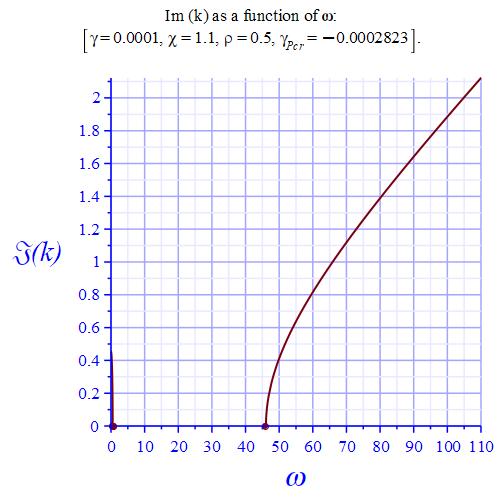}\hspace{0.3cm}\includegraphics[scale=0.32]{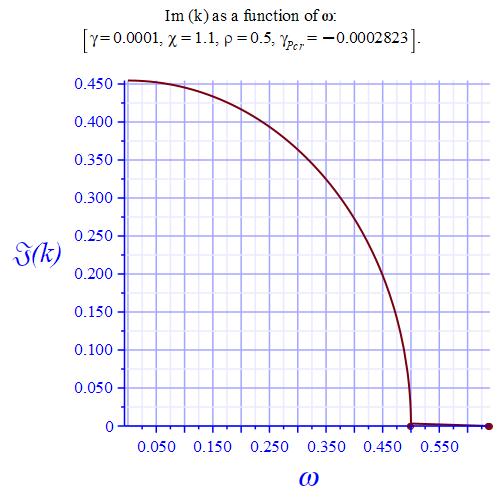}\hspace{0.3cm}\includegraphics[scale=0.32]{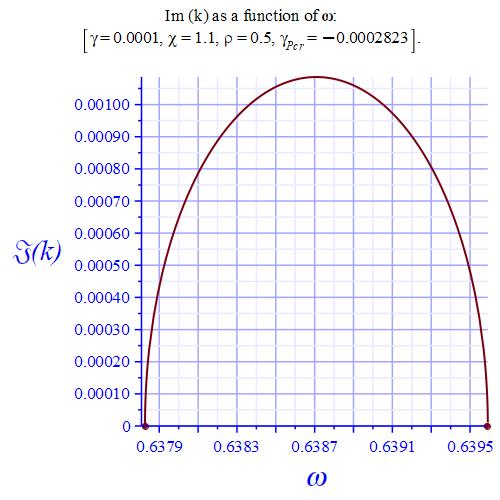}
\par\end{centering}
\centering{}\hspace{1cm}(a)\hspace{5.5cm}(b)\hspace{5.5cm}(c)\caption{\label{fig:fac-Imk3} The plot (a) shows the imaginary part $\Im\left\{ k\right\} $
of the wavenumber $k=k\left(\omega\right)$ as a function of frequency
$\omega$ for $\gamma=0.0001<\gamma_{\mathrm{Pcr}}\protect\cong0.0003121$,
$\chi=1.1>1$ and $\rho=0.5$. The plots (b) and (c) are fragments
of the plot (a). Diamond square dots (red) on the $\omega$-axis mark
the location of the low and the high frequency cutoffs. Note that
in the case there is no high frequency cutoff.}
\end{figure}

\begin{figure}[h]
\begin{centering}
\hspace{-1cm}\includegraphics[scale=0.4]{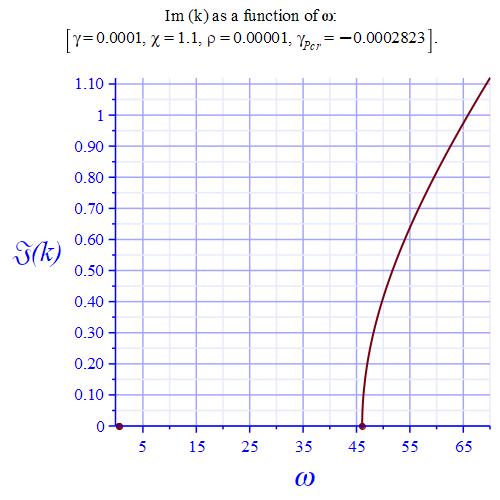}\hspace{2cm}\includegraphics[scale=0.4]{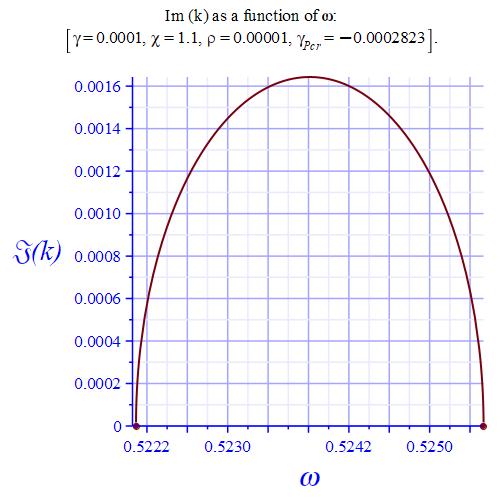}
\par\end{centering}
\centering{}\hspace{0.5cm}(a)\hspace{8.5cm}(b)\caption{\label{fig:fac-Imk4} The plot (a) shows the imaginary part $\Im\left\{ k\right\} $
of the wavenumber $k=k\left(\omega\right)$ as a function of frequency
$\omega$ for $\gamma=0.0001<\gamma_{\mathrm{Pcr}}\protect\cong0.0003121$,
$\chi=1.1>1$ and $\rho=0.5$. The plot (b) is fragments of the plot
(a). Diamond square dots (red) on the $\omega$-axis mark the location
of the low and the high frequency cutoffs. Note that in the case there
is no high frequency cutoff.}
\end{figure}

Figures \ref{fig:fac-Imk1}, \ref{fig:fac-Imk2}, \ref{fig:fac-Imk3}
and \ref{fig:fac-Imk4} show the dependence of the imaginary part
$\Im\left\{ k\right\} $of the wavenumber $k$ on frequency $\omega$.
Note that according to Figures \ref{fig:fac-Imk3} and \ref{fig:fac-Imk4}
operational frequencies and amplitudes of $\Im\left\{ k\right\} $
are much higher in the case of $\chi=1.1>1$ compare to the of $\chi=0.9<1$
shown in \ref{fig:fac-Imk1} and \ref{fig:fac-Imk2}.

\section{Cross-point model for factorized dispersion relations\label{sec:crospomod}}

Our spectral analysis of TWTs suggests that some aspects of the TWT
factorized dispersion relations may apply to other systems that might
have factorized dispersion relations. With that in mind let us consider
a system composed of some two interacting components. Suppose that
the dispersion relations for each of this components when they don't
interact in known and the corresponding dispersion relations are defined
by the following equations:

\begin{equation}
G_{1}\left(k,\omega\right)=0,\quad G_{2}\left(k,\omega\right)=0.\label{eq:GGgamG1a}
\end{equation}
Motivated by the expressions (\ref{eq:disomfac1c}), (\ref{eq:dispdim2c})
of the factorized TWT dispersion relations let us assume that a system
composed of some two interacting components satisfying equations (\ref{eq:GGgamG1a})
has the dispersion relation that can be written as follows:
\begin{equation}
G_{1}\left(k,\omega\right)G_{2}\left(k,\omega\right)=\gamma G_{\mathrm{c}}\left(k,\omega\right),\label{eq:GGgamG1b}
\end{equation}
where $\gamma$ is the \emph{coupling coefficient} and to $G_{\mathrm{c}}\left(k,\omega\right)$
as the \emph{coupling function}. It seems that a special form of equation
(\ref{eq:GGgamG1b}) for $G_{\mathrm{c}}\left(k,\omega\right)=1$,
that was setup at hoc, appeared in literature but we could not point
to a specific reference.

Assuming that the system dispersion relations (\ref{eq:GGgamG1b})
hold we proceed with making more specific assumptions about representation
of functions $G_{1}\left(k,\omega\right)$, $G_{2}\left(k,\omega\right)$
and $G_{\mathrm{c}}\left(k,\omega\right)$. Once again, inspired by
the factorized TWT dispersion equations (\ref{eq:disomfac1a})-(\ref{eq:disomfac1c})
and (\ref{eq:dispdim1a}), (\ref{eq:dispdim1b}) we assume that these
functions can be completely factorized, that is
\begin{equation}
G_{j}\left(k,\omega\right)\stackrel{\mathrm{def}}{=}\prod_{m=1}^{N_{j}}\left(\omega-\omega_{jm}\left(k\right)\right),\quad j=1,2;\quad G_{\mathrm{c}}\left(k,\omega\right)=\prod_{m=1}^{N_{\mathrm{c}}}\left(\omega-\omega_{\mathrm{c}m}\left(k\right)\right),\label{eq:GGgamG1ba}
\end{equation}
where $N_{1}$, $N_{2}$ and $N_{\mathrm{c}}$ are positive integers,
and $\omega_{jm}\left(k\right)$ and $\omega_{\mathrm{c}m}\left(k\right)$
are smooth functions presumed to be known.

The first question that comes to mind is whether one can formulate
conditions under which a physical system defined by its Lagrangian
would have the dispersion relations that can be written in the factorized
form (\ref{eq:GGgamG1b}), (\ref{eq:GGgamG1ba}). We admit that answering
this question require considerable efforts, and it is outside of the
scope of this paper. The reason for introducing general factorized
dispersion relations (\ref{eq:GGgamG1b}) is rather for setting up
a general framework that can be used to construct a simple approximation
to the TWT dispersion equations (\ref{eq:dispdim1a}), (\ref{eq:dispdim1b})
in a vicinity of a cross-point as explained in the following section.

\subsection{Cross-point model of factorized dispersion relations\label{subsec:cropofac}}

Suppose that $\left(\omega_{0},k_{0}\right)$ is a \emph{``cross-point''}
of the graphs of functions $G_{1}$ and $G_{2}$, that a point satisfying
the system of two dispersion relations (\ref{eq:GGgamG1a}), namely
\begin{equation}
G_{1}\left(\omega_{0},k_{0}\right)=0,\quad G_{2}\left(\omega_{0},k_{0}\right)=0.\label{eq:GGgamG1c}
\end{equation}
Suppose also coupling parameter $\gamma$ to be small, and consider
solutions to equation (\ref{eq:GGgamG1b}) in a small vicinity of
point $\left(\omega_{0},k_{0}\right)$, that is
\begin{equation}
\left(k,\omega\right)=\left(k_{0}+\delta,\omega_{0}+\delta\right),\quad\left|\delta\right|,\left|\kappa\right|\ll1,\label{eq:GGgamG1d}
\end{equation}
Assuming that equations (\ref{eq:GGgamG1c}) and (\ref{eq:GGgamG1c})
hold and that variables $\gamma$, $\delta$ and $\delta$ are small,
that is
\begin{equation}
\left|\gamma\right|\ll1,\quad\left|\delta\right|\ll1,\quad\left|\delta\right|\ll1,\label{eq:GGgamG1s}
\end{equation}
 we arrive at the following \emph{cross-point approximation} to the
dispersion equation (\ref{eq:GGgamG1b})
\begin{equation}
\left(g_{1\omega}\delta+g_{1k}\kappa\right)\left(g_{2\omega}\delta+g_{2k}\kappa\right)=\gamma g_{\gamma},\label{eq:GGgamG2a}
\end{equation}
where the constants $g_{j\omega}$, $g_{jk}$ and $g_{\gamma}$ are
defined by
\begin{equation}
g_{j\omega}=\left(\partial_{\omega}G_{j}\right)\left(\omega_{0},k_{0}\right),\quad g_{jk}=\left(\partial_{k}G_{j}\right)\left(\omega_{0},k_{0}\right),\quad j=1,2;\quad g_{\gamma}=G_{\mathrm{c}}\left(\omega_{0},k_{0}\right).\label{eq:GGgamG2b}
\end{equation}

Suppose now that the values of coefficients $g_{j\omega}$, $g_{jk}$
and $g_{\gamma}$ are generic in the sense that functions $g_{1\omega}\delta+g_{1k}\kappa$
and $g_{2\omega}\delta+g_{2k}\kappa$ are linearly independent. Under
this assumption we can transform equation (\ref{eq:GGgamG2a}) into
a simple special form by the following change of coordinates
\begin{equation}
g_{1\omega}\delta+g_{1k}\kappa=\delta^{\prime}+\kappa^{\prime},\quad g_{2\omega}\delta+g_{2k}\delta=\delta^{\prime}-\kappa^{\prime}.\label{eq:GGgamG2c}
\end{equation}
 Indeed equation (\ref{eq:GGgamG2a}) can be recast in terms these
variables as
\begin{equation}
\delta^{\prime2}-\kappa^{\prime2}=\gamma g_{\gamma}.\label{eq:GGgamG2d}
\end{equation}
Note now that the graph of equation (\ref{eq:GGgamG2d}) is a hyperbola
implying that the graph of original equation (\ref{eq:GGgamG2a})
is a linear transformation of the hyperbola associated with special
form (\ref{eq:GGgamG2d}).

\emph{In summary, we conclude that generically if the coupling parameter
$\gamma$ is small then the graph of the dispersion relations of two
interacting systems in a vicinity of the relevant cross-point is a
linear transformation of a hyperbola.}

In case when $g_{1\omega}\neq0$ and $g_{2\omega}\neq0$ we can divide
both sides of equation (\ref{eq:GGgamG2a}) by $g_{1\omega}g_{2\omega}$
obtaining the following equivalent equation 
\begin{gather}
\left(\delta+g_{1}\kappa\right)\left(\delta+g_{2}\kappa\right)=\gamma g_{\gamma},\quad g_{1}\stackrel{\mathrm{def}}{=}\frac{g_{1k}}{g_{1\omega}},\quad g_{2}\stackrel{\mathrm{def}}{=}\frac{g_{2k}}{g_{2\omega}},\quad g_{\gamma}\stackrel{\mathrm{def}}{=}\frac{g_{\mathrm{c}}}{g_{1\omega}g_{2\omega}}.\label{eq:GGgamG2e}\\
\delta=\omega-\omega_{0},\quad\kappa=k-k_{0}.\nonumber 
\end{gather}
\emph{The equations (\ref{eq:GGgamG2e}) can be viewed as an approximation
to the TWT dispersion relation in a small vicinity of point $\left(k_{0},\omega_{0}\right)$}.

Motivated by equations ((\ref{eq:GGgamG2e}) we introduce now the
following dispersion relations
\begin{equation}
\left(\omega+g_{1}k\right)\left(\omega+g_{2}k\right)=\gamma g_{\gamma},\label{eq:GGgamG2f}
\end{equation}
 where $g_{1},$$g_{2}$ and $g_{\gamma}$ are given real constant.
We refer to dispersion equation (\ref{eq:GGgamG2f}) as the \emph{cross-point
model dispersion relations}. Equation (\ref{eq:GGgamG2f}) can be
recast into the following quadratic with respect $\omega$ and $k$
equation
\begin{equation}
\omega^{2}-\left(g_{1}+g_{2}\right)k\omega+g_{1}g_{2}k^{2}-\gamma g_{\gamma}=0.\label{eq:GGgamG2h}
\end{equation}
Note that to obtain dispersion relations (\ref{eq:GGgamG2e}) from
dispersion relations (\ref{eq:GGgamG2f}) and (\ref{eq:GGgamG2h})
evidently we need to apply the following translation transformation
in the $k\omega$-plane 
\begin{equation}
\omega\rightarrow\omega-\omega_{0},\quad k\rightarrow k-k_{0}.\label{eq:GGgamG2j}
\end{equation}

Quadratic equation (\ref{eq:GGgamG2h}) can readily solved for $\omega$
and $k$ yielding respectively the following representation for cross-model
dispersion relations (\ref{eq:GGgamG2f}) and (\ref{eq:GGgamG2h}):
\begin{equation}
\omega=\Omega_{\pm}\left(\gamma,k\right)\stackrel{\mathrm{def}}{=}\frac{1}{2}\left[-k\left(g_{1}+g_{2}\right)\pm\sqrt{\left(g_{1}-g_{2}\right)^{2}k^{2}+4\gamma g_{\gamma}}\right],\label{eq:Omgamk1a}
\end{equation}
\begin{equation}
k=K_{\pm}\left(\gamma,\omega\right)\stackrel{\mathrm{def}}{=}\frac{1}{2g_{1}g_{2}}\left[-\omega\left(g_{1}+g_{2}\right)\pm\sqrt{\left(g_{1}-g_{2}\right)^{2}\omega^{2}+4\gamma g_{1}g_{2}g_{\gamma}}\right].\label{eq:Omgamk1b}
\end{equation}

Note that definitions (\ref{eq:Omgamk1a}) and (\ref{eq:Omgamk1b})
of $\Omega_{\pm}$ and $K_{\pm}$ readily imply the following scaling
properties of these functions:
\begin{equation}
\Omega_{\pm}\left(s^{2}\gamma,sk\right)=s\Omega_{\pm}\left(\gamma,k\right),\quad K_{\pm}\left(s^{2}\gamma,s\omega\right)=sK_{\pm}\left(\gamma,\omega\right),s>0.\label{eq:Omgamk1c}
\end{equation}
These definitions imply also the following inversion symmetry properties
\begin{equation}
\Omega_{+}\left(\gamma,k\right)=-\Omega_{-}\left(\gamma,-k\right),\quad K_{+}\left(\gamma,\omega\right)=-K_{-}\left(\gamma,-\omega\right).\label{eq:Omgamk1d}
\end{equation}
Scaling properties (\ref{eq:Omgamk1c}) imply the following identities:
\begin{equation}
\Omega_{\pm}\left(\gamma,k\right)=\sqrt{\gamma}\Omega_{\pm}\left(1,\frac{k}{\sqrt{\gamma}}\right),\quad,K_{\pm}\left(\gamma,\omega\right)=\sqrt{\gamma}K_{\pm}\left(1,\frac{\omega}{\sqrt{\gamma}}\right).\label{eq:Omgamk2a}
\end{equation}
Equations (\ref{eq:Omgamk2a}) in turn imply that being given $\gamma>0$
the cross-model dispersion relations (\ref{eq:GGgamG2f}) and (\ref{eq:GGgamG2h})
can be written in the following alternative form:
\begin{equation}
\frac{\omega}{\sqrt{\gamma}}=\Omega_{\pm}\left(1,\frac{k}{\sqrt{\gamma}}\right),\quad\frac{k}{\sqrt{\gamma}}=K_{\pm}\left(1,\frac{\omega}{\sqrt{\gamma}}\right).\label{eq:Omgamk2b}
\end{equation}
Note that equations (\ref{eq:Omgamk2b}) manifest a simple scaling
property for the graphs of the cross-model dispersion relation as
$\gamma$ varies. Indeed, according to equations (\ref{eq:Omgamk2b})
the graph of the cross-model dispersion relation for arbitrary positive
$\gamma$ can be obtain from its graph for $\gamma=1$ by the following
simple linear scaling transformation by factor $\sqrt{\gamma}$ (see
Figures \ref{fig:dis-crpo3}, \ref{fig:dis-crpo-instab1} and \ref{fig:dis-crpo-instab2}):
\begin{equation}
\omega\rightarrow\sqrt{\gamma}\omega,\quad k\rightarrow\sqrt{\gamma}k\label{eq:Omgamk2c}
\end{equation}

\begin{rem}[scaling of the cross-point model]
\label{rem:crospo-scal} Observe that scaling formula (\ref{eq:Omgamk2c})
shows that two branches of the cross-model dispersion relations written
in either form (\ref{eq:Omgamk1a}) or (\ref{eq:Omgamk1b}) are separated
by a distance proportional to $\sqrt{\gamma}$. Since evidently $\sqrt{\gamma}\gg\gamma$
for small $\gamma$, one can think of using small coupling of any
two systems with a cross point for enhanced sensing, see Remark \ref{rem:tran-instab}.
\end{rem}

Equations (\ref{eq:Omgamk1a}) and (\ref{eq:Omgamk1b}) are useful
for our studies of respectively the absolute and the convection instabilities.
Consequently, the requirement for $\Omega_{\pm}\left(\gamma,k\right)$
and $K_{\pm}\left(\gamma,k\right)$ defined by equations (\ref{eq:Omgamk1a})
and (\ref{eq:Omgamk1a} to be real is immediately satisfied if respectively
$g_{\gamma}\geq0$ and $g_{1}g_{2}g_{\gamma}\geq0$. But in the case
when $g_{\gamma}<0$ or $g_{1}g_{2}g_{\gamma}<0$ the $\Omega_{\pm}\left(\gamma,k\right)$
and $K_{\pm}\left(\gamma,k\right)$ are respectively real if and only
if the expression under the square root an non-negative, that is
\begin{equation}
\Omega_{\pm}\left(\gamma,k\right)\text{ is real if and only if }\left|k\right|\geq\frac{2\sqrt{\gamma\left|g_{\gamma}\right|}}{\left|g_{1}-g_{2}\right|},\quad g_{\gamma}<0,\label{eq:Omgamk3a}
\end{equation}
\begin{equation}
K_{\pm}\left(\gamma,\omega\right)\text{ is real if and only if }\left|\omega\right|\geq\frac{2\sqrt{\gamma\left|g_{1}g_{2}g_{\gamma}\right|}}{\left|g_{1}-g_{2}\right|},\quad g_{1}g_{2}g_{\gamma}<0,\label{eq:Omgamk3b}
\end{equation}
Using relations (\ref{eq:Omgamk1a}), (\ref{eq:Omgamk3a}) we may
conclude that
\begin{equation}
\Im\left\{ \Omega_{\pm}\left(\gamma,k\right)\right\} \neq0,\text{ for real }k,\text{if and only if }g_{\gamma}<0\text{ and }\left|k\right|\leq\frac{2\sqrt{\gamma\left|g_{\gamma}\right|}}{\left|g_{1}-g_{2}\right|}.\label{eq:Omgamk3c}
\end{equation}
In addition to that the following representation holds for complex-valued
$\Omega_{\pm}\left(\gamma,k\right)$:
\begin{gather}
\Omega_{\pm}\left(\gamma,k\right)=\frac{1}{2}\left[-k\left(g_{1}+g_{2}\right)\pm\mathrm{i}\sqrt{\left|\left(g_{1}-g_{2}\right)^{2}k^{2}+4\gamma g_{\gamma}\right|}\right],\label{eq:Omgamk3d}\\
\text{ for real }k,\text{ if and only if }g_{\gamma}<0\text{ and }\left|k\right|\leq\frac{2\sqrt{\gamma\left|g_{\gamma}\right|}}{\left|g_{1}-g_{2}\right|}.\nonumber 
\end{gather}
Evidently relations (\ref{eq:Omgamk3c}) and (\ref{eq:Omgamk3d})
are associated with the absolute instability for wavenumbers in interval
$\left|k\right|\leq\frac{2\sqrt{\gamma\left|g_{\gamma}\right|}}{\left|g_{1}-g_{2}\right|}$.

Using relations (\ref{eq:Omgamk1b}), (\ref{eq:Omgamk3b}) we may
conclude that
\begin{equation}
\Im\left\{ K_{\pm}\left(\gamma,\omega\right)\right\} \neq0,\text{ for real }\omega,\text{if and only if }g_{1}g_{2}g_{\gamma}<0\text{ and }\left|\omega\right|\leq\frac{2\sqrt{\gamma\left|g_{1}g_{2}g_{\gamma}\right|}}{\left|g_{1}-g_{2}\right|}.\label{eq:Omgamk4a}
\end{equation}
In addition to that the following representation holds for complex-valued
$K_{\pm}\left(\gamma,\omega\right)$:
\begin{gather}
K_{\pm}\left(\gamma,\omega\right)=\frac{1}{2g_{1}g_{2}}\left[-\omega\left(g_{1}+g_{2}\right)\pm\mathrm{i}\sqrt{\left|\left(g_{1}-g_{2}\right)^{2}\omega^{2}+4\gamma g_{1}g_{2}g_{\gamma}\right|}\right],\label{eq:Omgamk4b}\\
\text{ for real }\omega,\text{ if and only if }g_{1}g_{2}g_{\gamma}<0\text{ and }\left|\omega\right|\leq\frac{2\sqrt{\gamma\left|g_{1}g_{2}g_{\gamma}\right|}}{\left|g_{1}-g_{2}\right|}.\nonumber 
\end{gather}
Evidently relations (\ref{eq:Omgamk4a}) and (\ref{eq:Omgamk4b})
are associated with the convection instability for frequencies in
interval $\left|\omega\right|\leq\frac{2\sqrt{\gamma\left|g_{1}g_{2}g_{\gamma}\right|}}{\left|g_{1}-g_{2}\right|}$.

\begin{figure}[h]
\begin{centering}
\hspace{-0.5cm}\includegraphics[scale=0.4]{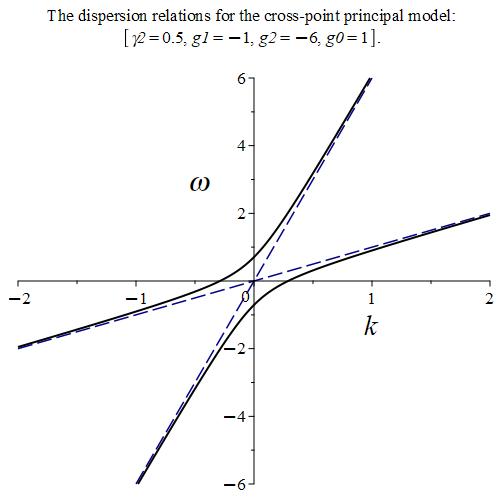}\hspace{0.5cm}\includegraphics[scale=0.4]{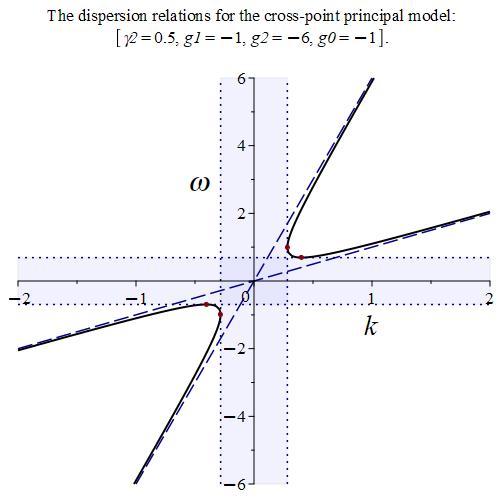}
\par\end{centering}
\centering{}(a)\hspace{7cm}(b)\caption{\label{fig:dis-crpo1} Plots of the cross-point model dispersion relations
(\ref{eq:GGgamG2e}) for $\gamma=0.5$, $g_{1}=-1$, $g_{2}=-6$ and
(a) $g_{\gamma}=1>0$; (b) $g_{\gamma}=-1<0$. Solid (black) curves
represent the dispersion curves for $\gamma=0.5$ whereas dashed (blue)
straight lines represent the dispersion curves for $\gamma=0$ as
a reference, that is the case when the two subsystems do not interact.
The doted (blue) lines identify instability edges and shaded area
represent points $\left(k,\omega\right)$ for which at least one of
$k$ and $\omega$ becomes complex-valued according to relations (\ref{eq:Omgamk3d})
and (\ref{eq:Omgamk4b}). The circular dots (red) identify the transition
to instability points.}
\end{figure}

\begin{figure}[h]
\begin{centering}
\hspace{-0.5cm}\includegraphics[scale=0.4]{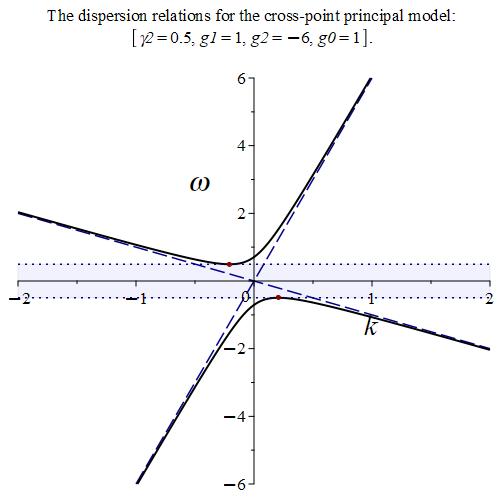}\hspace{0.5cm}\includegraphics[scale=0.4]{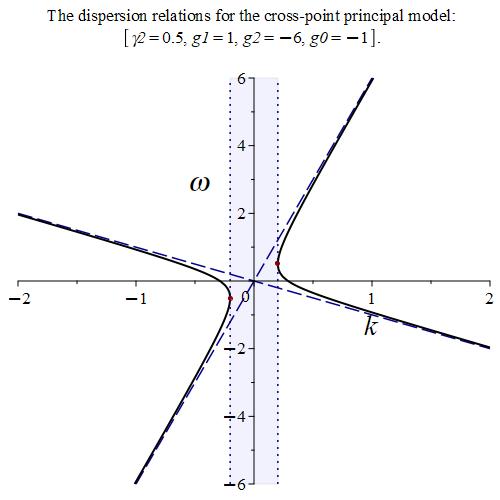}
\par\end{centering}
\centering{}(a)\hspace{7cm}(b)\caption{\label{fig:dis-crpo2} Plots of the cross-point model dispersion relations
(\ref{eq:GGgamG2e}) for $\gamma=0.5$, $g_{1}=1$, $g_{2}=-6$ and
(a) $g_{\gamma}=1>0$; (b) $g_{\gamma}=-1<0$. Solid (black) curves
represent the dispersion curves for $\gamma=0.5$ whereas dashed (blue)
straight lines represent the dispersion curves for $\gamma=0$ as
a reference, that is the case when the two subsystems do not interact.
The doted (blue) lines identify gap edges and shaded area represent
points $\left(k,\omega\right)$ for which at least one of $k$ and
$\omega$ becomes complex-valued according to relations (\ref{eq:Omgamk3d})
and (\ref{eq:Omgamk4b}). The circular dots (red) identify the transition
to instability points.}
\end{figure}

\begin{figure}[h]
\begin{centering}
\hspace{-0.5cm}\includegraphics[scale=0.4]{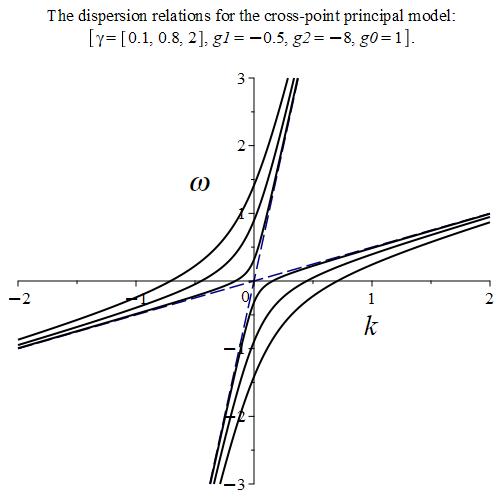}\hspace{0.5cm}\includegraphics[scale=0.4]{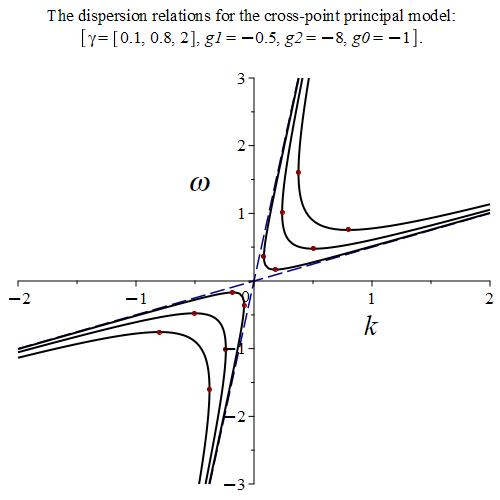}
\par\end{centering}
\centering{}(a)\hspace{7cm}(b)\caption{\label{fig:dis-crpo3} Plots of the cross-point model dispersion relations
(\ref{eq:GGgamG2e}) for $\gamma=0.1,0.8,2$, $g_{1}=-0.5$, $g_{2}=-8$
and (a) $g_{\gamma}=1>0$; (b) $g_{\gamma}=-1<0$. Solid (black) curves
represent the dispersion curves for $\gamma=0.1,0.8,2$ whereas dashed
(blue) straight lines represent the dispersion curves for $\gamma=0$
as a reference, that is the case when the two subsystems do not interact.
The solid curves that are closer to the dashed straight lines correspond
to smaller values of coupling coefficient $\gamma$. The circular
dots (red) identify the transition to instability points.}
\end{figure}

Figures \ref{fig:dis-crpo1}- \ref{fig:dis-crpo3} show the plots
of the cross-point dispersion relations (\ref{eq:GGgamG2f}) and (\ref{eq:GGgamG2h})
and shaded areas there represent points $\left(k,\omega\right)$ for
which at least one of $k$ and $\omega$ becomes complex-valued according
to relations (\ref{eq:Omgamk3d}) and (\ref{eq:Omgamk4b}). Figures
\ref{fig:dis-crpo1}- \ref{fig:dis-crpo3} assume efficiently that
$\omega_{0}=0$ and $k_{0}=0$ with understanding that in the case
of arbitrary real $\omega_{0}$ and $k_{0}$ we need to apply translation
transformation (\ref{eq:GGgamG2f}) in the $k\omega$-plane to the
relevant graphs.

Compare Fig. \ref{fig:dis-crpo3} with Figures \ref{fig:dis-Lev2}(c),
\ref{fig:dis-Lev3}(c), \ref{fig:disp-Lev5}(b), \ref{fig:dis-Lev5f}(c),
\ref{fig:disp-Lev6}(b), \ref{fig:dis-Lev6f}(b), (c), \ref{fig:dis-Lev7}
(c).

\subsection{Convection and absolute instability branches\label{subsec:cropobra}}

\begin{figure}[h]
\begin{centering}
\includegraphics[scale=0.6]{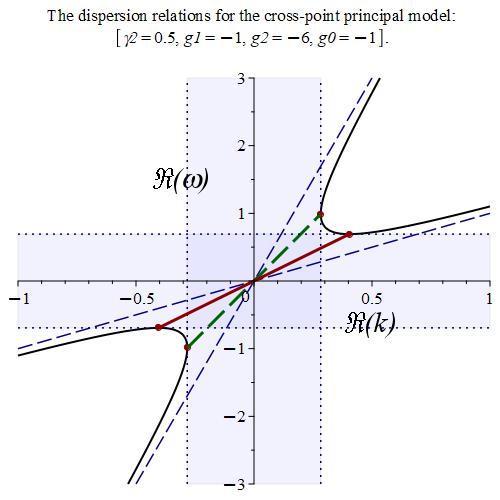}
\par\end{centering}
\centering{}\caption{\label{fig:dis-crpo-instab1} Dispersion-instability graph of the
cross-point model dispersion relations (\ref{eq:GGgamG2e}) for $\gamma=0.5$,
$g_{1}=-1$, $g_{2}=-6$ and $g_{\gamma}=-1<0$. Solid (black) curves
represent the dispersion curves for $\gamma=0.5$ whereas dashed (blue)
straight lines represent the dispersion curves for $\gamma=0$ as
a reference, that is the case when the two subsystems do not interact.
The doted (blue) lines identify instability edges and shaded area
represent points $\left(k,\omega\right)$ for which at least one of
$k$ and $\omega$ becomes complex-valued according to relations (\ref{eq:Omgamk3d})
and (\ref{eq:Omgamk4b}). The circular dots (red) identify points
of transition from stability to instability. The bold, solid (red)
segment represents a branch of points $\left(\Re\left\{ k\right\} ,\omega\right)$
with real $\omega$ and $\Im\left\{ k\right\} \protect\neq0$ which
are points of the convection instability described by relations (\ref{eq:Omgamk4b}).
The bold dashed (green) segment represents points $\left(k,\Re\left\{ \omega\right\} \right)$
with $k$ real and $\Im\left\{ \omega\right\} \protect\neq0$ which
are points of the absolute instability described by relations (\ref{eq:Omgamk3d}).
The circular dots (red) identify the transition to instability points.}
\end{figure}

\begin{figure}[h]
\begin{centering}
\hspace{-0.5cm}\includegraphics[scale=0.4]{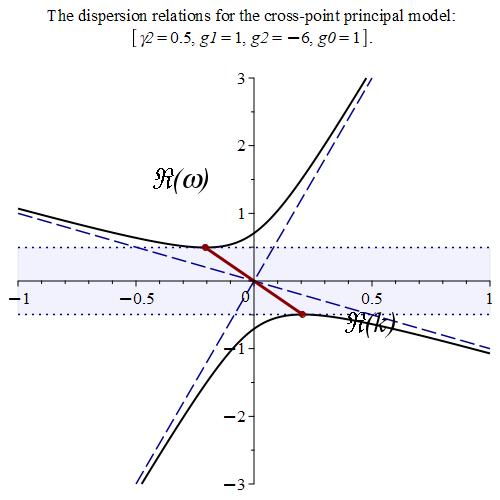}\hspace{0.5cm}\includegraphics[scale=0.4]{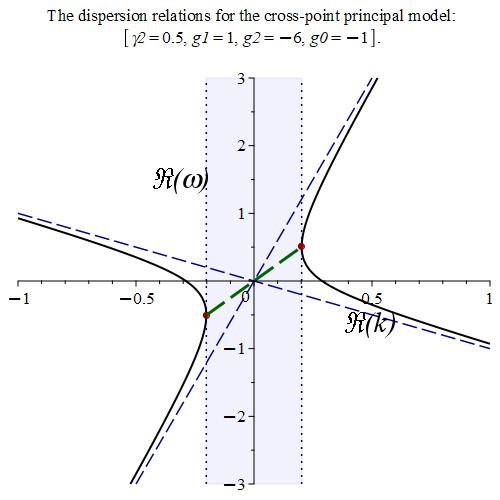}
\par\end{centering}
\centering{}(a)\hspace{7cm}(b)\caption{\label{fig:dis-crpo-instab2} Dispersion-instability graph of the
cross-point model dispersion relations (\ref{eq:GGgamG2e}) for $\gamma=0.5$,
$g_{1}=1$, $g_{2}=-6$ and (a) $g_{\gamma}=1>0$; (b) $g_{\gamma}=-1<0$.
Solid (black) curves represent the dispersion curves for $\gamma=0.5$
whereas dashed (blue) straight lines represent the dispersion curves
for $\gamma=0$ as a reference, that is the case when the two subsystems
do not interact. The doted (blue) lines identify instability edges
and shaded area represent points $\left(k,\omega\right)$ for which
at least one of $k$ and $\omega$ becomes complex-valued according
to relations (\ref{eq:Omgamk3d}) and (\ref{eq:Omgamk4b}). The circular
dots (red) identify points of transition from stability to instability.
The bold, solid (red) segment represents a branch of points $\left(\Re\left\{ k\right\} ,\omega\right)$
with real $\omega$ and $\Im\left\{ k\right\} \protect\neq0$ which
are points of the convection instability as described by described
by relations (\ref{eq:Omgamk4b}). The bold dashed (green) segment
represents points $\left(k,\Re\left\{ \omega\right\} \right)$ with
$k$ real and $\Im\left\{ \omega\right\} \protect\neq0$ which are
points of the absolute instability described by relations (\ref{eq:Omgamk3d}).
The circular dots (red) identify the transition to instability points.}
\end{figure}

Figures \ref{fig:dis-crpo-instab1} and \ref{fig:dis-crpo-instab2}
show the dispersion-instability plots that extend plots in Figures
\ref{fig:dis-crpo1}(b) and \ref{fig:dis-crpo1} by adding up there
the convection and absolute instability branches. Note that just in
case of Figures \ref{fig:dis-crpo1}- \ref{fig:dis-crpo3} \ref{fig:dis-crpo-instab1}
and \ref{fig:dis-crpo-instab2} assume efficiently that $\omega_{0}=0$
and $k_{0}=0$ with understanding that in the case of arbitrary real
$\omega_{0}$ and $k_{0}$ we need to apply translation transformation
(\ref{eq:GGgamG2f}) in the $k\omega$-plane to the relevant graphs.

\subsection{Lagrangian framework for the cross-point model\label{subsec:Lag-crpo}}

With cross-point model dispersion relations (\ref{eq:GGgamG2h}) in
mind let us consider the following form of the dispersion function
and the corresponding dispersion relations
\begin{equation}
G_{0}\left(k,\omega\right)\stackrel{\mathrm{def}}{=}A\omega^{2}+2\omega kB+Ck^{2}-D,\quad G_{0}\left(k,\omega\right)=0,\label{eq:GABC1a}
\end{equation}
where $A$, $B$, $C$ and $D$, are real-valued coefficients. The
choice of signs before coefficients in equations (\ref{eq:GABC1a})
is motivated by its applications to the GTL, the e-beam and other
physical systems. Note that equation (\ref{eq:GGgamG2h}) and hence
equation (\ref{eq:GABC1a}) tacitly assume that $\omega_{0}=0$ and
$k_{0}=0$.

It is natural and important to ask if the dispersion relations (\ref{eq:GABC1a})
can be associated with a ``real physical system'', that is with
the Euler-Lagrange equations of a Lagrangian? The answer to this question
is positive, and an expression for such a Lagrangian $\mathcal{L}_{G}$
is as follows:
\begin{equation}
\mathcal{L}_{G_{0}}\left(\partial_{t}Q,\partial_{z}Q,Q\right)\stackrel{\mathrm{def}}{=}\frac{Q^{2}}{2}G\left(\frac{\partial_{z}Q}{Q},\frac{\partial_{t}Q}{Q}\right)\equiv\frac{1}{2}\left[A\left(\partial_{t}Q\right)^{2}-2B\partial_{t}Q\partial_{z}Q+C\left(\partial_{z}Q\right)^{2}-DQ^{2}\right],\label{eq:GABC1b}
\end{equation}
where $Q=Q\left(z,t\right)$. Indeed, the EL equations for Lagrangian
$\mathcal{L}_{G}$ defined by equations (\ref{eq:GABC1b}) are
\begin{equation}
\left[A\partial_{t}^{2}-2B\partial_{t}\partial_{z}+C\partial_{z}^{2}+D\right]Q=0.\label{eq:GABC1c}
\end{equation}
To find the dispersion relations associated with the EL equation (\ref{eq:GABC1c})
we proceed in the standard fashion and consider the system eigenmodes
of the form
\begin{equation}
Q\left(z,t\right)=\hat{Q}\left(k,\omega\right)\mathrm{e}^{-\mathrm{i}\left(\omega t-kz\right)}.\label{eq:GABC1d}
\end{equation}
Plugging in expression (\ref{eq:GABC1c}) for $Q\left(z,t\right)$
in the EL equation (\ref{eq:GABC1c}) after elementary evaluations
we obtain
\begin{equation}
\mathrm{e}^{-\mathrm{i}\left(\omega t-kz\right)}\hat{Q}\left(k,\omega\right)\left[A\omega^{2}+2\omega kB+Ck^{2}-D\right]=0.\label{eq:GABC1e}
\end{equation}
Assuming naturally that $\hat{Q}\left(k,\omega\right)$ being an amplitude
of an eigenmode is not zero we recover from equation (\ref{eq:GABC1e})
the following dispersion relation associated with the EL equation
(\ref{eq:GABC1c}) 
\begin{equation}
A\omega^{2}+2\omega kB+Ck^{2}-D=0,\label{eq:GABC1f}
\end{equation}
which is evidently equivalent to the original dispersion relation
(\ref{eq:GABC1a}). Hence indeed the Lagrangian $\mathcal{L}_{G_{0}}$
defined by equation (\ref{eq:GABC1b}) yields indeed the EL equation
having the desired dispersion relation (\ref{eq:GABC1a}).

Motivated by the cross-point dispersion relations (\ref{eq:GGgamG2e})
we introduce cross-point dispersion relations
\begin{equation}
G_{\mathrm{crp}}\left(k,\omega\right)\stackrel{\mathrm{def}}{=}\left(\omega+g_{1}k\right)\left(\omega+g_{2}k\right)-\gamma g_{\gamma},\quad G_{\mathrm{crp}}\left(k,\omega\right)=0.\label{eq:GABC2a}
\end{equation}
Then according to formula (\ref{eq:GABC1b}) the corresponding to
dispersion relations (\ref{eq:GABC2a}) Lagrangian $\mathcal{L}_{\mathrm{crp}}$
is of the form
\begin{equation}
\mathcal{L}_{\mathrm{crp}}=\frac{1}{2}\left[\left(\partial_{t}Q+g_{1}\partial_{z}Q\right)\left(\partial_{t}Q+g_{2}\partial_{z}Q\right)-\gamma g_{\gamma}Q^{2}\right].\label{eq:GABC2b}
\end{equation}
The expression (\ref{eq:GABC2b}) can be also readily obtained from
the last expression of relations (\ref{eq:GABC1b}) by setting up
there the following values of coefficients:
\begin{equation}
A=1,\quad B=\frac{g_{1}+g_{2}}{2},\quad C=g_{1}g_{2},\quad D=\gamma g_{\gamma}.\label{eq:GABC2c}
\end{equation}

Recall that when setting up equation (\ref{eq:GABC1a}) we tacitly
assumed that $\omega_{0}=0$ and $k_{0}=0$. In case of arbitrary
real $\omega_{0}$ and $k_{0}$ representation (\ref{eq:GABC1a})
has to be replaced with
\begin{equation}
G\left(k,\omega\right)\stackrel{\mathrm{def}}{=}A\left(\omega-\omega_{0}\right)^{2}-2\left(\omega-\omega_{0}\right)\left(k-k_{0}\right)B-C\left(k-k_{0}\right)^{2}-D,\quad G\left(k,\omega\right)=0.\label{eq:GABC3a}
\end{equation}
We can ask the same question as before. Namely, is the dispersion
relations (\ref{eq:GABC3a}) can be associated with a ``real physical
system'', that is with the Euler-Lagrange equations of a Lagrangian?
To answer this question we note that the dispersion relation $G\left(k,\omega\right)=0$
turns into the relevant differential operator with respect to variables
$t$ and $z$ under substitution
\begin{equation}
\omega\rightarrow-\mathrm{i}\partial_{t}-\omega_{0},\quad k\rightarrow\mathrm{i}\partial_{z}-k_{0}.\label{eq:GABC3b}
\end{equation}
It is an elementary exercise to compute an expression of that differential
operator. It turns out that that expression contains terms $2\mathrm{i}\left(A\omega_{0}+Bk_{0}\right)\partial_{t}$
and $-2\mathrm{i}\left(B\omega_{0}+Ck_{0}\right)\partial_{z}$ which
evidently have pure imaginary coefficients, unless $A\omega_{0}+Bk_{0}=B\omega_{0}+Ck_{0}=0$.
The presence of these terms rules out a possibility of the existence
of a Lagrangian for arbitrary real $\omega_{0}$ and $k_{0}$ unless
$A=B=C=0$.

\section{Conclusions}

We extended here our previously constructed field theory for TWTs
as well the celebrated Pierce theory by replacing there the standard
transmission line (TL) with its generalization allowing for the low
frequency cutoff. We developed all the details of the extended TWT
field theory and using a particular choice of the TWT parameters we
derived a physically appealing factorized form of the TWT dispersion
relations. This form has two factors that represent exactly the dispersion
functions of non-interacting generalized transmission line (GTL) and
the electron beam (e-beam). We found that the factorized dispersion
relations comes with a number of interesting features including: (i)
focus points that belong to each dispersion curve as TWT principle
parameter varies; (ii) formation of ``hybrid'' branches of the TWT
dispersion curves parts of which can be traced to non-interacting
GTL and the e-beam. We also introduced and studied a simple ``cross-point
model dispersion relation''. This models accounts for the TWT dispersion
relation behavior near the cross-points of the dispersion functions
of non-interacting GTL and the e-beam when coupling between the GTL
and e-beam is small.

\textbf{\vspace{0.2cm}
}

\textbf{ACKNOWLEDGMENT:} This research was supported by AFOSR MURI
Grant FA9550-20-1-0409 administered through the University of New
Mexico. The author is grateful to E. Schamiloglu for sharing his deep
and vast knowledge of high power microwave devices and inspiring discussions.\textbf{\vspace{0.2cm}
}

\section{Appendix}

\subsection{Fourier transform\label{sec:four}}

The several common variations different in signs and constants. Our
preferred form of the \emph{Fourier transform} $\widehat{f}=f^{\land}$
of $f$ and the \emph{inverse Fourier transform} $f^{\land}$ of $f$
follows to \citep[1.1.7]{AdamHed}, \citep[20.2]{ArfWeb}, \citep[Notations]{DauLio1},
\citep[7.2, 7.5]{Foll}, \citep[25]{TreB}:
\begin{gather}
\widehat{f}\left(k\right)\stackrel{\mathrm{def}}{=}\int_{-\infty}^{\infty}f\left(z\right)e^{-\mathrm{i}kz}\,dz,\quad f\left(z\right)=\left[\widehat{f}\left(k\right)\right]^{\lor}=\frac{1}{2\pi}\int_{-\infty}^{\infty}\widehat{f}\left(k\right)\mathrm{e}^{\mathrm{i}kz}\,\mathrm{d}k\label{eq:fouri1a}
\end{gather}
\begin{equation}
\widehat{f}\left(\omega\right)\stackrel{\mathrm{def}}{=}\int_{-\infty}^{\infty}f\left(t\right)e^{\mathrm{i}\omega t}\,\mathrm{d}t,\quad f\left(t\right)=\frac{1}{2\pi}\int_{-\infty}^{\infty}\widehat{f}\left(\omega\right)\mathrm{e}^{-\mathrm{i}\omega t}\,\mathrm{d}\omega,\label{eq:fouri1b}
\end{equation}
\begin{gather}
\widehat{f}\left(k,\omega\right)\stackrel{\mathrm{def}}{=}\int_{-\infty}^{\infty}f\left(z,t\right)e^{\mathrm{i}\left(\omega t-kz\right)}\,dz\mathrm{d}t,\label{eq:fouri1c}\\
f\left(z,t\right)=\left[\widehat{f}\left(k,\omega\right)\right]^{\lor}=\frac{1}{\left(2\pi\right)^{2}}\int_{-\infty}^{\infty}\widehat{f}\left(k,\omega\right)\mathrm{e}^{-\mathrm{i}\left(\omega t-kz\right)}\,\mathrm{d}k\mathrm{d}\omega.\nonumber 
\end{gather}
\emph{Note the difference of the choice of the sign for time $t$
and spacial variable $z$ in the above formula. It is motivated by
the desire to have ``wave'' form for exponential $\mathrm{e}^{-\mathrm{i}\left(\omega t-kz\right)}$
when both variables $t$ and $z$ are present}.

For multi-dimensional space variable $x\in\mathbb{R}^{n}$ the Fourier
transform $\widehat{f}$ of $f$ and the inverse Fourier transform
$f^{\land}$ of $f$ are defined by, \citep[1.1.7]{AdamHed}, \citep[Notations]{DauLio1},
\citep[7.5]{Foll}: 
\begin{equation}
\widehat{f}\left(k\right)\stackrel{\mathrm{def}}{=}\int_{\mathbb{R}^{n}}\widehat{f}\left(x\right)\mathrm{e}^{-\mathrm{i}k\cdot x}\,\mathrm{d}x,\quad f\left(x\right)=\left[\widehat{f}\left(k\right)\right]^{\lor}=\frac{1}{\left(2\pi\right)^{n}}\int_{\mathbb{R}^{n}}\widehat{f}\left(k\right)\mathrm{e}^{\mathrm{i}k\cdot x}\,\mathrm{d}k,\quad k,x\in\mathbb{R}^{n},\label{eq:fourier1d}
\end{equation}
which is consistent with equations (\ref{eq:fouri1a}). Then the Plancherel-Parseval
formula reads, \citep[4.3.1]{Evans}, \citep[7.5]{Foll}, \citep[0.26]{FolPDE}:
\begin{gather}
\left(f,g\right)=\left(2\pi\right)^{-n}\left(\widehat{f},\widehat{g}\right),\quad\left\Vert f\right\Vert =\left(2\pi\right)^{-n/2}\left\Vert \widehat{f}\right\Vert ,\label{eq:forier1e}\\
\left(f,g\right)\stackrel{\mathrm{def}}{=}\int_{\mathbb{R}^{n}}\overline{f\left(x\right)}g\left(x\right)\,\mathrm{d}x,\quad\left\Vert f\right\Vert \stackrel{\mathrm{def}}{=}\sqrt{\left(f,f\right)}.\nonumber 
\end{gather}

This preference was motivated by the fact that the so-defined Fourier
transform of the convolution of two functions has its simplest form.
Namely, the convolution $f\ast g$ of two functions $f$ and $g$
is defined by \citealt[4.3.1]{Evans}, \citep[7.2, 7.5]{Foll},
\begin{gather}
\left[f\ast g\right]\left(t\right)=\left[g\ast f\right]\left(t\right)=\int_{-\infty}^{\infty}f\left(t-t^{\prime}\right)g\left(t^{\prime}\right)\,\mathrm{d}t^{\prime},\label{eq:fourier2a}\\
\left[f\ast g\right]\left(z,t\right)=\left[g\ast f\right]\left(z,t\right)=\int_{-\infty}^{\infty}f\left(z-z^{\prime},t-t^{\prime}\right)g\left(z^{\prime},t^{\prime}\right)\,\mathrm{d}z^{\prime}\mathrm{d}t^{\prime}.\label{eq:fourier2b}
\end{gather}
 Then its Fourier transform as defined by equations (\ref{eq:fouri1a})-(\ref{eq:fouri1c})
satisfies the following properties:
\begin{gather}
\left[f\ast g\right]^{\land}\left(\omega\right)=\widehat{f}\left(\omega\right)\widehat{g}\left(\omega\right),\label{eq:fourier3a}\\
\left[f\ast g\right]^{\land}\left(k,\omega\right)=\widehat{f}\left(k,\omega\right)\widehat{g}\left(k,\omega\right).\label{eq:fourier3b}
\end{gather}

\subsection{The Pierce theory concise review\label{subsec:pier-rev}}

This section is essentially an excerpt from our analysis in \citet{SchaFig}.
The dispersion relation of the Pierce model written in terms of propagation
constant $\beta$ reads, \citet[Sec. 12.2.2, 12.2.3]{Gilm1}, \citep[2.2, 2.3]{PierTWT}
\begin{equation}
\frac{2C_{\mathrm{P}}^{3}\beta_{\mathrm{e}}}{\left(\beta-\beta_{\mathrm{e}}\right)^{2}}\frac{\beta^{2}\beta_{\mathrm{c}}}{\beta^{2}-\beta_{\mathrm{c}}^{2}}+1=0,\quad\beta=\frac{\omega}{u},\quad\beta_{\mathrm{e}}=\frac{\omega}{\mathring{v}},\quad\beta_{\mathrm{c}}=\frac{\omega}{w},\label{eq:pierbet1a}
\end{equation}
where $C_{\mathrm{P}}$ is the so-called \emph{Pierce gain parameter},
$\beta$, $\beta_{\mathrm{e}}$ and $\beta_{\mathrm{c}}$ are wavenumbers
(propagation constants) of the eigenmode, the electron beam and the
``circuit'', that is, the transmission line, respectively, and $u$,
$\mathring{v}$ and $w$ are the corresponding phase velocities. We
refer to $\beta$ and to the corresponding 
\begin{equation}
\frac{2C_{\mathrm{P}}^{3}\widetilde{\beta}_{\mathrm{e}}}{\left(\beta-\beta_{\mathrm{e}}\right)^{2}}\frac{\beta^{2}\beta_{\mathrm{c}}}{\beta^{2}-\beta_{\mathrm{c}}^{2}}+1=0,\quad\beta=\frac{\omega}{u},\quad\beta_{\mathrm{e}}=\frac{\omega}{\mathring{v}},\quad\beta_{\mathrm{c}}=\frac{\omega}{w},\label{eq:pierbet1aa}
\end{equation}

Equation (\ref{eq:pierbet1a}) can be readily transformed into
\begin{equation}
\frac{2\mathring{v}wC_{\mathrm{P}}^{3}}{w^{2}-u^{2}}+\frac{\left(u-\mathring{v}\right)^{2}}{u^{2}}=0,\quad u=\frac{\omega}{\beta},\label{eq:pierbet1b}
\end{equation}
and we refer to its solutions $u$ as\emph{ characteristic velocities}.
\emph{The TWT characteristic equation can be viewed as an equivalent
representation of the TWT dispersion relations in terms of the phase
velocity associated with the relevant TWT eigenmode.}

The dimensionless form of equation (\ref{eq:pierbet1b}) is
\begin{equation}
\frac{\gamma}{\chi^{2}-u^{\prime2}}+\frac{\left(u^{\prime}-1\right)^{2}}{u^{\prime2}}=0,\quad u^{\prime}=\frac{u}{\mathring{v}},\quad\chi=\frac{w}{\mathring{v}},\quad\gamma=2\chi C_{\mathrm{P}}^{3}.\label{eq:pierbet1c}
\end{equation}
The Pierce theory emerges from the TWT field theory as its high-frequency
limit, that is when $\omega\rightarrow\infty$, \citet[Chap. 4.2, 29, 62]{FigTWTbk},
with the following characteristic equation in dimensionless form:
\begin{gather}
\mathscr{D}\left(u^{\prime}\right)=\frac{\gamma}{\chi^{2}-u^{\prime2}}+\frac{\left(u^{\prime}-1\right)^{2}}{u^{\prime2}}=0,\quad\gamma^{\prime}=\frac{\gamma}{\mathring{v}{}^{2}},\quad u^{\prime}=\frac{u}{\mathring{v}},\quad\chi=\frac{w}{\mathring{v}}.\label{eq:pierbet1d}
\end{gather}
It can be readily verified that equation (\ref{eq:pierbet1d}) is
equivalent to equation (\ref{eq:pierbet1c}) of the Pierce theory.
Note also that equations (\ref{eq:pierbet1c}) and (\ref{eq:pierbet1d})
are identical if the following relation holds between our TWT principal
parameter $\gamma$ and the Pierce gain parameter $C_{\mathrm{P}}$,
\citet[Chap. 62]{FigTWTbk}:
\begin{gather}
\gamma=\frac{b^{2}}{C}\frac{e^{2}}{m}R_{\mathrm{sc}}^{2}\sigma_{\mathrm{B}}\mathring{n}=2\mathring{v}wC_{\mathrm{P}}^{3}\;\text{ or }\gamma^{\prime}=\frac{\gamma}{\mathring{v}{}^{2}}=2\frac{w}{\mathring{v}}C_{\mathrm{P}}^{3}=2\chi C_{\mathrm{P}}^{3}.\label{eq:pierbet1e}
\end{gather}
Note that $C_{\mathrm{P}}^{3}$ is defined as one-quarter of the ratio
of the circuit (transmission line) impedance $Z_{0}$ to the e-beam
impedance $\frac{V_{0}}{I_{0}}$, \citep[Sec. 2.3]{PierTWT}, \citep[Sec. 12.2.2, 12.2.3]{Gilm1},
\citet[Chap. 62]{FigTWTbk}:
\begin{equation}
C_{\mathrm{P}}^{3}=\frac{Z_{0}I_{0}}{4V_{0}},\label{eq:pierbet1f}
\end{equation}
where $I_{0}$ and $V_{0}$ are the current and the voltage associated
with the e-beam. The quantities used above satisfy the following equations
\begin{gather}
Z_{0}=\sqrt{\frac{L}{C}},\quad I_{0}=e\sigma_{\mathrm{B}}\mathring{n},\quad eV_{0}=\frac{m\mathring{v}^{2}}{2},\quad w=\frac{1}{\sqrt{LC}}.\label{eq:pierbet1h}
\end{gather}
A straightforward verification confirms the consistency of equations
(\ref{eq:pierbet1f}) and (\ref{eq:pierbet1h}) with equation (\ref{eq:pierbet1e}). 

\subsubsection{Critical value of parameter $\gamma$ in the Pierce theory}

An equivalent form of the characteristic equation (\ref{eq:pierbet1c})
for the Pierce model is
\begin{equation}
F\left(u\right)=\frac{\left(u^{2}-{\it \chi}^{2}\right)\left(u-1\right)^{2}}{u^{2}}=\gamma>0.\label{eq:charP1a}
\end{equation}
Fig. provides graphical representation of function $F\left(u\right)$
defined by equations (\ref{eq:charP1a}) for two cases: $\chi=0.9<1$
and $\chi=1.1>1$. Fig. shows an important difference between the
two cases since in equations (\ref{eq:charP1a}) in $\gamma>0$.

An analysis of the characteristic equation (\ref{eq:charP1a}) shows
that in case when $\chi<1$ there exists a critical value $\gamma_{\mathrm{Pcr}}>0$
of the parameter $\chi$ such that:
\begin{enumerate}
\item for $0<\gamma<\gamma_{\mathrm{Pcr}}$ all solutions $u$ to equation
(\ref{eq:charP1a}) are real-valued, there is no amplification;
\item for $\gamma>\gamma_{\mathrm{Pcr}}$ there are exactly two different
real-valued solutions $u$ to equation (\ref{eq:charP1a}) and exactly
two different complex-valued solutions that are complex-conjugate,
that is there is an amplification.
\end{enumerate}
Consequently, in case when $\chi<1$ the amplification is possible
if only if $\gamma>\gamma_{\mathrm{Pcr}}$ and if that is the case
it occurs for all frequencies. The critical value $\gamma_{\mathrm{Pcr}}$
and the corresponding to it critical value $u_{\mathrm{Pcr}}$ satisfy
the following relations:
\begin{equation}
u_{\mathrm{Pcr}}\left(\chi\right)=\chi^{\frac{2}{3}},\quad\gamma_{\mathrm{Pcr}}\left(\chi\right)=\left.\frac{\left(u^{2}-{\it \chi}^{2}\right)\left(u-1\right)^{2}}{u^{2}}\right|_{u=u_{\mathrm{Pcr}}}=\left(1-\chi^{\frac{2}{3}}\right)^{3}=\left(1-u_{\mathrm{Pcr}}\left(\chi\right)\right)^{3}.\label{eq:charP1b}
\end{equation}
Equations (\ref{eq:charP1b}) imply that
\begin{equation}
u_{\mathrm{Pcr}}\left(\chi\right)=1-\gamma_{\mathrm{Pcr}}\left(\chi\right)^{\frac{1}{3}}.\label{eq:charP1ba}
\end{equation}
Formulas (\ref{eq:charP1b}) readily imply also that $\gamma_{\mathrm{Pcr}}>0$
if $\chi<1$ whereas $\gamma_{\mathrm{Pcr}}<0$ if $\chi>1,$and
\begin{equation}
\lim_{\chi\rightarrow1}\gamma_{\mathrm{Pcr}}\left(\chi\right)=0,\quad\lim_{\chi\rightarrow1}u_{\mathrm{Pcr}}\left(\chi\right)=1.\label{eq:charP1c}
\end{equation}
Hence in the case of $\chi>1$ we have $\gamma_{\mathrm{Pcr}}<0<\gamma$
implying that for any $\gamma>0$ there are always exactly two non-real
solutions $u$ to equation (\ref{eq:charP1a}) and consequently there
is always amplification. In contract, in the case $\chi<1$ the amplification
exists if and only if $\gamma>\gamma_{\mathrm{Pcr}}>0$.

\begin{figure}[h]
\begin{centering}
\hspace{-0.5cm}\includegraphics[scale=0.17]{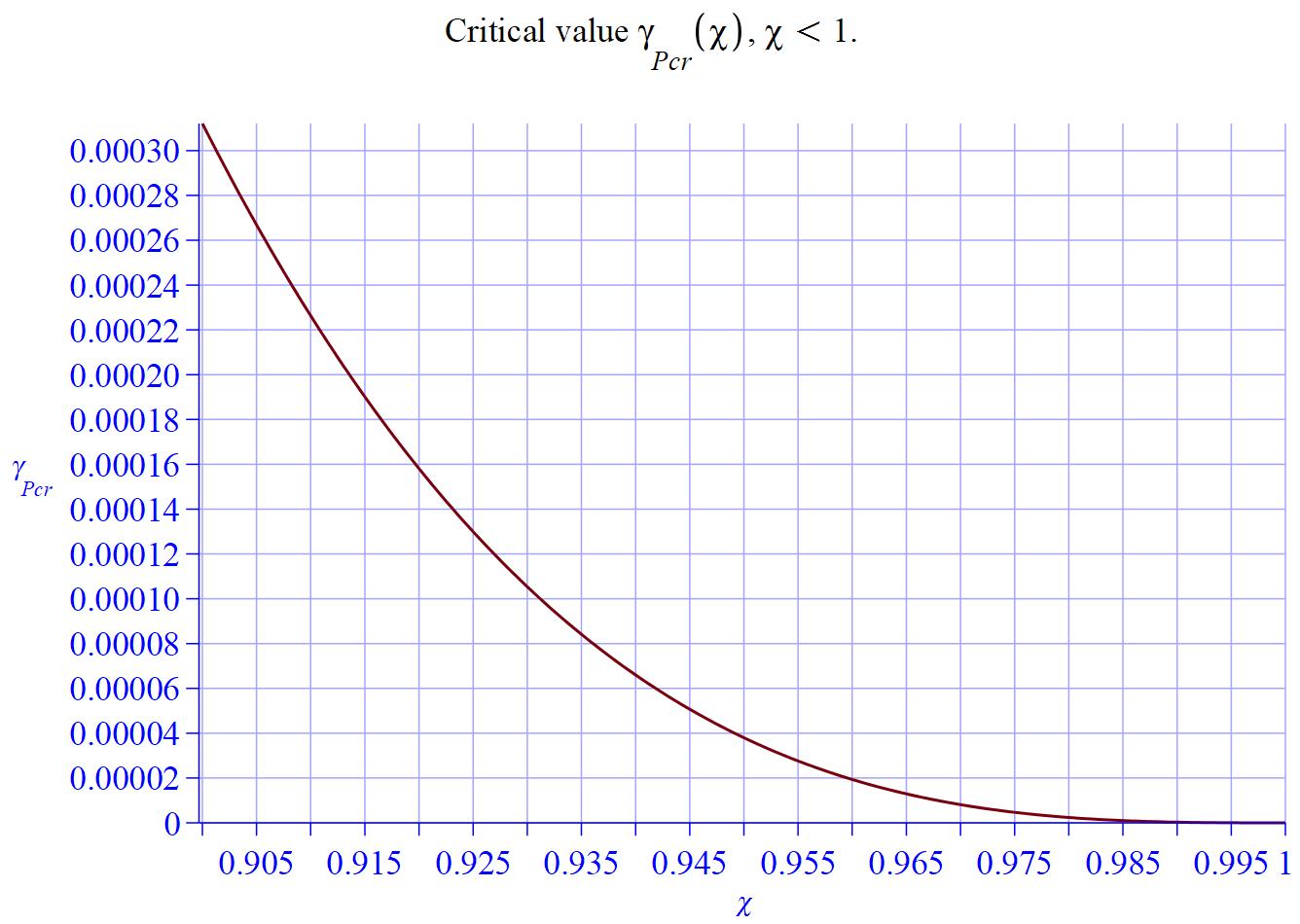}\hspace{1cm}\includegraphics[scale=0.17]{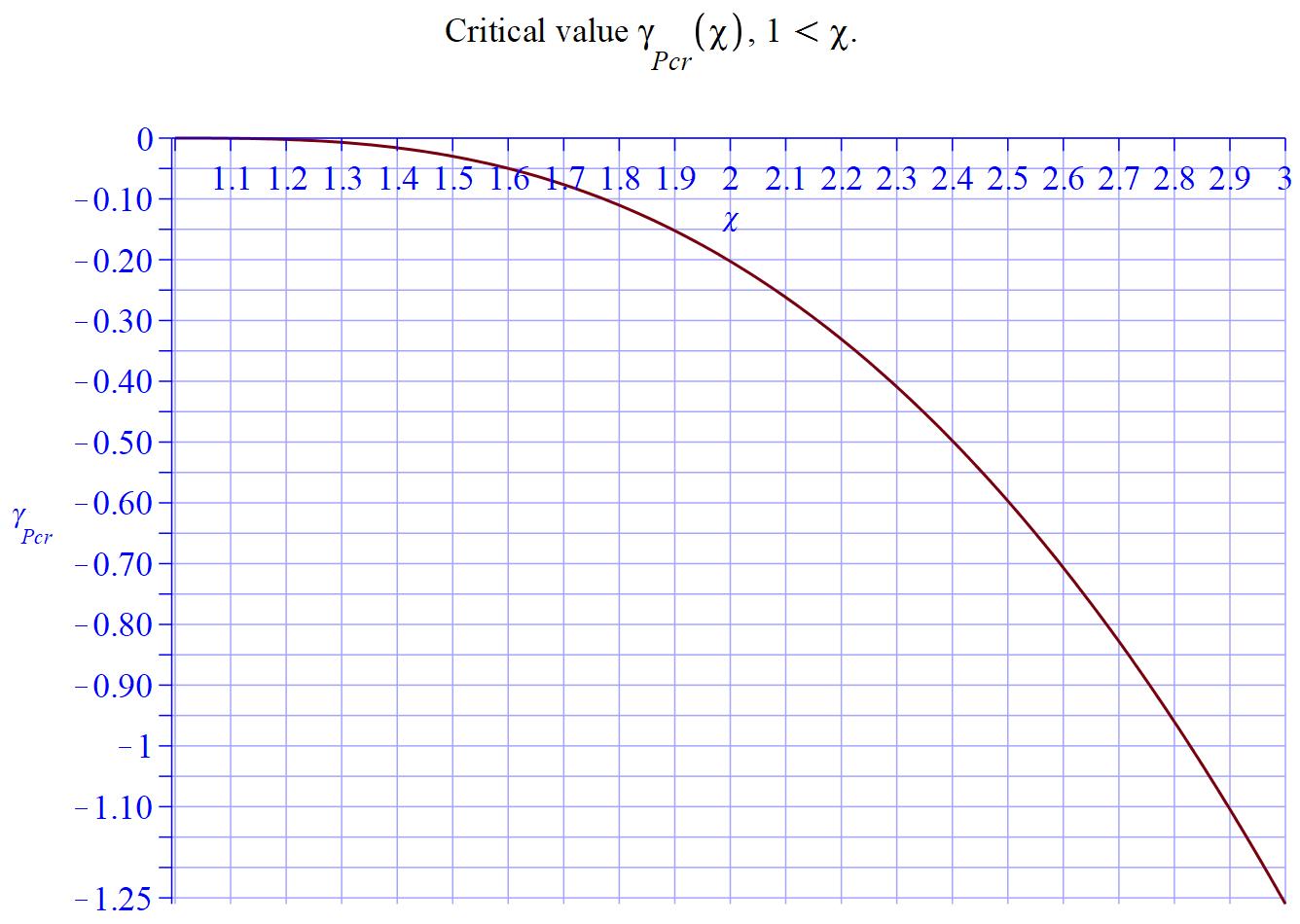}
\par\end{centering}
\centering{}(a)\hspace{8cm}(b)\caption{\label{fig:gam-Pcr} The plot of the critical value function $\gamma_{\mathrm{Pcr}}\left(\chi\right)$
defined by formula (\ref{eq:charP1b}) for : (a) $0.9<\chi<1$; (b)
$\chi>1$. Note that $\gamma_{\mathrm{Pcr}}\left(\chi\right)$ is
negative for $\chi>1$ implying that always $\gamma>0>\gamma_{\mathrm{Pcr}}\left(\chi\right)$.
We remind that the typical values of $\check{\gamma}$ are between
$2\cdot10^{-6}$ and $0.00675$ (see Remark \ref{rem:typical-gam}).}
\end{figure}

The value of $\gamma_{\mathrm{Pcr}}\left(\chi\right)$ can be used
as natural benchmark unit for given $\chi$. The Fig. \ref{fig:gam-Pcr}
tabulates the values if $\gamma_{\mathrm{Pcr}}$ for different values
of $\chi<1.$

\begin{rem}[typical values of the Pierce parameter and our TWT principal parameter]
\label{rem:typical-gam} For recent studies of the Pierce parameters
we refer the reader to \citep{Sim17}. In particular, according \citep[Sec. II, p. 3]{Sim17},
the typical values of the Pierce parameter $C_{\mathrm{P}}$ are between
$0.01$ and $0.15$. Then, in view of relations (\ref{eq:pierbet1e}),
the corresponding expected typical values of $\check{\gamma}$ are
between $2\cdot10^{-6}$ and $0.00675$.
\end{rem}

\textbf{\vspace{0.2cm}
}

\textbf{DATA AVAILABILITY:} The data that support the findings of
this study are available within the article.\textbf{\vspace{0.2cm}
}

\end{document}